\magnification1200
\input epsf

\rightline{KCL-MTH-98-55}
\rightline{hep-th/9811101}
\vskip 1cm
\centerline {\bf{{Supergravity, Brane Dynamics and String Duality}}${\
}^{\ddag}$}
\vskip 1cm
\centerline{\bf P.C. West }
\vskip .2cm
\centerline{Department of Mathematics}
\centerline{King's College, London, UK}

\vskip 1cm
\leftline{\sl Abstract}
 In this review we show  that a  Clifford algebra possesses  a
unique
 irreducible representation; the spinor representation.
 We discuss what types of spinors can exist in
Minkowski space-times and we explain how to construct all the
supersymmetry algebras that contain a given space-time Lie algebra.
After  deriving  the irreducible representations of the superymmetry
algebras,  we  explain how to use them to systematically construct
supergravity theories. We give the maximally
supersymmetric supergravity theories in
ten and eleven dimensions  and discuss their properties. We
find which superbranes can exist for a given supersymmetry algebra
  and we give the dynamics of the superbranes that occur in M theory.
Finally,  we discuss how the properties of supergravity theories and
superbranes provide evidence for string duality.
\par
In effect, we present a continuous chain of argument that begins  with
Clifford algebras and leads via  supersymmetry algebras and their
irreducible representations to  supergravity theories,
 string duality, brane dynamics  and M theory.

\vskip5cm

${\ }^{\ddag}$
This material is based on lectures presented at   the
EU conference on {\it Duality and Supersymmetric Theories}, the Issac
Newton Institute, Cambridge, UK and at the  TASI 1997 Summer School,
Boulder, Colorado, USA.

\vfill\eject

\medskip
\centerline {\bf { Contents}}
\bigskip
{\bf {0 Introduction}}
\medskip
\bigskip
{\bf {1 Clifford Algebras and Spinors}}
\medskip
\medskip
 {\bf {2 The Supersymmetry Algebra in Four dimensions}}
\medskip
\medskip
 {\bf {3 Irreducible Representations of Supersymmetry}}
\medskip
\medskip
{\bf {4 Three Ways to Construct a Supergravity Theory}}
\medskip
\medskip
{\bf {5 Eleven Dimensional Supergravity}}
\medskip
\medskip
 {\bf {6 IIA and IIB Supergravity}}
\medskip
\medskip
\ {\bf {7  Brane Dynamics}}
\medskip
\medskip
 {\bf {8  String Duality} }
\bigskip
\vfill\eject

\medskip
\centerline {\bf { 0. Introduction}}
\bigskip
In this review we begin with an account of Clifford algebras and
show that they each possess only one irreducible representation,
the spinor representation.
We discuss the types of spinors can exist in  Minkowski space-times.
Equipped with this knowledge, we show how to systematically construct
the supersymmetry algebras that contain a given space-time Lie group,
such as the Poincare group. In the context of the four-dimensional
supersymmetry algebras, we illustrate   how to find all irreducible
representations of the supersymmetry algebras and so arrive at a
listing of all possible supersymmetric theories.
\par
We then turn to the construction of theories of local supersymmetry,
that is supergravity theories and show how to derive
these theories from a knowledge of  the corresponding
irreducible representation of the supersymmetry algebra, (i.e. on-shell
states). Of the three   methods given for the construction of
supergravity theories two are rather systematic in that they always lead
to the desired theory with very little  additional information.
In section 5 and we give the unique supergravity theory in eleven
dimensions while in section 6 we describe that two maximally
supersymmetric theories in ten dimensions, the IIA and IIB
supergravity theories.  In addition to explaining how one constructs
these theories according to the methods given in section 4, we
describe the properties of these theories. These include the $SL(2,R)$
invariance of the IIB theory and the derivation of the IIA
supergravity theory from the eleven-dimensional supergravity theory by
compactification on a circle.  The ten-dimensional IIA and IIB
supergravity theories are the  low energy effective actions of IIA and
IIB string theories and  we discuss the consequences of this
relationship for string theories.  The eleven-dimensional
supergravity theory is thought to be the low energy effective action
of a yet to be clearly defined  theory called M theory.
\par
The supergravity theories admit solitonic  solutions to their
classical field  equations that correspond to static p-branes. A
p-brane is an object which sweeps out a $p+1$-dimensional surface as
it moves through space-time. They generalise strings which
 are 1-branes.  However,  from  the string theory perspective
p-branes for
$p>1$ are non-perturbative objects. In section 7, we give the
dynamics of p-branes  and discuss which p-branes can occur in the IIA,
IIB and M theories. In particular, we find that in M theory we can
have only twobranes and fivebranes and we give the equations of
motion of these branes.
\par
The supergravity theories in ten and eleven dimensions form the basis
for most discussions of string duality and in section 8 we  outline
some of these arguments. In particular we discuss the
non-perturbative string duality symmetries and the way they rotate
perturbative string states into the p-branes discussed in section 7.
\par
In effect this review traces a continuous chain of argument that
begins with Clifford algebras and leads via supersymmetry algebras
and  their representations to supergravity theories, superbranes  and
then to  string dualities and M theory.
\bigskip

{\centerline {\bf 1.Clifford Algebras and Spinors}}
\medskip
In this section we define a Clifford algebra in an arbitrary dimension and
find its irreducible representations and their properties. This enables us to
find which types of spinors are allowed in a given  Minkowski space-time.
\par
The starting point for the construction of
  supersymmetric  theories is    the supersymmetry algebra which underlies
it. Supersymmetry algebras contain supercharges which
transform as spinors under the appropriate Lorentz group.
Hence, even to construct the supersymmetry
algebras, as we do in the  section two, we must first find out
what types of spinors are possible in a given dimension and what
 are their properties. We will find in subsequent sections that
 supersymmetric algebras and the supersymmetric theories
on which they are based rely for their existence in an essential way on
 the detailed properties of
Clifford algebras, that we will derive in this section.
\par
As far as I am aware the first discussion of spinors in
arbitrary dimensions
was given in [100]
and many of the steps in this section
are taken from this paper. Use has also been made of  the
 reviews of references [101],  [102] and [194].
\medskip

{\bf 1.1. Clifford Algebras}

A Clifford algebra in $D$ dimensions is defined as a set
containing  $D$ elements   $\gamma_m$   which  satisfy the relation
$$\{\gamma_m,\gamma_n\}=\equiv \gamma_m \gamma_n + \gamma_n \gamma_m
=
2\eta_{mn}
\eqno(1.1.1)$$
where the labels  $m,n,\dots$ take
$D$ values and  $\eta_{mn}$ is the flat metric in $R^{s,t}\ (s+t=D)$;
that is the metric
$\eta_{mn}$ is a diagonal matrix whose first $t$ entries down the diagonal
are $-1$ and whose last $s$ entries are $+1$. We can raise and lower the
$m,n,\dots$
indices using the metric $\eta_{mn}= \eta^{mn}$ in the usual way.
\par
Under  multiplication the  $D$ elements  $\gamma_n$ of the
Clifford algebra  generate a    finite
group denoted  $C_D$ which  consists  of the elements
$${C}_D=\{\pm1,\pm\gamma_m,\pm\gamma_{m_1,m_2},
\dots,\pm\gamma_{m_1\dots m_D}\}
\eqno(1.1.2)$$
The  $\gamma_{m_1m_2\dots}$ is non-vanishing only if all indices
$m_1,m_2,\ldots$ are different in which case it equals
$$\gamma_{m_1m_2\dots}=\gamma_{m_1}\gamma_{m_2}\dots
\eqno(1.1.3)$$
The  set of matrices $\gamma_{m_1m_2\dots m_p}$ for all possible
different values of the $m$'s contains
$${D!\over(D-p)!p!}=\left({D\atop p}\right)
$$
different
elements. Hence the group $C_D$ generated by the $\gamma_m$  has order
$$2\sum^D_{p=0}\left({D\atop p}\right)=2(1+1)^D=2^{D+1}
\eqno(1.1.4)$$
\medskip
{\bf 1.2 Clifford Algebras in Even Dimensions }
\medskip
To find the representations of $C_D$ is a standard exercise
in representation theory of finite groups [116].
We will first consider the case of even $D$.
\par
The number
 of irreducible representations of any finite dimensional group, $G$
equals the number of its conjugacy classes. We recall that
the conjugacy class $[a]$ of $a\in G$ is given by
$$[a]=\{gag^{-1}\quad \forall \ g\in G\}
\eqno(1.2.1)$$
For even $D$ it is
straightforward to show, using equation (1.1.1), that the conjugacy classes
of $C_D$ are given by
$$[+1],[-1],[\gamma_m],[\gamma_{m_1m_2}],\dots,[\gamma_{m_1\dots m_D}]
\eqno(1.2.2)$$
Hence for $D$ even there are  $2^D+1$ inequivalent  irreducible
representations of $C_D$.
\par
Next we use the fact that the number of
inequivalent one-dimensional representations of any finite group $G$ is
equal to the order of
$G$ divided by the order of the commutator group
of $G$. We denote the commutator group of $G$ by   $Com(G)$. It is defined
to be the group $Com(G)= aba^{-1}b^{-1},\  \forall \ a,b\ \in \ G$.
For $D$ even
the commutant of $C_D$ is just the elements $\pm1$ and so has order 2.
As a result, the number of inequivalent irreducible one-dimensional
representations of $C_D$ is $2^D$. Since the total number of irreducible
representations is  $2^D+1$, we conclude that there is only one
irreducible representation whose dimension is greater that one.
\par
Finally, we make use of the theorem that if we denote the order of any
 finite group  by
$ord G$ and it has $p$ irreducible inequivalent representations of
dimension $n_p$ then
$$ ord G= \sum_p(n_p)^2
\eqno(1.2.3)$$
Applying this theorem to $C_D$ we find that
$$2^{D+1}=1^2 2^D+n^2
\eqno(1.2.4)$$
where $n$ is the dimension of the only irreducible representation
whose
dimension  is greater than one.
We therefore conclude that $n= 2^{D\over2}$.
These results are summarised in the following theorem.
\par
 {\bf {Theorem}}$\ \ $
For $D$ even the group $C_D$ has  $2^D+1$  inequivalent irreducible
representations.  Of these irreducible representations
$2^D$ are one-dimensional and the remaining representation has dimension
 $2^{D\over2}$.
\par
This means that we can represent the $\gamma_m$ as $2^{D\over2}$ by
$2^{D\over2}$ matrices for the irreducible representation with
dimension greater than one.  Our next task is to find the properties
of this representation under complex conjugation and transpose.
\par
That the above are  irreducible representations of the group $C_D$
means that they provide a  representation of the group which consists
of the elements  given in equation (1.1.2) together with a group
composition  law which is derived from  the Clifford algebra
relations using  only the operation  of multiplication. In
particular,  the group operations do not include the operations of
addition and subtraction which  also occur in the Clifford algebra
defining condition of equation  (1.1.1). Hence,  the irreducible
representations of $C_D$  are not necessarily
irreducible representations of the
Clifford algebra itself.   In fact,  all the one-dimensional
irreducible representations  of $C_D$ do not
extend to be also representations of the Clifford algebra as they do
not obey the rules for addition and subtraction. As such, the only
representation of $C_D$ and the
 Clifford algebra is the
 unique  irreducible representation of dimension greater than one
described above. It follows that  Clifford
algebra itself has only one irreducible representation and this has
dimension
$2^{{D\over 2}}$.
It is of course the well known spinor representation.
In fact, the one-dimensional representations are not
 faithful representations of $C_D$ and we shall not consider them in
what follows.
\par
Given an irreducible representation of the Clifford algebra, also denoted
$\gamma_m$, with dimension greater than one we can take its complex conjugate.
Denoting  the complex conjugate of the representation by
$\gamma_m^*$, it is obvious that $\gamma_m^*$ also satisfies equation (1.1.1)
and so form a representation of the same Clifford algebra.
It follows that they also form a representation
of $C_D$. However, there is only one  irreducible representation of $C_D$
of dimension greater than one and as a result, the
complex conjugate representation and the
original representation must be equivalent. Consequently,   there
exist a matrix $B$ such that
$$\gamma^*_m= B\gamma_mB^{-1}
\eqno(1.2.5)$$
We can choose the scale of $B$ such that $|\det B|=1$.
Taking the complex conjugate
of equation (1.2.5) we find that
$$\gamma_m=(\gamma^*_m)^*=+B^*B\gamma^mB^{-1}B^{-1*}
\eqno(1.2.6)$$
Hence, we conclude that $B^*B$ commutes with the
irreducible representation and by Schur's Lemma must be a constant
times the identity matrix, i.e.
$$B^*B=\epsilon I
\eqno(1.2.7)$$
Taking the complex conjugate of the above relation we find that
 $BB^*=\epsilon^*I$ and so
$BB^*BB^{-1}=\epsilon^*I=\epsilon I$ thus  $\epsilon=\epsilon^*$
i.e. $\epsilon$ is real. Since we have chosen $|\det B|
=1$ we conclude that  that $|\epsilon|=1$ and so $\epsilon=\pm1$.
\par
We can also consider the transpose of the irreducible representation  $\gamma_m$
 that has dimension greater than one. Denoting the transpose of $\gamma_m$ by
 $\gamma_m^T$, we find, using a very similar argument, that   $\gamma_m$ and
 $\gamma_m^T$ are equivalent representations and so
there exists a matrix $C$, called the charge conjugation
matrix, such that
$$\gamma^T_m=-C\gamma_mC^{-1}
\eqno(1.2.8)$$
\par
We denote  the Hermitian conjugate of $\gamma_m$ by
$\gamma_m^\dagger=\gamma^{*T}_m$.
We can relate $C$ to $B$ if we know the Hermiticity properties of the
$\gamma_ m$. For simplicity, and because this is the case of most
interest to us, from now on we assume that we are in a  Lorentzian
space-time whose
 metric $\eta_{mn}$  is given by
$\eta=$ diag $(-1,+1,+1,\dots,+1)$.  Any finite-dimensional
representation of a  finite group $G$  can be chosen to be unitary.
Making this choice for our group $C_D$ we have $\gamma_m \gamma_m
^\dagger=1$. Taking into account the relationship $\gamma_n \gamma_n
=\eta_{nn}$ we conclude that
$${\gamma_0}^{\dagger}=-\gamma_0,\quad\gamma^\dagger_m=\gamma_m;
\quad m=1,\dots,D-1
\eqno(1.2.9)$$
We could,  as  some texts do, regard this equation as part of the
definition of  the Clifford algebra.  We may rewrite  equation (1.2.9)
as
$$\gamma^\dagger_m=\gamma_0\gamma_m\gamma_0
\eqno(1.2.10)$$
We may take $C$ to be given by $C=- B^T\gamma_0$ as  then
$$
C\gamma_mC^{-1}= B^T\gamma_0\gamma_m(-\gamma_0 {((B)^T)}^{-1})
=- B^T\gamma_m^\dagger {(B)^T}^{-1}=-{(B^{-1}\gamma_m^*B)}^T
=-\gamma_m^T
\eqno(1.2.11)$$
as required.
\par
Further restrictions on $B$ can be found by
computing $\gamma_m^T$ in two ways:  we see that
$\gamma_m^T={(\gamma^ *_m)}^\dagger=(\gamma_m^\dagger)^*$ implies
$${(B^{-1})}^\dagger \gamma_0\gamma_m\gamma_0 B^\dagger
=B\gamma_0\gamma_m\gamma_0B^{-1}
\eqno(1.2.12)$$
Using Schur's Lemma we deduce that $-\gamma_0 B^\dagger B\gamma_0$
is proportional to the unit matrix and as a result so is
$B^\dagger B$, i.e. $B^\dagger B=\mu I$. Since $|\det B|=1$ we find
that
$|\mu|=1$, but taking the matrix element of
$B^\dagger B=\mu I$ with any vector we conclude that $\mu$ is real and
positive.
Hence $\mu=1$ and consequently  $B$ is unitary,
i.e. $B^\dagger B=I$.
This result and the previously derived equation $BB^*=\epsilon I$
 imply that
$$ B^T=\epsilon B,\quad C^T=-\epsilon C
\eqno(1.2.13)$$
\par
We now wish to determine $\epsilon$ in terms of the space-time dimension $D$.
Consider the set of matrices
$$I,\gamma_m,\gamma_{m_1m_2m_3},\gamma_{m_1m_2m_3},\gamma_{m_1\dots
m_D}.
\eqno(1.2.14)$$
There are $2^D=\sum_p\left({D\atop p}\right)=(1+1)^D$ such matrices and as they
are linearly independent they form a basis for the space of all $2^{D\over2}$
by $2^{D\over2}$ matrices.
Using equation (1.1.1) we can relate $\gamma_{m_1\dots m_p}$ to
$\gamma_{m_p\dots m_1}$, to find that the sign change required to
reverse the order of
the indices is given by
$$\gamma_{m_1\dots m_p}=(-1)^{p{(p-1)\over2}}\gamma_{m_p\dots m_1}.
\eqno(1.2.15)$$
This equation together with equation (1.2.11) imply that
$$
C\gamma_{m_1\dots m_p}C^{-1}=(-1)^p(-1)^{p{(p-1)\over2}}
{\gamma_{m_1\dots m_p}}^T
\eqno(1.2.16)$$
or equivalently
$${(C\gamma_{m_1\dots m_p})}=\epsilon(-1)^{(p-1)(p-2)\over2}{(C\gamma_{m_1\dots
m_p})}^T
\eqno(1.2.17)$$
Using this result we can calculate the
 number of anti-symmetric matrices in
the complete set of equation (1.2.14)
 when multiplied from the left by $C$;
It is given by
 is given by
$$\sum^D_{p=0}{1\over2}\left(1-\epsilon(-1)^{(p-1)(p-2)\over2}\right)\left({D
\atop p}\right)
\eqno(1.2.18)$$
Using the relationship
$$(-1)^{(p-1)(p-2)\over2}=-{1\over2}\big[(1+i)i^n+(1-i)(-i)^n\big],
\eqno(1.2.19)$$
we can carry out the sum in equation (1.2.18).
 We know, however, that the number of anti-symmetric
matrices is $2^{{D\over 2}}
(2^{{D\over 2}}-1){1\over 2}$. Equating these two methods of evaluating the
number of antisymmetric matrices  we  find that
$$\epsilon=-\sqrt2\cos{\pi\over4}(D+1).
\eqno(1.2.20)$$
Put another way  $\epsilon=+1$ for $D=2,4$; mod 8
and $\epsilon=-1$ for $D=6,8$; mod 8.
It follows from equation (1.2.13) that for  $D=2,4$; mod 8, $B$
is a symmetric unitary matrix. Writing $B$ in terms of its
real and imaginary parts $B=B_1+iB_2$ where $B_1$ and $B_2$ are symmetric and
real, the unitarity condition becomes $B^2_1+B^2_2=1$ and $[B_1,B_2]=0$. Under
a change of basis of the $\gamma_m$ matrices;
$\gamma^{m\prime}=A\gamma^nA^{-1}$ we
find that the matrix $B$ changes as $B^\prime=A^*BA^{-1}$.
In fact, we can use $A$ to diagonalize $B_1$ and $B_2$
which, still  being unitary, must be   of the form
$B= diag(e^{i\alpha_1},\dots,e^{i\alpha_D})$. Carrying
out another $A$ transformation of the form $A= diag (e^{i{\alpha_1\over2}},
\dots,e^{i{\alpha_D\over2}})$ we find the new $B$ equals one. Hence if
$D=2,4$; mod 8 the $\gamma_m$ matrices can be chosen to be real and
$C=\gamma^0$.
\medskip

{\bf 1.3 Spinors in Even Dimensions}
\medskip
By definition a  spinor $\lambda$ transforms under $Spin(1,D-1)$ as
$$\delta\lambda={1\over4}w^{mn}\gamma_{mn}\lambda
\eqno(1.3.1)$$
where $w^{mn}=-w^{nm}$ are the parameters of the Lorentz transformation.
The group $Spin(1,D-1)$ is by definition the group generated by
${1\over4}w^{mn}\gamma_{mn}$ and it is the covering group of
$SO(1,D-1)$. The Dirac
conjugate denoted $\bar\lambda^D$ must transform such that
$\bar\lambda^{D}
\lambda \equiv \bar\lambda^{D\alpha}
\lambda_\alpha$ is invariant and so transforms under a Lorentz
transformation as
$$\delta\bar\lambda^D=\bar\lambda^D\left(-{1\over4}w^{mn}\gamma_{mn}\right).
\eqno(1.3.2)$$
Using the relation $\gamma_{mn}^\dagger=-\gamma_0\gamma_{nm}\gamma_0
=\gamma_0\gamma_{mn}\gamma_0$ we find that
$$\delta(\lambda^+\gamma^0)=(\lambda^+\gamma^0)\left(-{1\over4}w^{mn}\gamma_
{mn}\right)
\eqno(1.3.3)$$
Consequently, we can take the Dirac conjugate to be defined  by
$$\bar\lambda^D\equiv\lambda^\dagger\gamma^0
\eqno(1.3.4)$$
The Majorana conjugate, denoted $\bar\lambda^M$ is defined by
$$\bar\lambda^m=\lambda^T C.
\eqno(1.3.5)$$
Using the relationship
$\gamma_{mn}^T=C\gamma_n\gamma_mC^{-1}=-C\gamma_{mn}C^{-1}$
we find that
$$\delta\bar\lambda^M= \lambda^T{1\over4}w^{mn}\gamma_{mn}^T C
=-\lambda^T
C\left({1\over4}w^{mn}\gamma_{mn}\right)
=\bar\lambda^M\left(-{1\over4}w^{mn}\gamma_{mn}\right)
\eqno(1.3.6)$$
Hence the  Majorana conjugate transforms like the Dirac conjugate
 under $Spin(1,D-1)$ transformations and as a result we can define
a  Majorana spinor to be one is  whose Dirac and Majorana conjugates are equal:
$$\bar\lambda^D=\bar\lambda^M
\eqno(1.3.7)$$
The above
condition can be rewritten as $\lambda^*=-{(\gamma^0)}^T C^T\lambda$
and,  using the relation $C= B^T\gamma^0$, it  becomes
$$\lambda^*=B\lambda
\eqno(1.3.8)$$
We could have directly verified that $B^{-1}\lambda^*$ transforms
under $Spin(1,D-1)$  in the same way as
 $\lambda$
by using the equation
$B\gamma_{mn}B^{-1}= {(\gamma_{mn})}^*$
and as a result have imposed this Majorana condition without any mention of
 the Dirac conjugate.
\par
We are finally in a position to discover the  dimensions in which
Majorana spinors exist. If we impose the relationship $\lambda^*=B\lambda$,
then, taking the complex conjugate we find that it implies the relationship
$\lambda=B^*\lambda^*$. Substituting this condition into the first relation
we find that
$$\lambda=B^*B\lambda=\epsilon\lambda
\eqno(1.3.9)$$
since $B^*B=\epsilon I$. Consequently, Majorana spinors can only exist if
$\epsilon=+1$, which is the case only in the dimensions $D=2,4$; mod
8, i.e.
 $D=2,4,10,12\dots$.
\par
In an even-dimensional space-time we can construct the matrix
$$\gamma^{D+1}=\gamma_0\gamma_1\dots\gamma_{D-1}=
\gamma_{0 1 \dots D-1}
\eqno(1.3.10)$$
This matrix anticommutes with $\gamma_m$ and so commutes with
the generators
$\left(-{1\over4}\gamma_{mn}\right)$ of spin$(1,D-1)$,
the covering group of
$SO(1,D-1)$. Hence $\gamma^{D+1}\chi$ transforms like a spinor
if $\chi$
does. A
straightforward calculation shows that
$$(\gamma^{D+1})^2=(-1)^{D(D-1)\over2}(-1)=(-1)^{{D\over2}-1}
\eqno(1.3.11)$$
Hence  $(\gamma^{D+1})^2=1$ for $D=2$ mod 4 while  $(\gamma^{D+1
})^2=-1$ for $D=4$ mod 4. In either case we can define Weyl spinors
$$\gamma^{D+1}\chi=\pm\chi\qquad{\rm if}\ D=2\ {\rm mod}\ 4
\eqno(1.3.12)$$
and
$$i\gamma^{D+1}\chi=\pm\chi\qquad{\rm if}\ D=4\ {\rm mod}\ 4
\eqno(1.3.13)$$
We can now consider when Majorana-Weyl spinors exist.
We found that Majorana
spinors (i.e. $\chi^*=B\chi$ ) exist if $\epsilon=1$, i.e. when
$D=2,4$; mod 8. Taking the complex conjugate of the above  Weyl
conditions and using the relationship ${(\gamma^{D+1})}^*=
B\gamma^{D+1}B^{-1}$ we find we get a
non-vanishing solution  only   if  $D=2$ mod 4.
Hence Majorana-Weyl spinors only exist if $D=2$ mod 8 i.e.
$D=2,10,18,26,\dots$. The  factor of $i$ is necessary for $D=4$ mod 4 as the
chirality condition must have an operator that squares to one, however it is
this same
factor of $i$ that gets a minus sign under
 complex conjugation and so rules out the possibility of having
Majorana-Weyl spinors in these dimensions.
 We note that
these are the  dimensions in which self-dual Lorentzian
lattices exist and, except for 18 dimensions,
these  are the dimensions in which
critical strings exist.
\par
Corresponding to the above chiral conditions
  we can defined projectors onto the spaces of
positive and negative chiral spinors. These projectors are given by
$ P_{\pm}= {1\over 2}(1\mp a \gamma ^{D+1})$ where $a=1$ if $D=2 $ mod 4 and
$a=i$ if $D=4 $ mod 4. It is easy to verify that they are
indeed projectors;
 i.e.
$P_{\pm}P_{\mp}=0, \
P_{\pm}^2= P_{\pm}, \ P_{\mp}^2 = P_{\mp}$ and $P_{\pm} + P_{\mp}=1$.
Under complex conjugation the projectors transform as
$$ P_{\pm}^*= \cases {BP_{\pm}B^{-1}, & if $D=2$ mod 4
    \cr BP_{\mp}B^{-1}, & if $D=4$ mod 4\cr}
\eqno(1.3.14)$$
This equation  places restrictions on the form that the matrices
$B$ and the
chiral projectors can take. For example, let us write  the
 $2^{D\over 2}$ by $2^{D\over 2}$ $\gamma$-matrices in terms of
$2^{{D\over 2}-1}$ by $2^{{D\over 2}-1}$ blocks. We also  choose our
basis of spinor such that the projection operators are diagonal and such that
$P_{+}$ has only its upper diagonal block non-vanishing and
equal to the identity matrix
and $P_{-}$ with only its lower diagonal block non-zero and equal to the
identity matrix space.  Applying  equation (1.3.14), we find
 that  if $D=2$ mod 4, the matrix $B$ has only its
two diagonal blocks non-zero
 and  if $D=0$ mod 4, only its  off-diagonal blocks non-zero.
\par
 Under  complex conjugation the chiral spinors transform as
$$ B^{-1}{( P_{\pm}\lambda )}^*= \cases {P_{\pm}B^{-1}\lambda ^*,
& if $D=2$ mod 4 \cr
P_{\mp}B^{-1}\lambda^*, & if $D=4$ mod 4\cr}
\eqno(1.3.15)$$
Hence, complex conjugation and multiplication by $B^{-1}$
relates the same chirality spinors if
$D=2$ mod 4 and opposite chirality spinors  if $D=4$ mod 4. As such, in the
dimensions $D=2,4$; mod 8 where we can define the Majorana  spinors,
the Majorana condition relates
 same chirality spinors if $D=2$ mod 4 and opposite chirality spinors
if $D=4$ mod 4.
\par
We now investigate how a matrix transformation  on $\lambda $ acts on its
 chiral components.
 Under the matrix transformation $\lambda \to A\lambda $ we find  that
$B^{-1}\lambda^* \to (B^{-1}A^*B)B^{-1}\lambda^* $ and so the equivalent
transformation on $B^{-1}\lambda^* $
is $B^{-1}A^*B$. Clearly if $A$
is a polynomial in the $\gamma$-matrices with real coefficients
then this transformation is the same on $\lambda $ and
$\lambda ^*$. As
we have already discussed this is the case with Lorentz transformations
which are generated by $J^{mn}= {1\over 2} \gamma ^{mn}$.
However, if $A=EP_{\pm}$ where $E$
is a polynomial in the $\gamma$-matrices with real coefficients then
the transformation becomes
$$ B^{-1}{(E P_{\pm})}^*B= \cases {CP_{\pm},
& if $D=2$ mod 4 \cr
P_{\mp}, & if $D=4$ mod 4\cr}
\eqno(1.3.16)$$
As an example of the latter let us consider the chiral projections of
 Lorentz transformations  which are given by
$J^{mn}_{\pm}\equiv   {1\over 2} \gamma ^{mn}P_{\pm}$.  In this case the
above equation becomes
$B^{-1}{ J^{mn}_{\pm}}^*B =  J^{mn}_{\pm}$   for  $D=2$ mod 4 and
$B^{-1}{ J^{mn}_{\pm}}^*B =  J^{mn}_{\mp}$   for  $D=4$ mod 4.
Hence for  $D=2$ mod 4
we find that the representation  generated
by ${ J^{mn}_{\pm}}$ generate  a representation which for a given chirality
is conjugate to the  complex conjugate representation
of the same chirality.
 On the other hand,  for  $D=4$ mod 4 we find
that the complex conjugate of the
chiral Lorentz transformations of a given chirality is conjugate
to that for the opposite chirality.
 In  dimensions $D=2,4$; mod 8 we can choose $B=1$ and it is
straightforward to interpret the corresponding constraints.
For example, if $D=2$ mod 8 then ${(J^{mn}_{\pm})}^*=
J^{mn}_{\pm}$ and the representation they generate is contained
in the group $SL(2^{{D\over 2}-1},{\bf R})$.
\par
 However, for   $D=6,8$; mod 8 the matrix $B$
is anti-symmetric and so can not be chosen to be one. If  in addition
we take $D=2$ mod 4 that is  $D=6$ mod 8 then  then we conclude that the chiral
Lorentz transformations are contained in  the group  $SU^*(2^{{D\over 2}-1})$.
The group $SU^*(N) $ is the group of $N$ by $N$ complex matrices
of determinant one that commute with the operation of complex conjugation
and multiplication by an anti-symmetric matrix   $B$
which  obeys $B^\dagger B=1$ [103].
That is if $A\in \ SU^*(N)$ then
$BAB^{-1}= A^*$. Taking such an infinitesimal transformation
$A= I+K$ we find that $SU^*(N)$
 has real dimension $N^2-1$.
The first such case is
 $D=6$, since $SU^*(4)$ and $Spin(1,5)$ both have dimension 15
we must conclude that  the group of six-dimensional
chiral Lorentz transformations generated by $J^{mn}_{\pm}$ is
isomorphic to the group $SU^*(4)$.
\par
If the spinors carry internal spinor indices that transform  under a
pseudo-real representation of an internal group then we can also
 define
 a kind of Majorana
spinor when $D=6,8$; mod 8 by the condition
$${(\lambda _i)}^*= \Omega ^{ij}B\lambda _j
\eqno(1.3.17)$$
In this equation  $ \Omega ^{ij}$ is a real anti-symmetric
matrix which also obeys the relations
$\Omega ^{ij}=\Omega _{ij}$ and
$\Omega ^{ij}\Omega _{jk}=-\delta ^i_k$.
We call spinors that satisfy this type of  Majorana condition
 symplectic Majorana spinors.
 Taking the complex conjugate of this symplectic  Majorana
condition, using the above  relations and
 the fact that $B^*B=-1$
we find that it is indeed a consistent condition. The symplectic Majorana
condition  should also
be such that the internal group, which acts on the
internal indices $i,j,\dots$, acts in the same way on the
left and right hand sides of the symplectic Majorana condition. This requires
$\lambda_i$ to carry  a pseudo-real representation of the internal group.
\par
The  vector representation of the
group $USp(N)$ provides one of the most important examples of  a
pseudo-real representation.
The group $USp(N)$ consists of
unitary matrices that in addition preserve $ \Omega ^{ij}$;  that is matrices
$A$ which satisfy $A^\dagger A=1$ and $A^T\Omega A=\Omega$.
Taking such an infinitesimal
transformation we find that this group has dimension
${1\over 2}N(N+1)$. Under $\lambda \to A\lambda$, we find that
$\lambda^* \to A^*\lambda^*$, but $\Omega \lambda \to \Omega A\lambda
=-\Omega A\Omega \Omega \lambda$.
However, using the defining conditions of $USp(N)$ we can show that
$$-\Omega A\Omega = -{(A^T)}^{-1}\Omega \Omega = {(A^T)}^{-1}
= {(A^\dagger)}^{T}= A^{*}
\eqno(1.3.18)$$
which means that the symplectic Majorana condition preserves $USp(N)$.
We note that $USp(2)= SU(2)$.
\par
In addition to the symplectic
Majorana condition of equation (1.3.17) which requires
$D=6,8$; mod 8 we can also impose a Weyl constraint if $D=2$ mod 4.
That is dimensions $D=6$ mod 8 symplectic Majorana-Weyl spinors exist.
\par
An important example of a symplectic Majorana-Weyl spinors
are those in six dimensions that transform under $USp(4)$.
These spinors naturally arise when we reduce that Majorana
eleven-dimensional spinors  to six dimensions. The eleven-dimensional spinors
transform under $Spin(1,10)$ which under the reduction to six dimensions
becomes $Spin(1,5)\times Spin(5)$. The $Spin(5)$ which is isomorphic to
$USp(4)$ and becomes the internal group in six dimensions. In fact we
get  symplectic Majorana-Weyl spinors of both chirality each of which has
$4\times 4 =16$ components.
\medskip
{\bf 1.4 Clifford Algebras in Odd Dimensions}
\medskip
We now take the dimension of space-time $D$ to be odd.
The group $C_D$ of equation (1.1.2) is generated by the $\gamma_n$ and
has order $2^{D+1}$. The  irreducible representations
can be found using the same arguments as we did
for the case of a space-time of even dimensions.
However, there are some differences
which are a consequence of the fact that the conjugacy
classes are not
given by the obvious generalisation of those for
 even-dimensional case which were listed in equation (1.2.2).
From all the $\gamma$-matrices, $\gamma
_m,\ m=0,1,\dots,D-1$ we can form the matrix
$ \gamma_D \equiv \gamma_0\gamma_1\dots\gamma_{D-1}$. This matrix
commutes with all the $\gamma_m\ m=0,1,\dots,D-1$ and so all products
of the $\gamma_m$. As such, $\pm \gamma_D$ form conjugacy
classes by themselves and as a result  the full
list of  conjugacy classes
is given by
$$[1],[-1],[\gamma_m],[\gamma_{m_1m_2}],\dots, [\gamma_{m_1\dots m_D}],
[-\gamma_{m_1\dots m_D}]
\eqno(1.4.1)$$
There are  $2^{D}+2$ conjugacy classes and so $2^{D}+2$
inequivalent irreducible representations of $C_D$.
\par
The commutator group of $C_D$ is  given by $\{\pm1\}$ and so has order 2.
As such, the number of
inequivalent irreducible one-dimensional representations of $C_D$  is $2^{D}$.
Hence, in an odd-dimensional space-time we have two inequivalent irreducible
representations of $C_D$ of dimension greater than one.
In either of these two irreducible representations,
the matrix $\gamma_D$ commutes with the entire representation
and so by Schur's Lemma must be a multiple of the identity i.e.
$\gamma_D= a^{-1}I$ where $a$ is a constant. Multiplying both sides
by $\gamma_{D-1}$ we find the result
$$\gamma_{D-1}=a \gamma_0\gamma_1\dots\gamma_{D-2}=a\gamma_{01\ldots D-2}
\eqno(1.4.2)$$
Using equation (1.3.11), we conclude  that
$(\gamma_{01\ldots D-2})^2=-(-1)^{{(D-1)}\over 2}$ as the matrix
$\gamma_{01\ldots D-2}$ is the same as that denoted by $\gamma^{D+1}$ for the
even-dimensional space-time with one dimension lower.
However, as  $\gamma^2_{D-1}=+1$
we must conclude that $a=\pm1$  for $D=3$ mod 4  and $a=\pm i$ for $D=5$ mod 4.
The $\gamma_m;\ m=0,1,\dots,D-2$  generate an even-dimensional Clifford
algebra and we recall that the corresponding subgroup $C_{D-1}$  has
 a unique irreducible representation of dimension greater than one,
the dimension being $2^{{(D-1)\over 2}}$.
It follows that      the
two irreducible representations for $D$ odd which have dimension greater than
one must coincide with this irreducible
representation when restricted to $C_{D-1}$.
Hence, the two inequivalent irreducible representations for $D$ odd
are generated by the unique irreducible
representation for the $\gamma_m,\ m=0,1,\dots,D-2$, with the remaining
$\gamma$-matrix being given by
$\gamma_{D-1}=a \gamma_0\gamma_1\dots\gamma_{D-2}$.
The two possible choices of $a$,
given above, corresponding to the two inequivalent irreducible representations.
 Clearly, these two inequivalent irreducible representations both
 have dimension $2^{{(D-1)}\over2}$ as this is the dimension of the unique
irreducible representation with dimension greater than one in
the space-time with one dimension less. We can check that this is consistent
with the relationship
between the order, $2^{D+1}$, of the group and
the sum of the dimension
squared of all irreducible representations. The latter is given by
$1^2.2^{D}+(2^{{(D-1)}
\over2})^2+{(2^{({(D-1)}\over2})}^2=2^{D+1}$ as required.
\par
We now extend the complex conjugation and transpose
properties discussed previously for
even-dimensional
space-time to the  case of an odd-dimensional space-time.
Clearly, for the matrices
$\gamma_m,\ m=0,1,\dots,D-2$, these properties  are
the same and are given in equations (2.1.5) and (2.1.8).
It only remains to consider
$\gamma_{D-1}=a\gamma_0\gamma_1\dots\gamma_{D-2}
\equiv a \gamma_{01\ldots D-2}$. It follows
from the previous section that
$$
\gamma_{01\ldots D-2}^*=B\gamma_{01\ldots D-2}B^{-1}
\eqno(1.4.3)$$
and
$$C\gamma_{01\ldots D-2} C^{-1}=(-1)^{{{(D-1)}\over2}}\gamma_{01\ldots D-2}^T
\eqno(1.4.4)$$
We may also write this last equation as
$$C \gamma_{01\ldots D-2}
=-\epsilon(-1)^{{{(D-1)}\over2}}{(C\gamma_{01\ldots D-2})}^T
\eqno(1.4.5)$$
as $C^T=-\epsilon C$.
Taking into account the different possible values of $a$ discussed
above
 we conclude that
$$
\gamma^*_{D-1}=-(-1)^{{(D-1)}\over2}B\gamma_{D-1}B^{-1}
\eqno(1.4.6)$$
and
$$\gamma_{D-1}^T=(-1)^{{{(D-1)}\over2}}C\gamma_{D-1}C^{-1}
\eqno(1.4.7)$$
As we did for the even-dimensional case we can adopt the choice
 $C= B^T\gamma^0$, whereupon
we find that $\gamma^\dagger_{D-1}=\gamma^0\gamma_{D-1}
\gamma^0$. The representation is automatically unitary as a consequence of
being
unitary on the $C_{D-1}$ subgroup.
\par
For $D=3$ mod 4 $\gamma_{D-1}$ has the same relationships under complex
conjugation and transpose as do the $\gamma_m,\ m=0,1,\dots,D-2$ and as
 a result for  $D=3$ mod 4
$${(C\gamma_{m_1\dots m_p})}^T
=\epsilon(-1)^{(p-1)(p-2)\over2}(C\gamma_{m_1\dots
m_p})\quad m_1,\dots,m_p=0,\dots,D-1
\eqno(1.4.8)$$
For $D=5$ mod 4, we get an additional minus sign in this relationship
if one of the $m_1,\dots,m_p$ takes the value $D-1$.
\par
Let us now consider which types of spinors can exist in odd
dimensional space-times. Clearly in odd-dimensional space-times
the Weyl condition
is not a Lorentz invariant condition and so one cannot define such spinors.
 However, we can ask which odd-dimensional
space-times have Majorana spinors, which we take to be defined by
$\chi^*=B\chi$. Since  either of
the two inequivalent irreducible
representations  coincides
with the unique   irreducible representation of dimension greater than one
when restricted to the subgroup $C_{D-1}$, the matrix $B$   is
the same  as in the even-dimensional case. It follows that
 we require $\epsilon=+1$ which is the case for D=3,5 mod 8.
 We must, however, verify that the Majorana condition
is preserved by all
Lorentz transformations. Those that  are generated by
${1\over2}\gamma_{mn},\ m, n =0,1\ldots ,D-2 $ are guaranteed to
work; however,
carrying out the Lorentz transformation
$\delta\chi={1\over 4}\gamma_{m{D-1}}\chi, m=0,1\ldots ,D-2$ we find
it  preserves the Majorana constraint only
if $D=3$ mod 4.
Hence, Majorana spinors exist in  odd $D$-dimensional space-time if
$D=3$ mod 8. We note that these odd dimensions are precisely
one dimension  higher  than  those where
Majorana-Weyl spinors exist.
 This is not a coincidence as the reduction of a Majorana spinor in
$D=3$ mod 8 dimensions leads to two Majorana Weyl spinors of opposite
chirality and, since the resulting matrix $\gamma_{D+1}$, is real we may Weyl
project to find a Majorana-Weyl spinor.
\medskip
{\bf 1.5 Central Charges}
\medskip
One important application of the above theory is to
find what central charges
can appear in a supersymmetry algebra. That is what generators can appear
in the anti-commutator $\{Q_\alpha,\ Q_\beta\}$ where $Q_\alpha$ is the
generator of supersymmetry transformations. As we shall see, the
result  depends on the dimension of space-time and on
 whether  the  spinor $Q_\alpha$ is Weyl or Majorana-Weyl.
To begin with we take the dimension of space-time to be even.
\par
The right-hand side of the anti-commutator of the supercharges takes
the form [117]
$$ \{Q_\alpha,\ Q_\beta\} = (\gamma_{m}C^{-1})_{\alpha \beta}P^m+
\sum (\gamma_{m_1\dots m_p}C^{-1})_{\alpha \beta}
Z^{m_1\dots m_p}
\eqno(1.5.1)$$
where $Z^{m_1\dots m_p}$ are the central charges and  $P^m$ is
  the generator of translations. The sum is over all possible
central terms.
Clearly, the matrix $(\gamma_{m_1\dots m_p}C^{-1})_{\alpha \beta}$ must be
symmetric
in $\alpha, \beta$. Examining equation (1.2.17) we find that this will
be the case if
$$\epsilon {(-1)}^{(p-1)(p-2)\over 2}=1
\eqno(1.5.2)$$
\par
For $D=2,4$; mod 8, $\epsilon=1$ and so we find central charges for
$p=1,2$; mod 4. In these dimensions we can define Majorana spinors and
adopting this constraint still allows these central charge
although the Majorana condition will place reality conditions on
them.
For $D=6,8$; mod 8 $\epsilon=-1$ and so we find central charges of
rank $p$ for
$p=3,4$ mod 4.
\par
Let us now consider the case when the spinors are Weyl spinors that is satisfy
${(P_{\pm}Q)}_\alpha =0$. in this case we must modify
the terms on the right hand-side of the anti-commutator to be given by
$$ \{Q_\alpha,\ Q_\beta\}
= (\gamma_{m_1\dots m_p}P_{\pm}C^{-1})_{\alpha \beta}
Z^{m_1\dots m_p}
\eqno(1.5.3)$$
In this case $ (\gamma_{m_1\dots m_p}C^{-1})_{\alpha \beta}$ and
$ (\gamma_{m_1\dots m_p}\gamma^{D+1}C^{-1})_{\alpha \beta}$ must be symmetric.
However, the latter matrix is equal to $ \epsilon$ times
$ (\gamma_{m_1\dots m_{D-p}}C^{-1})_{\alpha \beta}$. Hence central charges of
 rank $p$ are possible if $p=1,2$; mod 4 and $D-p=1,2$; mod 4.
\par
Finally, we can consider Majorana-Weyl spinors which only exist in
$D=2$ mod 8. In this case we have the condition on $p$ of
the Weyl case above,
which must be taken with $D=2+8n$,for $n\in\  {\bf Z}$, allows only
central charges of rank $p=1$ mod 4. An example of this latter case is
provided by
$N=1$ $D=10$ which, if we take Majorana-Weyl spinors, is the algebra
that underlies the $N=1$ Yang-Mills theory and the type I supergravity
theory
 that exists in
ten dimensions. The algebra
is then given by
$$ \{Q_\alpha,\ Q_\beta\} = (\gamma_{m}P_{\pm}C^{-1})_{\alpha \beta}
P^{m}+
(\gamma_{m_1\dots m_5}P_{\pm}C^{-1})_{\alpha \beta}
Z^{m_1\dots m_5}+
(\gamma_{m_1\dots m_9}P_{\pm}C^{-1})_{\alpha \beta}
Z^{m_1\dots m_9}
\eqno(1.5.4)$$
Clearly $P^m$ is the usual generator of translations.
\par
The above discussion can be generalised to the case of odd-dimensional
space-times although in this case we do not have Weyl spinors. For the
case of $D=3$ mod 4, $\gamma_{D-1}$
behaves exactly like the other $\gamma$-matrices under complex conjugation and
transpose and as a result we find from equation (1.4.8) precisely
the same condition for
the existence of central charges i.e. equation (1.5.2) with the value of
$\epsilon$ being the same as  that in one dimension less.
The case of $D=5$ mod 4 can be deduced in a similar way by taking
into account the discussion below equation (1.4.8).
\par
A very important
example is the supersymmetry algebra in eleven dimensions
for Majorana supercharges.
 This algebra underlies the supergravity theory in this dimension.
For this algebra   we can have central charges
of rank $p$ with $p=1,2$; mod 4 and so the supersymmetry algebra takes
the form
 [117].
$$
\{Q_\alpha,Q_\beta\}
=(\gamma^mC^{-1})_{\alpha\beta}P_m+(\Gamma^{mn}C^{-1})_{\alpha\beta}
Z_{mn}+(\Gamma^{mnpqr}C^{-1})_{\alpha\beta}Z_{mnpqr}
\eqno(1.5.5)$$
We need only go up to rank five thanks to the identity
$$\epsilon^{m_1\dots m_p m_{p+1}\dots m_{D-1}}\gamma_{n_{p+1}\dots n_{D-1}}
\propto\gamma^{m_1\dots m_p}
$$
which is true in all odd-dimensional spaces.
\par
To find the possible central charges is the obvious part of constructing the
supersymmetry algebra we must also find their relations with the rest of the
generators of the algebra. This involves enforcing the super Jacobi identities
as explained in the next section.
\par
It is straightforward to extend the discussion to supersymmetry algebras
which contain supersymmetry generators $Q_\alpha^i$
that carry a representation of
an internal algebra. The anti-commutator $\{ Q_\alpha^i,Q_\beta^j\}$ is now
symmetric
under interchange in $\alpha, i$ and $\beta, j$.
\medskip
For an account of  Clifford algebras and spinors in non-Lorentzian
space-times the reader is encouraged to consult reference [102].
\bigskip

{\centerline {\bf {2 The Supersymmetry Algebra in Four dimensions}}}
\medskip

The starting point for the construction of
 any   supersymmetric  theory which is invariant under either rigid or
local supersymmetry  is the supersymmetry algebra which underlies it.
  In section this section
we demonstate how to systematically
 construct supersymmetry algebras from some very mild assumptions.
\par
Supersymmetry algebras contain  generators, called supercharges,
that
are Grassmann odd, transform under the Lorentz group as  spinors and
obey anti-commutation relations.   It is a remarkable fact that
supersymmetric algebras   can be constructed from very little
information.
 We must specify the    spinorial character  of the
supercharges  and what space-time Lie algebra is contained in
 the supersymmetry algebra. In  section one,  we deduced
the  possible spinors that  can exist in a given dimension
and we will see that the different choices lead to different
superysymmetry algebras. The most important of the space-time Lie
subalgebras is the Poincare algebra, but other possible choices
include  the de-Sitter algebra  or the conformal algebra.  Given a
 choice of the spinorial character of the supercharges and the
space-time Lie subalgebra,  the deduction of the corresponding
supersymmetry algebra relies on a generalisation of the Jacobi
identites that occur in Lie algebras.
\par
In this section, we show how to systematically construct  the
supersymmetry algebras in four dimensions. In particular,  we
demonstrate how to construct all supersymmetry algebras that contain
the Poincar\' e Lie algebra. We also state  the  supersymmetry algebras
that contain the conformal and de Sitter Lie algebras.
The generalisation to the sytematic construction of supersymmetry
algebras in higher dimensions   is
straightforward. The only difference is that the no-go theorem
must be used with more caution as its proof contains a
number of assumptions that are not valid for higher-dimensional
theories.
\medskip
This material is essentially the same as
that given in reference  [0] and we have kept the same equation
numbers as in that reference. We thank World Scientific publishing
for their kind permission to reproduce this material.
\medskip
In the 1960's, with the growing awareness of the significance of
internal symmetries such as $SU(2)$ and larger groups, physicists
attempted to
 find a symmetry which would combine in a non-trivial way the
space-time Poincar\'e group with an internal symmetry group. After
much effort it was shown that such an attempt was impossible within
the context of a Lie group. Coleman and Mandula$^4$ showed on very
general assumptions that any Lie group which contained the
Poincar\'e group $P$, whose generators $P_a$ and $J_{ab}$ satisfy the
 relations
$$\eqalignno{[P_a,P_b]&=0\cr
[P_a,J_{bc}]&=(\eta_{ab}P_c-\eta_{ac}P_b)\cr
[J_{ab},J_{cd}]&=-(\eta_{ac}J_{bd}+\eta_{bd}J_{ac}-\eta_{ad}J_{bc}-\eta_{bc}J_
{ad})&(2.1)}$$
and an internal symmetry group $G$ with generators $T_s$ such that
$$[T_r,T_s]=f_{rst}T_t\eqno(2.2)$$
must be a direct product of $P$ and $G$; or in other words
$$[P_a,T_s]=0=[J_{ab},T_s]\eqno(2.3)$$
They also showed that $G$ must be of the form of a semisimple group
with additional $U(1)$ groups.

It is worthwhile to make some remarks concerning the status of this
no-go theorem. Clearly there are Lie groups that contain
the Poincar\'e group and internal symmetry groups in a non-trivial
manner; however, the theorem states that these groups lead to trivial
physics. Consider, for example, two-body scattering; once
we have imposed conservation of angular momentum and momentum the
scattering angle is the only unknown quantity. If there were a Lie
group that had a non-trivial mixing with the Poincar\'e group then
there would be further generators associated with
space-time. The  resulting conservation laws will
further constrain, for example,  two-body scattering, and so the
scattering angle can only take on  discrete values. However, the
scattering process is expected to be  analytic in  the scattering
angle,
$\theta$, and hence we must  conclude  that the process does not
depend on $\theta$ at all.

Essentially the theorem shows that if one used a Lie group that
contained an internal group which mixed in a non-trivial manner with
the Poincar\'e group then the S-matrix for all processes would be zero.
The theorem assumes among other things, that the S-matrix exists
and is non-trivial, the vacuum is non-degenerate and that there are
no   massless particles. It is important to realise that the
theorem only applies to symmetries that act on S-matrix elements and
not on all the other many symmetries that occur in quantum field
theory. Indeed it is not uncommon to find examples of the latter
symmetries. Of course, no-go theorems are only as strong as the
assumptions required to prove them.

In a remarkable paper Gol'fand and Likhtman$^1$ showed that provided
one generalised the concept of a Lie group one could indeed find a
symmetry that included the Poincar\'e group and an internal symmetry
group in a non-trivial way. In this section we will discuss this
approach to the supersymmetry group; having adopted a more general
notion of a group, we will show that one is led, with the aid of
the Coleman-Mandula theorem, and a few assumptions, to the known
supersymmetry group. Since the structure of a Lie group, at least in
some local region of the identity, is determined entirely by its Lie
algebra, it is necessary to adopt a more general notion than a Lie
algebra. The vital step in discovering the supersymmetry algebra
is to introduce generators $Q_\alpha^i$, which satisfy
anti-commutation relations, i.e.
$$\eqalignno{\{Q_\alpha^i,Q_\beta^j\}&=Q_\alpha^iQ_\beta^j
+Q_\beta^jQ_\alpha^i\cr
&=\ {\rm some\ other\ generator}&(2.4)}$$
The significance of the $i$ and $\alpha$ indices will become
apparent   shortly. Let us therefore assume that the supersymmetry
group involves  generators $P_a,\ J_{ab},\ T_s$ and possibly some
other  generators which satisfy commutation relations, as well as the
generators $Q_\alpha^i\ (i=1,2,\dots,N)$. We will call the former
generators which satisfy Eqs. (2.1), (2.2) and (2.3)  even and
those satisfying Eq. (2.4)  odd generators.

Having let the genie out of the bottle we promptly replace the
stopper and demand that the supersymmetry algebras have a $Z_2$
graded   structure. This simply means that the even and odd generators
must  satisfy the rules:

$$\eqalignno{
[{\rm even},{\rm even}]&={\rm even}\cr
\{{\rm odd},{\rm odd}\}&={\rm even}\cr
[{\rm even},{\rm odd}]&={\rm odd}&(2.5)}$$
We must still have the relations
$$[P_a,T_s]=0=[J_{ab},T_s]\eqno(2.6)$$
since the even (bosonic) subgroup  must obey the Coleman-Mandula
theorem.

Let us now investigate the commutator between $J_{ab}$ and
$Q_\alpha^i$. As a result of Eq.(2.5) it must be of the form
$$[Q_\alpha^i,J_{ab}]=(b_{ab})_\alpha^\beta Q_\beta^i\eqno(2.7)$$
since by definition the $Q_\alpha^i$ are the only odd generators. We
take the $\alpha$ indices to be those rotated by $J_{ab}$. As in a
Lie  algebra we have some generalised Jacobi identities. If we
denote   an even generator by
$B$ and an odd generator by $F$, we find that
$$\eqalignno{
\big[[B_1,B_2],B_3\big]+\big[[B_3,B_1],B_2\big]+\big[[B_2,B_3],B_1\big]&=0\cr
\big[[B_1,B_2],F_3\big]+\big[[F_3,B_1],B_2\big]+\big[[B_2,F_3],B_1\big]&=0\cr
\{[B_1,F_2],F_3\}+\{[B_1,F_3],F_2\}+[\{F_2,F_3\},B_1]&=0\cr
[\{F_1,F_2\},F_3]+[\{F_1,F_3\},F_2]+[\{F_2,F_3\},F_1]&=0&(2.8)}$$
The reader may verify, by expanding each bracket, that these
relations  are indeed identically true.

The identity
$$\big[[J_{ab},J_{cd}],Q_\alpha^i\big]+\big[[Q_\alpha^i,J_{ab}],J_{cd}\big]+\big
[[J_{cd},Q_\alpha^i],J_{ab}\big]=0\eqno(2.9)$$
upon use of Eq. (2.7) implies that
$$[b_{ab},b_{cd}]_\alpha^\beta=-\eta_{ac}(b_{bd})_\alpha^\beta-\eta_{bd}(b_{ac
})_\alpha^\beta+\eta_{ad}(b_{bc})_\alpha^\beta
+\eta_{bc}(b_{ad})_\alpha^\beta\eqno(2.10)$$
This means that the $(b_{cd})_\alpha^\beta$ form a representation of
the Lorentz algebra or in other words that the $Q_\alpha^i$ carry a
representation of the Lorentz group. We will select $Q_\alpha^i$ to
be  in the
$(0,{1\over2})\oplus({1\over2},
0)$ representation of the Lorentz group, i.e.
$$[Q_\alpha^i,J_{ab}]={1\over2}(\sigma_{ab})_\alpha^\beta
Q_\beta^i\eqno(2.11)$$
We can choose $Q_\alpha^i$ to be a Majorana spinor, i.e.
$$Q_\alpha^i=C_{\alpha\beta}\bar Q^{\beta i}\eqno(2.12)$$
where $C_{\alpha\beta}=-C_{\beta\alpha}$ is the charge conjugation
matrix (see
Appendix A). This does not represent a loss of
generality since, if the algebra admits complex conjugation as an
involution we can always redefine the supercharges so as to satisfy
(2.12) (see Note 1 at the end of this chapter).

The above calculation reflects the more general result that the
$Q_\alpha^i$
must belong to a realization of the even (bosonic) subalgebras of the
supersymmetry
group. This is a simple consequence of demanding that the algebra
be $Z_2$ graded.
The commutator of any even generator $B_1$, with $Q_\alpha^i$ is of
the form
$$[Q_\alpha^i,B_1]=(h_1)_{\alpha j}^{i\beta}Q_\beta^j\eqno(2.13)$$
The generalised Jacobi identity
$$\big[[Q_\alpha^i,B_1],B_2\big]+\big[[B_1,B_2],Q_\alpha^i\big]+\big[[B_2,Q_
\alpha^i],B_1\big]=0\eqno(2.14)$$
implies that
$$[h_1,h_2]_{\alpha
j}^{i\beta}Q_\beta^j=\big[Q_\alpha^i[B_1,B_2]\big]\eqno(2.15)$$
or in other words the matrices $h$ represent the Lie algebra of the
even generators.

The above remarks imply that
$$[Q_\alpha^i,T_r]=(l_r)^i_jQ_\alpha^j+(t_r)^i_j(i\gamma_5)_\alpha^\beta
Q_\beta^j\eqno(2.16)$$
where $(l_r)^i_j+i\gamma_5(t_r)^i_j$ represent the Lie algebra of
the internal
symmetry group. This results from the fact that
$\delta_\beta^\alpha$   and $(\gamma_5)_\alpha^\beta$ are the only
invariant tensors which  are  scalar and pseudoscalar.

The remaining odd-even commutator is $[Q_\alpha^i,P_a]$. A
possibility  that is allowed by the generalised Jacobi identities
that  involve the internal symmetry
group and the Lorentz group is
$$[Q_\alpha^i,P_a]=c(\gamma_a)_\alpha^\beta Q_\beta^i\eqno(2.17)$$
However, the $\big[[Q_\alpha^i,P_a],P_b\big]+\dots$ identity implies
that the constant $c=0$, i.e.
$$[Q_\alpha^i,P_a]=0\eqno(2.18)$$
More generally we could have considered
$(c\gamma_a+d\gamma_a\gamma_5)Q$, on the
right-hand side of (2.17); however  the above Jacobi identity
and  the Majorana
condition imply that $c=d=0$. (See Note 2 at the end of this
chapter).  Let us
finally consider the $\{Q_\alpha^i,Q_\beta^j\}$ anticommutator. This
object must
be composed of even generators and must be symmetric under
interchange  of $\alpha
\ \leftrightarrow\ \beta$ and $i\ \leftrightarrow\ j$. The even
generators are those of the Poincar\'e group, the internal symmetry
group and other even generators which, from the Coleman-Mandula
theorem, commute with the Poincar\'e group, i.e.
they are scalar and pseudoscalar. Hence the most general possibility
is of the  form
$$\{Q_\alpha^i,Q_\beta^j\}=r(\gamma^aC)_{\alpha\beta}P_a\delta^{ij}+s(\sigma^{
ab}C)_{\alpha\beta}J_{ab}\delta^{ij}+C_{\alpha\beta}U^{ij}+(\gamma_5C)_{\alpha
\beta}V^{ij}\eqno(2.19)$$
We have not included a $(\gamma^b\gamma_5C)_{\alpha\beta}L_b^{ij}$
term  as the
$(Q,Q,J_{ab})$ Jacobi identity implies that $L_b^{ij}$ mixes
nontrivially with
the Poincar\'e group and so is excluded by the no-go theorem.

The fact that we have only used numerically invariant tensors under
the Poincar\'e
group is a consequence of the generalised Jacobi identities between
two odd and
one even generators.

To illustrate the argument more clearly, let us temporarily
specialise  to the  case $N=1$ where there is only one supercharge
$Q_\alpha$. Equation (2.19) then
reads
$$\{Q_\alpha,Q_\beta\}
=r(\gamma^aC)_{\alpha\beta}P_a+s(\sigma^{ab}C)_{\alpha\beta}J_{ab}.$$
Using the Jacobi identity
$$\{[P_a,Q_\alpha],Q_\beta\}+\{[P_a,Q_\beta],Q_\alpha\}
+[\{Q_\alpha,Q_\beta\},
P_a]=0,$$
we find that
$$0=s(\sigma^{cd}C)_{\alpha\beta}[J_{cd},P_a]
=s(\sigma^{cd}C)_{\alpha\beta}(-
\eta_{ac}P_d+\eta_{ad}P_c),$$
and, consequently, $s=0$. We are free to scale the generator $P_a$
in   order to
bring $r=2$.

Let us now consider the commutator of the generator of the internal
group and
the supercharge. For only one supercharge, Eq. (2.16) reduces to
$$[Q_\alpha,T_r]=l_rQ_\alpha+i(\gamma_5)^\beta_\alpha t_rQ_\beta.$$
Taking the adjoint of this equation, multiplying by $(i\gamma^0)$ and
using the
definition of the Dirac conjugate given in Appendix A, we find that
$$[\bar Q^\alpha,T_r]=l^*_r\bar Q^\alpha+\bar
Q^\beta(it^*_r)(\gamma_5)_\beta^
\alpha.$$
Multiplying by $C_{\gamma\alpha}$ and using Eq. (2.12), we arrive at
the equation
$$[Q_\alpha,T_r]=l^*_rQ_\alpha+it^*_r(\gamma_5)_\alpha^\beta Q_\beta.$$
Comparing this equation with the one we started from, we therefore
conclude that
$$l^*_r=l_r,\qquad t^*_r=t_r.$$
The Jacobi identity
$$[\{Q_\alpha,Q_\beta\},T_r]+\big[[T_r,Q_\alpha],Q_\beta\big]
+\big[[T_r,Q_\beta
],Q_\alpha\big]=0$$
results in the equation
$$\eqalignno{
[0+\big(l_r\delta_\alpha^\gamma+it_r(\gamma_5)\alpha^\gamma\big)
2(\gamma_aC)_{
\gamma\beta}P_a]+(\alpha\ \leftrightarrow\ \beta)&=\cr
2P_a\{l_r(\gamma_aC)_{\alpha\beta}+it_r(\gamma_5\gamma_aC)_{\alpha\beta}\}+(
\alpha\ \leftrightarrow\ \beta)&=0.}$$
Since $(\gamma_aC)_{\alpha\beta}$ and
$(\gamma_5\gamma_aC)_{\alpha\beta}$ are
symmetric and antisymmetric in $\alpha,\ \beta$ respectively, we
conclude  that $l_r=0$ but $t_r$ has no constant placed on it.
Consequently, we  find that we have
only one internal generator $R$ and we may scale it such that
$$[Q_\alpha,R]=i(\gamma_5)_\alpha^\beta Q_\beta.$$
The $N=1$ supersymmetry algebra is summarised in Eq. (2.27).

Let us now return to the extend supersymmetry algebra. The even
generators $U^
{ij}=-U^{ji}$ and $V^{ij}=-V^{ji}$ are called central charges$^5$ and
are often
also denoted by $Z$. It is a consequence of the generalised Jacobi
identities
\big($(Q,Q,Q)$ and $(Q,Q,Z)$\big) that they commute with all other
generators
including themselves, i.e.
$$[U^{ij},{\rm anything}]=0=[V^{ij},{\rm anything}]\eqno(2.20)$$
We note that the Coleman-Mandula theorem allowed a semi-simple group
plus $U(1)$ factors. The details of the calculation are given in  note
5 at the end of the chapter. Their role in supersymmetric
theories will emerge in later  chapters.

In general, we should write, on the right-hand side of (2.19),
$$(\gamma^aC)_{
\alpha\beta}\omega^{ij}P_a+\dots,$$
where $\omega^{ij}$
is an
arbitrary  real symmetric
matrix. However, one can show that it is possible to redefine (rotate
and rescale)
the supercharges, whilst preserving the Majorana condition, in such a
way as to
bring $\omega^{ij}$ to the form $\omega^{ij}=r\delta^{ij}$ (see Note
3  at the
end of this chapter). The $[P_a,\{Q_\alpha^i,Q_\beta^j\}]+\dots=0$
identity implies
that $s=0$ and we can normalise $P_a$ by setting $r=2$ yielding the
final result
$$\{Q_\alpha^i,Q_\beta^j\}=2(\gamma_aC)_{\alpha\beta}\delta^{ij}P^a+C_{\alpha
\beta}U^{ij}+(\gamma_5C)_{\alpha\beta}V^{ij}\eqno(2.21)$$
In any case $r$ and $s$ have different dimensions and so it would
require the
introduction of a  parameter with  a non-zero dimension  in
order that they were both non-zero.

Had we chosen another irreducible Lorentz representation for
$Q_\alpha^i$ other
than $(j+{1\over2},j)\oplus(j,j+{1\over2})$ we would not have been
able to put
$P_a$, i.e. a $({1\over2},{1\over2})$ representation, on the
right-hand side of
Eq. (2.21). The simplest choice is $(0,{1\over2})\oplus({1\over2},0)$.
In fact
this is the only possible choice (see Note 4).

Finally, we must discuss the constraints placed on the internal
symmetry group
by the generalised Jacobi identity. This discussion is complicated by
the particular
way the Majorana constraint of Eq. (2.12) is written. A two-component
version of this constraint is
$$\bar Q_{\dot Ai}=(Q_A^i)^*;\quad A,\dot A=1,2\eqno(2.22)$$
(see Appendix A for two-component notation). Equation (2.19) and
(2.16) then become
$$\eqalignno{
\{Q_A^i,\bar Q_{\dot Bj}\}&=-2i(\sigma^a)_{A\dot B}\delta^i_jP_a\cr
\{Q_A^i,Q_B^j\}&=\varepsilon_{AB}(U^{ij}+iV^{ij})\cr
[Q_A^i,J_{ab}]&=+{1\over2}(\sigma_{ab})_A^BQ_B^i&(2.23)\cr
\noalign{and}
[Q_A^i,T_r]&=(l_r+it_r)^i_jQ_A^j&(2.24)}$$
Taking the complex conjugate of the last equation and using the
Majorana condition
we find that
$$[Q_{\dot Ai},T_r]=Q_{\dot Ak}(U_r^\dagger)^k_i\eqno(2.25)$$
where $(U_r)^i_j=(l_r+it_r)^i_j$. The $(Q,\bar Q,T)$ Jacobi identity
then implies
that $\delta^i_j$ be an invariant tensor of $G$, i.e.
$$U_r+U_r^\dagger=0\eqno(2.26)$$
Hence $U_r$ is an anti-Hermitian matrix and so represents the
generators of the
unitary group $U(N)$. However, taking account of the central charge
terms in the
$(Q,Q,T)$ Jacobi identity one finds that there is for every central
charge an
invariant antisymmetric tensor of the internal group and so the
possible internal
symmetry group is further reduced. If there is only one central
charge, the internal
group is $Sp(N)$ while if there are no central charges it is $U(N)$.

To summarise, once we have adopted the rule that the algebra be $Z_2$
graded and
contain the Poincar\'e group and an internal symmetry group then the
generalised
Jacobi identities place very strong constraints on any possible
algebra. In fact,
once one makes the further assumption that $Q_\alpha^i$ are spinors
under the
Lorentz group then the algebra is determined to be of the form of
equations
(2.1), (2.6), (2.11), (2.16), (2.18) and (2.21).

The simplest algebra is for $N=1$ and takes the form
$$\eqalignno{
\{Q_\alpha,Q_\beta\}&=2(\gamma_aC)_{\alpha\beta}P^a\cr
[Q_\alpha,P_a]&=0\cr
[Q_\alpha,J_{cd}]&={1\over2}(\sigma_{cd})_\alpha^\beta Q_\beta\cr
[Q_\alpha,R]&=i(\gamma_5)_\alpha^\beta Q_\beta&(2.27)}$$
as well as the commutation relations of the Poincar\'e group. We note
that there
are no central charges (i.e. $U^{11}=V^{11}=0$), and the internal
symmetry group
becomes just a chiral rotation with generator $R$.

We now wish to prove three of the statements above. This is done here
rather than
in the above text, in order that the main line of argument should not
become
obscured by technical points. These points are best clarified in
two-component
notation.

{\bf {Note 1:}}
\quad Suppose we have an algebra that admits a complex
conjugation
as an involution; for the supercharges this means that
$$(Q_A^i)^*=b_i^{\ j} Q_{\dot Aj};\quad(Q_{\dot
Aj})^*=d^j_{\ k}Q_A^k$$  There is no mixing of the Lorentz indices
since $(Q_A^i)^*$ transforms  like $Q_{\dot Ai}$, namely in the
$(0,{1\over2})$ representation of the  Lorentz group,  and not like
$Q_A^i$ which is in the $({1\over2},0)$ representation.  The lowering
of the $i$ index under $*$ is at this point purely a notational
device. Two
successive $*$ operations yield the unit operation and this implies
that
$$(b_i^{\ j})^*d^j_{\ k}=\delta^i_k\eqno(2.28)$$
and in particular that $b_i^{\ j}$ is an invertible matrix. We now
make  the redefinitions
$$
{Q^\prime}_A^i=Q_A^i,\
{Q^\prime}_{\dot Ai}=b_i^{\ j}Q_{\dot A j}
\eqno(2.29)$$
Taking the complex conjugate of ${Q^\prime}_{ Ai}$, we find
$$({Q^\prime}_A^i)^*=(Q_A^i)^*=b_i^{\ j}Q_{\dot Aj}={Q^\prime}_{\dot
Aj}$$
while
$$({Q^\prime}_{\dot Ai})^*=(b_i^{\ j})^*(Q_{\dot
Aj})^*=(b_i^{\ j})^*d^j_{\ k}Q_A^k=Q_A^i\eqno(2.30)$$
using Eq. (2.28).

Thus the ${Q^\prime}_A^i$ satisfy the Majorana condition, as
required.  If the
$Q$'s do not initially satisfy the Majorana condition, we may simply
redefine
them so that they do.

{\bf {Note 2:}}\quad Suppose the $[Q_A,P_a]$ commutator were of the
form
$$[Q_A,P_a]=e(\sigma_a)_{A\dot B}Q^{\dot B}\eqno(2.31)$$
where $e$ is a complex number and for simplicity we have suppressed
the $i$ index.
Taking the complex conjugate (see Appendix A), we find that
$$[Q_{\dot A},P_a]=-e^*(\sigma_a)_{B\dot A}Q^B\eqno(2.32)$$
Consideration of the $\big[[Q_A,P_a],P_b\big]+\dots=0$ Jacobi
identity  yields
the result
$$-|e|^2(\sigma_a)_{A\dot B}(\sigma^b)^{C\dot B}-(a\ \leftrightarrow
b)=0\eqno(2.33)$$
Consequently $e=0$ and we recover the result
$$[Q_A,P_a]=0\eqno(2.34)$$

{\bf {Note 3:}}\quad The most general form of the $Q^{Ai},Q^{\dot B}_j$
anticommutator
is
$$\{Q^{Ai},Q^{\dot B}_j\}=-2iU^i_j(\sigma^m)^{A\dot B}P_m+{\rm terms\
involving
\ other\ Dirac\ matrices}\eqno(2.35)$$
Taking the complex conjugate of this equation and comparing it with
itself, we
find that $U$ is a Hermitian matrix
$$(U^i_j)^*=U^j_i\eqno(2.36)$$
We now make a field redefinition of the supercharge
$${Q^\prime}^{Ai}=B^i_jQ^{Aj}\eqno(2.37)$$
and its complex conjugate
$${Q^\prime}^{\dot A}_i=(B^i_j)^*Q^{\dot A}_j\eqno(2.38)$$
Upon making this redefinition in Eq. (2.35), the $U$ matrix becomes
replaced by
$${U^\prime}^i_j=B^i_kU^k_l(B^j_l)^*\quad{\rm or}\quad
U^\prime=BUB^\dagger\eqno(2.39)$$
Since $U$ is a Hermitian matrix, we may diagonalise it in the form
$c_i\delta^
i_j$ using a unitarity matrix $B$. We note that this preserves the
Majorana
condition on $Q^{Ai}$. Finally, we may scale $Q^i\rightarrow(1/\sqrt
c^i)Q^i$
to bring $U$ to the form $U=d_i\delta^i_j$, where $d_i=\pm1$. In
fact,  taking
$A=B=1$ and $i=j=k$, we realise that the right-hand side of Eq.
(2.35)  is a positive-definite operator
and since the energy $-iP_0$ is
assumed positive definite, we
can only find $d_i=+1$. The final result is
$$\{Q^{Ai},Q^{\dot B}_j\}=-2i\delta^i_j(\sigma^m)^{A\dot
B}P_m\eqno(2.40)$$

{\bf {Note 4:}}
\quad Let us suppose that the supercharge $Q$ contains
an irreducible
representation of the Lorentz group other than
$(0,{1\over2})\oplus({1\over2},
0)$, say, the representation $Q_{A_1\dots A_n,\dot B_1\dots\dot B_m}$
where the
$A$ and $B$ indices are understood to be separately symmetrised and
$n+m$ is odd
in order that $Q$ be odd and $n+m>1$. By projecting the
$\{Q,Q^\dagger\}$ anti-commutator
we may find the anti-commutator involving $Q_{A_1\dots A_n,\dot
B_1\dots\dot B
_m}$ and its Hermitian conjugate. Let us consider in particular the
anti-commutator
involving $Q=Q_{11\dots1,\dot 1\dot 1\dots \dot 1}$, this must result
in an object of spin $n +m>1$. However, by the Coleman-Mandula no-go
theorem no such generator  can occur
in the algebra and so the anti-commutator must vanish, i.e.
$QQ^\dagger+Q^\dagger
Q=0$.

Assuming the space on which $Q$ acts has a positive definite norm, one
such example
being the space of on-shell states, we must conclude that $Q$
vanishes. However
if $Q_{11\dots1,\dot 1\dot 1\dots \dot 1}$ vanishes, so must
$Q_{A_1\dots A_n,\dot  B_1\dots\dot
B_m}$ by its Lorentz properties, and we are left only with the
$(0,{1\over2})
\oplus({1\over2},0)$ representation.

{\bf { Note 5:}}\quad We now return to the proof of equation (2.20).
Using the (Q,Q,Z) Jocobi identity it is straightforward to show that
the supercharge $Q$ commutes with the central charges $Z$. The
(Q,Q,U)  Jacobi identity then implies that the central charges
commute with themselves. Finally, one considers the $(Q,Q,T_r)$ Jacobi
identity, which shows that the commutator of $T_r$ and $Z$
takes the generic form $[T_r,\ Z]= \dots Z$. However, the generators
$T_r$ and $Z$ form the internal symmetry group of the supersymmetry
algebra and from the no-go theorem we know that this group must be a
semisimple Lie group times
$U(1)$ factors. We recall that a semisimple Lie group is one that has
no normal Abelian subgoups other that the group itself and the identity
element.  As such, we must conclude that $T_r$ and $Z$  commute,
and hence our final result that the central charges commute with all
generators, that is they really are central.

Although the above discussion started with the Poincar\'e group, one
could equally
well have started with the conformal or (anti-) de Sitter groups and
obtained the
superconformal and super (anti-)de Sitter algebras. For completeness,
we now list
these algebras. The superconformal algebra which has the generators
$P_n,J_{mn},
D,K_n,A,Q^{\alpha i}$, $S^{\alpha i}$ and the internal symmetry
generators $T_r$ and $A$  is given by the Lorentz group  plus:
$$\eqalignno{
[J_{mn},P_k]&=\eta_{nk}P_m-\eta_{mk}P_n\cr
[J_{mn},K_k]&=\eta_{nk}K_m-\eta_{mk}K_n\cr
[D,P_K]&=-P_K,\quad[D,K_K]=+K_K\cr
[P_m,K_n]&=-2J_{mn}+2\eta_{mn}D\quad[K_n,K_m]=0,\quad[P_n,P_m]=0\cr
[Q_\alpha^i,J_{mn}]&={1\over2}(\gamma_{mn})_\alpha^\beta
Q_\beta^i,\quad[S_\alpha
^i,J_{mn}]={1\over2}(\gamma_{mn})_\alpha^\beta S^{\beta i}\cr
\{Q_\alpha^i,Q_\beta^j\}&=-2(\gamma^nC^{-1})_{\alpha\beta}P_n\delta^{ij}\cr
\{S_\alpha^i,S_\beta^j\}&=+2(\gamma^nC^{-1})_{\alpha\beta}K_n\delta^{ij}\cr
[Q_\alpha^i,D]&={1\over2}Q_\alpha^i,\quad[S_\alpha^i,D]=-{1\over2}S_\alpha^i\cr
[Q_\alpha^i,K_n]&=-(\gamma_n)_\alpha^\beta
S_\beta^i,\quad[S_\alpha^i,P_n]=(\gamma
_n)_\alpha^\beta Q_\beta^i\cr
[Q_\alpha^i,T_r]&=\big( \delta_\alpha^\beta(\tau_{r_1})^i_j
 +(\gamma_5)_\alpha^
\beta(\tau_{r_2})^i_j\big) Q_\beta^j\cr
[S_\alpha^i,T_r]&=\big(\delta_\alpha^\beta(\tau_{r_1})^i_j-(\gamma_5)_\alpha^
\beta(\tau_{r_2})^i_j\big)Q_\beta^j\cr
[Q_\alpha^i,A]&=-i(\gamma_5)_\alpha^\beta
Q_\beta^i\left({4-N\over4N}\right)\cr
[S_\alpha^i,A]&={4-N\over4N}i(\gamma_5)_\alpha^\beta S_\beta^i\cr
\{Q_\alpha^i,S_\beta^j\}&=-2(C^{-1}_{\alpha\beta})
D\delta^{ij}+(\gamma^{mn}C^{-
1})_{\alpha\beta}J_{mn}\delta^{ij}+4i(\gamma_5C^{-1}_{\alpha\beta})
A\delta^{ij}\cr
&\quad-2(\tau_{r_1})^{ij}(C^{-1})_{\alpha\beta}+\big((\tau_{r_2})^{ij}(\gamma_
5C^{-1})_{\alpha\beta}\big)T_r&(2.41)}$$
The $T_r$ and $A$ generate $U(N)$, and $\tau_1+\gamma_5\tau_2$ are in
the fundamental
representation of $SU(N)$.

The case of $N=4$ is singular and one can have either
$$[Q_\alpha^i,A]=0\quad{\rm
or}\quad[Q_\alpha^i,A]=-i(\gamma_5)_\alpha^\beta Q
_\beta^i$$
and similarly for $S_\alpha^i$ and $A$. One may verify that both
possibilities
are allowed by the $N=4$ Jacobi identities and so form acceptable
superalgebras.

The anti-de Sitter superalgebra has generators $M_{mn},\
T_{ij}=-T_{ji}$ and
$Q^{\alpha i}$, and is given by
$$\eqalignno{
[M_{mn},M_{pq}]&=\eta_{np}M_{mq}+3\ {\rm terms}\cr
[M_{mn},T_{ij}]&=0\quad[Q_\alpha^i,M_{mn}]={1\over2}
(\gamma_{mn})_\alpha
^{\ \beta} Q_\beta^i\cr
[Q_\alpha^i,T^{jk}]&=-2i(\delta^{ij}Q_\alpha^k-\delta^{ik}Q_\alpha^j)\cr
\{Q_\alpha^i,Q_\beta^j\}&=\delta^{ij}(\gamma_{mn}C^{-1})_{\alpha\beta}iM_{mn}+
(C^{-1})_{\alpha\beta}T^{ij}\cr
[T^{ij},T^{kl}]&=-2i(\delta^{jk}T^{il}+3\ {\rm terms})&(2.42)}$$
\bigskip

{\centerline {\bf {3. Irreducible Representations of Four-Dimensional
Supersymmetry}}}
\medskip
It is a relatively
straightforward proceedure to find the   irreducible
representations of any  supersymmetry algebra. These representations
tell us which supersymmetric theories are possible for a given
supersymmetry algebra. To be precise,  they provide a list of all the
particles that occur in each of the   theories that have
a given supersymmetric algebra as a symmetry. In this section, we carry
out this proceedure  for the four-dimensional supersymmetry algebras.
Using a very similar procedure one can find the irreducible
representations  of  the supersymmetry algebras in other dimensions.
The reader is referred to reference  [160] where this is procedure is
sketched in an arbitary dimension and where lists of irreducible
representations are given.
\par
One of the most interesting features of the construction of these
irreducible representations is when the central charges of the
supersymmetry algebra are non-trivial. In this case, and when the
central charges take particular values, the massive representations
contain many fewer states that those contained in the generic massive
representation. When this occurs the states in the representation are
called BPS states and they play an important role in discussions
of string duality.
\medskip
The first part  of this section  is taken from reference [0]
and we have kept the equation numbers the same as in that
reference.
\medskip

In this chapter we wish to find the irreducible representations of
supersymmetry [11], or, put another way, we want to know what is the
possible particle content of supersymmetric theories. As is well
known  the irreducible representations of the Poincar\'e group are
found by  the Wigner method of induced representations [12]. This
method  consists of finding a representation of a subgroup of the
Poincar\'e  group and boosting it up to a representation of the full
group. In  practice, one adopts the following recipe: we choose a
given momentum
$q^\mu$ which satisfies $q^\mu q_\mu =0$ or $q^\mu q_\mu =-m^2$
depending which  case we are considering. We find the subgroup $H$
which leaves
$q^\mu$  intact and find a representation of $H$  on the
$|q^\mu\rangle$ states.  We then induce this representation to the
whole of the Poincar\'e  group $P$, in the usual way. In this
construction there is a  one-to-one correspondence between points of
$P/H$ and four-momentum  which satisfies $P_\mu P^\mu=0$ or
$P_\mu P^\mu=-m^2$. One can show that the  result
is independent of the
choice of momentum $q^\mu$ one starts  with.

In what follows we will not discuss the irreducible representations
in  general, but only that part applicable to the rest frame, i.e. the
representations of $H$ in the states at rest. We can do this safely
in  the knowledge that once the representation of $H$ on the
rest-frame  states is known then the representation of $P$ is uniquely
given and  that every irreducible representation of the Poincar\'e
group can be  obtained by considering every irreducible representation
of $H$.

In terms of physics, the procedure has a simple interpretation,
namely,  the properties of a particle are determined entirely by its
behaviour  in a given frame (i.e. for given $q^\mu$). The general
behaviour is  obtained from the given
$q^\mu$ by boosting either the observer or the frame with momentum
$q^\mu$ to one with arbitrary momentum.

The procedure outlined above for the Poincar\'e group can be
generalised to any group of the form $S\otimes_s T$ where the symbol
$\otimes_s$ denotes the semi-direct product of the groups $S$ and $T$
where $T$ is Abelian. It also applies to the supersymmetry group and
we  shall take it for granted that the above recipe is the correct
procedure and does in fact yield all irreducible representations of
the supersymmetry group.

Let us first consider the massless case $q_\mu q^\mu=0$, for which we
choose the standard momentum $q^\mu_s=(m,0,0,m)$ for our "rest
frame".  We must now find $H$ whose group elements leave
$q^\mu_s=(m,0,0,m)$ intact. Clearly this contains $Q_\alpha^i,\ P_\mu$
and $T_s$, since  these generators all commute with $P_\mu$ and so
rotate the states with
$q^\mu_s$ into themselves. As we will see in the last section one can
not have non-vanishing  central charges for the massless  case.

Under the Lorentz group the action of the generator
${1\over2}\Lambda^{\mu\nu} J_{\mu\nu}$ creates an infinitesimal
transformation $q^\mu\rightarrow\Lambda^
\mu_\nu q^\nu+q^\mu$. Hence $q^\mu_s$ is left invariant provided the
parameters obey the relations
$$\Lambda_{30}=0,\quad\Lambda_{10}+\Lambda_{13}=0,
\quad\Lambda_{20}+\Lambda_{2
3}=0\eqno(8.1)$$
Thus the Lorentz generators in $H$ are
$$T_1=J_{10}+J_{13},\quad T_2=J_{20}+J_{23},\quad J=J_{12}\eqno(8.2)$$
These generators form the algebra
$$\eqalignno{
[T_1,J]&=-T_2\cr
[T_2,J]&=+T_1\cr
[T_1,T_2]&=0&(8.3)}$$
The reader will recognise this to be the Lie algebra of $E_2$, the
group of translations and rotations in a two-dimensional plane.

Now the only unitary representations of $E_2$ which are finite
dimensional have $T_1$ and $T_2$ trivially realised, i.e.
$$T_1|q^\mu_s\rangle=T_2|q^\mu_s\rangle=0
\eqno(8.4)$$
This results from the theorem that all non-trivial unitary
representations of  noncompact groups are infinite-dimensional. We
will  assume we require finite-dimensional representations of $H$.

Hence for the Poincar\'e group, in the case of massless particles,
finding representations of $H$ results in finding representations of
$E_2$ and consequently for the generator $J$ alone. We choose our
states so that
$$J|\lambda\rangle=i\lambda|\lambda\rangle
\eqno(8.5)$$
Our generators are anti-Hermitian. In fact, $J$ is the helicity
operator and we select $\lambda$ to be integer or half-integer (i.e.
$J={\underline q}\cdot {\underline  J}/|{\underline q}|$ evaluated at
${\underline q}=(0,0,m)$ where
$J_i=\varepsilon_{ijk}J_{jk},\ i,j=1,2,3)$.

Let us now consider the action of the supercharges $Q_\alpha^i$ on
the  rest-frame states, $|q^\mu_s\rangle$. The calculation is easiest
when  performed using the two-component formulation of the
supersymmetry  algebra of Eq. (2.23). On rest-frame states we find
that
$$\eqalignno{
\{Q^{Ai},Q^{\dot B}_j\}&=-2\delta^i_j(\sigma_\mu)^{A\dot B}q^\mu_s\cr
&=-2\delta^i_j(\sigma_0+\sigma_3)^{A\dot
B}m=+4m\delta^i_j\left(\matrix{
0&0\cr 0&1\cr}\right)^{A\dot B}&(8.6)}$$
In particular these imply the relations
$$\eqalignno{
\{Q^{1i},Q^{\dot 1}_j\}&=0\cr
\{Q^{2i},Q^{\dot2}_j\}&=4m\delta^i_j\cr
\{Q^i_i,Q^{2j}\}=\{Q^1_i,Q^{\dot2j}\}=0&(8.7)}$$
The first relation implies that
$$\langle
q^\mu_s|\big(Q^{1i}(Q^{1i})^*+(Q^{1i})^*Q^{1
i}\big)|q^\mu_s\rangle=0\eqno(8.8)$$ Demanding that the norm on
physical states be positive definite and
vanishes only if the state vanishes, yields
$$Q_2^i|q^\mu_s\rangle=Q_{\dot2i}|q^\mu_s\rangle=0\eqno(8.9)$$
Hence, all generators in $H$ have zero action on rest-frame states
except $J,\ T_s,\ P_\mu,\ Q_1^i$ and $Q_{\dot1i}$. Using Eq. (2.23)
we  find that
$$\eqalignno{
[Q_1^i,J]&={1\over2}(\sigma_{12})_1^1Q_1^i\cr
&=-{i\over2}Q_1^i&(8.10)}$$
Similarly, we find that complex conjugation implies
$$[(Q_1^i)^*,J]=+{i\over2}(Q_1^i)^*\eqno(8.11)$$
The relations between the remaining generators summarised in Eqs.
(8.7), (8.10), (8.11) and (2.24) can be summarised by the statement
that $Q_1^i$ and $(Q_1^i)^*$ form a Clifford algebra, act as raising
and lowering operators for the helicity operator $J$ and transform
under the $N$ and $\bar N$ representation of $SU(N)$.

We find the representations of this algebra in the usual way; we
choose a state of given helicity, say $\lambda$, and let it be the
vacuum state for the operator $(Q_1^i)^*$, i.e.
$$\eqalignno{
Q_1^i|\lambda\rangle&=0\cr
J|\lambda\rangle&=i\lambda|\lambda\rangle&(8.12)}$$
The states of this representation are then
$$\eqalignno{
|\lambda\rangle&=|\lambda\rangle\cr
|\lambda-{1\over2},i\rangle&=(Q_1^i)^*|\lambda\rangle\cr
|\lambda-1,[ij]\rangle&=(Q_1^i)^*(Q_1^j)^*|\lambda\rangle&(8.13)}$$
etc. These states have the helicities indicated and belong to the
$[ijk\dots]$
anti-symmetric representation of $SU(N)$. The series will terminate
after the helicity $\lambda-(N/2)$, as the next state will be an
object  antisymmetric in $N+1$ indices. Since there
are only
$N$ labels, this object vanishes identically. The states have
helicities from $\lambda$ to
$\lambda-(N/2)$, there being $N!/\big(m!(N-m)!\big)$ states with
helicity
$\lambda-(m/2)$.

To obtain a set of states which represent particles of both
helicities  we must add to the above set the representations with
helicities from
$-\lambda$ to $-\lambda+(N/2)$. The exception is the so-called CPT
self-conjugate sets of states which automatically contain both
helicity  states.

The representations of the full supersymmetry group are obtained by
boosting the above states in accordance with the Wigner method of
induced representations.

Hence the massless irreducible representation of $N=1$ supersymmetry
comprises only the two states
$$\eqalignno{&\quad|\lambda\rangle\cr
&|\lambda-{1\over2}\rangle=(Q_1)|\lambda\rangle&(8.14)}$$
with helicities $\lambda$ and $\lambda-{1\over2}$ and since
$$Q_1Q_1|\lambda\rangle=0\eqno(8.15)$$
there are no more states.

To obtain a CPT-invariant theory we must add states of the opposite
helicities, i.e. $-\lambda$ and $-\lambda+{1\over2}$. For example, if
$\lambda={1\over2}$
we get on-shell helicity states 0 and ${1\over2}$ and their CPT
conjugates with helicities $-{1\over2},0$, giving a theory with two
spin 0 and one Majorana spin ${1\over2}$. Alternatively, if
$\lambda=2$ then we get on-shell helicity
states $3/2$ and 2 and their CPT self conjugates with helicity $-3/2$
and $-2$; this results in a theory with one spin 2 and one spin $3/2$
particles. These on-shell states are those of the Wess-Zumino model
and
$N=1$ supergravity respectively. Later in this discussion we will
give  a complete account of these theories.

For $N=4$ with $\lambda=1$ we get the massless states
$$|1\rangle,\ |{1\over2},i\rangle,\ |0,[ij]\rangle,\
|-{1\over2},[ijk]\rangle, \ |-1,[ijkl]\rangle\eqno(8.16)$$
This is a CPT self-conjugate theory with one spin 1, four
spin ${1\over2}$ and six spin 0 particles.

Table 8.1 below gives the multiplicity for massless irreducible
representations which have maximal helicity 1 or less.
\vfil\eject
\centerline{Table 8.1\quad Multiplicities for massless irreducible}
\centerline{representations with maximal helicity 1 or less}
$$\offinterlineskip\halign{\hfil$# $\hfil&\quad\strut#&\vrule\quad
\hfil$# $\hfil&\quad\hfil$# $\hfil&\quad\hfil$# $\hfil&\quad\hfil$#
$\hfil&\quad# \cr
\noalign{\hrule}
\qquad N&&&&&&\cr
{\rm Spin}&&1&1&2&2&4\cr
\noalign{\hrule}
{\rm Spin}\ 1&&-&1&1&-&1\cr
{\rm Spin}\ {1\over2}&&1&1&2&2&4\cr
{\rm Spin}\ 0&&2&-&2&4&6\cr
\noalign{\hrule}}$$
We see that as $N$ increases, the multiplicities of each spin and
the   number of different types of spin increases. The simplest
theories  are  those for $N=1$.
The one in the first column in the Wess-Zumino model and the one in
the second column is the $N=1$ supersymmetric Yang-Mills theory. The
latter contains one spin 1 and one spin ${1\over 2}$, consistent with
the formula for the lowest helicity $\lambda-(N/2)$, which in this
case  gives $1-{1\over 2}={1\over 2}$. The $N=4$ multiplet is CPT self
conjugate, since in this case we have $\lambda-(N/2)=1-4/2=-1$. The
Table stops at $N$ equal to 4 since when $N$ is greater than 4 we
must have particles of spin greater than 1. Clearly, $N>4$ implies
that $\lambda-(N/2)=1-(N/2)<-1$. This leads us to the well-known
statement that the $N=4$ supersymmetric theory is the maximally
extended Yang-Mills theory.

The content for massless on-shell representations with a maximum
helicity 2 is given in Table 8.2.
\par
\centerline{Table 8.2\quad Multiplicity for massless on-shell
representations with maximal helicity 2.}
$$\offinterlineskip\halign{\hfil$# $\hfil&\quad\strut#&\vrule\quad
\hfil$# $\hfil&\quad\hfil$# $\hfil&\quad\hfil$# $\hfil&\quad\hfil$#
$\hfil&
\quad\hfil$# $\hfil&\quad\hfil$# $\hfil&\quad\hfil$# $\hfil&\quad# \cr
\noalign{\hrule}
\qquad N&&&&&&&&&\cr
{\rm Spin}&&1&2&3&4&5&6&7&8\cr
\noalign{\hrule}
{\rm Spin}\ 2&&1&1&1&1&1&1&1&1\cr
{\rm Spin}\ {3\over2}&&1&2&3&4&5&6&8&8\cr
{\rm Spin}\ 1&&&1&3&6&10&16&28&28\cr
{\rm Spin}\ {1\over2}&&&&1&4&11&26&56&56\cr
{\rm Spin}\ 0&&&&&2&10&30&70&70\cr
\noalign{\hrule}}$$
The $N=1$ supergravity theory contains only one spin 2 graviton and
one spin $ 3/2$ gravitino. It is often referred to as simple
supergravity theory. For the $N=8$ supergravity theory,
$\lambda-(N/2)=2-{8\over2}=-2$. Consequently it is CPT self-conjugate
and contains all particles from spin 2 to spin 0. Clearly, for
theories  in which $N$ is greater than 8, particles of spin
higher than
 2 will   occur. Thus, the $N=8$ theory is the maximally extended
supergravity    theory.

It has sometimes been  claimed that this theory is in fact the largest
possible   consistent supersymmetric  theory. This contention rests on
the widely held belief   that it is impossible to consistently couple
massless particles of spin
${5\over2}$ to other particles. In fact
 superstring theories do include spin ${5\over 2}$
 particles, but
these are massive.

We now consider the massive irreducible representations of
supersymmetry. We take our rest-frame momentum to
be
$$q_s^\mu=(m,0,0,0)\eqno(8.17)$$
The corresponding little group is then generated by
$$P_m,Q^{\alpha
i},T^r,Z_1^{ij},Z_2^{ij},J_m\equiv{1\over2}\varepsilon_{mnr}J^
{nr}\eqno(8.18)$$
where $m,n,r=1,2,3$ for the present discussion. The $J_m$ generate
the  group $SU(2)$. Let us first consider the case where the central
charges  are trivially realised.

When acting on the rest-frame states the supercharges obey the
algebra
$$\eqalignno{\{Q^{Ai},(Q^{Bj})^*\}&=2\delta^A_B\delta^i_jm\cr
\{Q^{Ai},Q^{Bj}\}&=0&(8.19)}$$
The action of the $T^r$ is that of $U(N)$ with the $SU(2)$ rotation
generators satisfy
$$\eqalignno{[J_m,J_n]&=\varepsilon_{mnr}J_r\cr
[Q^{Ai},J_m]&=i(\sigma_m)^A_{\ B}Q^{Bi}&(8.20)}$$
where $(\sigma_m)$ are the Pauli matrices. We note that as far as
$SU(2)$ is
concerned the dotted spinor $Q^{\dot Ai}$ behaves like the undotted
spinor $Q_{Ai}$.

We observe that unlike the massless case none of the supercharges
are   trivially realised and so the Clifford algebra they form has
$4N$   elements, that is, twice as many as those for the massless
case. The   unique irreducible representation of the Clifford algebra
is found in   the usual way. We define a Clifford vacuum
$$Q_A^i|q^\mu_s\rangle=0,\quad A=1,2\ ,i=1,\dots , N
\eqno(8.21)$$
and the representation is carried by the states
$$|q^\mu_s\rangle,\ (Q_A^i)^*|q^\mu_s\rangle,\
(Q_A^i)^*(Q_B^j)^*|q^\mu_s\rangle,\dots\eqno(8.22)$$
Thanks to the anti-commuting nature of the $(Q_A^j)^*$ this series
terminates when  $Q^*$  is applied $2N+1$ times.

The structure of the above representation is not particularly
apparent  since it is not clear how many particles of a given spin it
contains.  The properties of the Clifford algebra are more easily
displayed by  defining the real generators
$$\eqalignno{\Gamma^i_{2A-1}&={1\over2m}\big(Q^{Ai}+(Q^{Ai})^*\big)\cr
\Gamma^i_{2A}&={i\over2m}\big(Q^{Ai}-(Q^{Ai})^*\big)&(8.23)\cr
\noalign{where the}
\Gamma^i_p&=(\Gamma^i_1,\Gamma^i_2,\Gamma^i_3,\Gamma^i_4)&(8.24)}$$
are Hermitian. The Clifford algebra of Eq. (8.19) now becomes
$$\{\Gamma_p^i,\Gamma_q^j\}=\delta^{ij}\delta_{pq}\eqno(8.25)$$
The $4N$ elements of the Clifford algebra carry the group $SO(4N)$ in
the standard manner' the $4N(4N-1)/2$ generators of $SO(4N)$ being
$$O^{ij}_{mn}={1\over2}[\Gamma_m^i,\Gamma_n^j]\eqno(8.26)$$
As there are an even number of elements in the basis of the Clifford
algebra, we may define a "parity" $(\gamma_5)$ operator
$$\Gamma_{4N+1}=\prod^4_{p=1}\prod^N_{i=1}\Gamma_p^i\eqno(8.27)$$
which obeys the relations
$$\eqalignno{
(\Gamma_{4N+1})^2&=+1\cr
\{\Gamma_{4N+1},\Gamma^i_p\}&=0&(8.28)}$$
Indeed, the irreducible representation of Eq. (8.22) is of dimension
$2^{2N}$ and transforms according to an irreducible representation
of
$SO(4N)$ of dimension $2^{2N-1}$ with $\Gamma_{4N+1}=-1$ and another
of  dimension $2^{2N-1}$ with $\Gamma_{4N+1}=+1$. Now any linear
transformation of the $Q$,
$Q^*$ (for example $\delta Q=rQ$) can be represented by a
generator   formed from the commutator
 of the $Q$ and $Q^*$ (for example, $r[Q,Q^*])$. In particular
the
$SU(2)$ rotation generators are given by
$$s_k=-{i\over4m}(\sigma_k)^A_{\ B}[Q^{jB},(Q^{jA})^*]\eqno(8.29)$$
One may easily verify that
$$[Q^{jA},s_k]=i(\sigma_k)^A_BQ^{Bj}\eqno(8.30)$$
The states of a given spin will be classified by that subgroup of
$SO(4N)$ which commutes with the appropriate $SU(2)$ rotation
subgroup  of $SO(4N)$. This will be the group generated by all
generators  bilinear in $Q,Q^*$ that have their two-component index
contracted,  i.e.
$$\eqalignno{
\Lambda^i_j&={i\over2m}[Q^{Ai},(Q^{j}_A)^*]\cr
k^{ij}&={i\over2m}[Q^{Ai},Q_A^j]&
(8.31)}$$
with $(k^{ij})^\dagger=k_{ij}$. It is easy to verify that the
$\Lambda^i_j,\ k^
{ij}$ and $k_{ij}$ generate the group $USp(2N)$ and so the states of
a  given spin are labelled by representations of $USp(2N)$. That the
group  is $USp(2N)$
is most easily seen by defining
$$Q_A^a=\left\{\matrix{
Q_A^i\delta_i^a\qquad&a=1,\dots,N\qquad\cr
\varepsilon_{AB}(Q^{Bi})^*&a=N+1,\dots,2N\cr}\right.\eqno(8.32)$$
for then the generators $\Lambda^i_j,\ k^{ij}$ and $k_{ij}$ are given
by
$$s^{ab}={i\over2m}[Q^{Aa},Q_A^b].\eqno(8.33)$$
Using the fact that
$$\{Q_A^a,Q_B^b\}=\varepsilon_{AB}\Omega^{ab}\eqno(8.34)$$
where
$$\Omega^{ab}=\left(\matrix{
0&1\cr -1&0\cr}\right)$$
we can verify that
$$[s^{ab},s^{cd}]=\Omega^{ac}s^{bd}+\Omega^{ad}s^{bc}
+\Omega^{bc}s^{ad}+\Omega^{bd}s^{ac}\eqno(8.35)$$
which is the algebra of $USp(2N)$.

The particle content of a massive irreducible representation is
given   by the following

{\bf Theorem} [21]: If our Clifford vacuum is a scalar under the
$SU(2)$ spin group and the internal symmetry group, then the
irreducible massive representation of supersymmetry has the following
content
$$2^{2N}=\left[{N\over2},(0)\right]+\left[{N-1\over2},(1)\right]
+\dots+\left[{
N-\kappa\over2},(\kappa)\right]+\dots+[0,(N)]\eqno(8.36)$$
where the first entry in the bracket denotes the spin and the last
entry, say $(k)$, denotes which $k$th-fold antisymmetric traceless
irreducible representation of $USp(2N)$ that this spin belongs to.

\centerline{Table 8.3 Some massive representations (without central
charges) labelled in}
\centerline{terms of the $USp(2N)$ representations.}
$$\offinterlineskip\halign{\hfil$# $\hfil&\quad\strut#&\vrule\quad
\hfil$# $\hfil&\quad\hfil$# $\hfil&\quad\hfil$# $\hfil&\quad\hfil$#
$\hfil&\quad\strut#&\vrule\quad\hfil$# $\hfil&\quad\hfil$#
$\hfil&\quad\hfil$# $\hfil
&\quad\strut#&\vrule\quad\hfil$# $\hfil&\quad\hfil$#
$\hfil&\quad\strut#&\vrule
\quad# \cr
\noalign{\hrule}
\qquad N&&&&&&&&&&&&&&\cr
{\rm Spin}&&&1&&&&&2&&&&3&&4\cr
\noalign{\hrule}
{\rm Spin}\ 2&&&&&1&&&&1&&&1&&1\cr
{\rm Spin}\ {3\over2}&&&&1&2&&&1&4&&1&6&&8\cr
{\rm Spin}\ 1&&&1&2&1&&1&4&5+1&&6&14+1&&27\cr
{\rm Spin}\ {1\over2}&&1&2&1&&&4&5+1&4&&14&14^\prime+6&&48\cr
{\rm Spin}\ 0&&2&1&&&&5&4&1&&14^\prime&14&&42\cr
\noalign{\hrule}}$$
\par
Consider an example with two supercharges. The classifying group is
$USp(4)$ and the $2^4$ states are one spin 1, four spin $1/2$, and
five  spin 0 corresponding to the ${\underline 1}-,\ {\underline 4}-$
and
${\underline 5}-$dimensional representations of $USp(4)$. For more
examples   see Table 8.3.
\par
Should the Clifford vacuum carry spin and belong to a non-trivial
representation of the internal group $U(N)$, then the irreducible
representation is found by taking the tensor product of the vacuum
and  the representation given in the above theorem.
\medskip

\centerline {\bf Massive Representations with a Central Charge}
\medskip
We now consider the case of particles that are massive, but which
also possess a central charge. We take the particles to be in their
rest-frame with
momentum  $q^\mu\equiv (M,0,0,0)$. The
isotropy group, H  contains $(P^a, Q^{i}_A,Q_i^{\dot A},
\underline J, T_r$ and $ Z^{ij})$. In  the rest frame of the
particles, that is for the  momentum $q^\mu$, the algebra
of the  supercharges
is given by
$$\{Q^{A i},\ {(Q^{B j})}^* \}= 2\delta ^A_B \delta ^i_j M
\eqno(1)$$
and
$$\{Q_A^i,\ Q_B^j \}= \epsilon _{AB} Z^{ij}
\eqno(2)$$
\par
To discover what is the particle content in a supermultiplet
we would like to rewrite the above algebra as a Clifford
algebra. The first step in this proceedure is to carry out a unitary
transformation on the internal symmetry index of the supercharges
i.e. $Q_A^i\to U^i_j Q_A^j$ or $ Q_A\to U Q_A^i$ with $U^\dagger
U=1$. Such a transformation preserves the form of the first relation
of equation (1). However,  the unitary transformation can be chosen
[104] such that the central charge
,which transforms as $Z\to UZU^T$, can be brought to the form of a
matrix which has all its entries zero except for the 2 by 2 matices
down  its diagonal. These 2 by 2 matrices are anti-symmetric as a
consequence of the anti-symmetric nature
of $Z^{ij}$ which is preserved
by the unitary transformation. This is  the closest one can come to
diagonalising an anti-symmetric matrix. Let us for simplicity restrict
our attention to
 even $N$.   We can  best write
down this matrix we replace the
$i,j=1,2,\ldots N$ internal indices  by $i=(a,m), \ j=(b,n),\
a,b=1,2,\  m,n=1,\ldots ,{N\over 2}$ whereupon
$$Z^{(a,m) (b,n)} =
2 \epsilon ^{ab}\delta ^{mn}Z_n
\eqno(3)$$
In fact one also show that    $Z_n\ge 0$. The supercharges in the
rest-frame satisfy the relations
$$ \{Q^{A (a m)},\ {(Q^{ B {(b n)}})}^*  \}= 2\delta ^A_B \delta
^a_b
\delta ^m_n M
\eqno(4)$$
and
$$\{Q_A^{(a m)},\ Q_B^{(b n) } \}= 2 \epsilon _{AB}
 \epsilon ^{ab}\delta ^{mn}Z_n
\eqno(5)$$
\par
We now define the supercharges
$$ S^{Am}_1 = {1\over \sqrt 2}(Q^{A1m}+ (Q^{B2m}\epsilon _{BA})^*)
\eqno(6)$$
$$
 S^{Am}_2 = {1\over \sqrt 2}(Q^{A1m}- (Q^{B2m}\epsilon _{BA})^*)
\eqno(7)$$
in terms of which all the anti-commutators vanish except for
$$
\{S^{Am}_1,\ {(S^{Bn}_1)}^*\}= 2 \delta ^{A B} \delta ^{mn}(M-Z_n)
\eqno(8)$$
$$
\{S^{Am}_2,\ {(S^{Bn}_2)}^*\}= 2 \delta ^{A B} \delta ^{mn}(M+Z_n)
\eqno(9)$$
This algebra is a  Clifford algebra
formed from the $2N$ operators $S^{Am}_1$ and
$S^{Am}_2$  and their complex conjugates. It follows from equation
(9) that if we take the same indices on each supercharge that
the right-hand side is positive definite and hence
$Z_n\le M $.
\par
To find the irreducible representation of supersymmetry we
follow a similar procedure to that which we followed  for  massive
and massless particles. The result crucially depends on whether
$Z_n<M,\ \forall\  n$ or if one or more values of n we
saturate the bound $Z_n=M$.
\par
Let us first consider  $Z_n<M,\ \forall\  n$. In this case,
the right-hand sides of both equations (8) and (9) are
non-zero. Taking
$S^{Am}_1$  and $S^{Am}_2$ to annihilate the vaccum
the physical states are given by the creation opperators
${(S^{Am}_1)}^*$  and
${(S^{Am}_2)}^*$ acting on the vacuum. The resulting representation
has
$2^{2N}$  states and has the same structure as for the massive case
in the absence of a central charge. The states are classified by
$USP(2N)$ as for the massive case.
\par
Let us now suppose  that  $q$ of the $Z_n$'s saturate the bound
i.e $Z_n=M$. For these values of $n$ the right-hand side of
equation (8) vanishes; taking the expectation value of this relation
for any physical state we find that
$$<phys| S^{An}_1 {(S^{An}_1)}^*|phys> +
<phys| {(S^{An}_1)}^*{(S^{An}_1)}|phys>=0
\eqno(10)$$
The scalar product on the space of physical states
satisfies all the axioms of a scalar product and hence we conclude
that both of the above terms vanish and as a result
$$ {(S^{Bn}_1)}^*|phys>=0={(S^{An}_1)}|phys>
\eqno(11)$$
This argument is the same as that used to eliminate
half of the supercharges and their complex conjugates in the massless
case; however in the case under consideration here
 it only eliminates
$q$ of the supercharges and their complex conjugates. There remain the
${N\over 2}$  supercharges ${(S^{Bm}_2)}$ and the ${N\over 2}-q $
supercharges
${(S^{Bm}_1)}$ for the values of $m$ for which $Z_m$ do not saturate
the bound as well as their complex conjugates. These supercharges
form a Clifford algebra and we can take the
${N\over 2}$  supercharges  ${(S^{Bm}_2)}$ and the
${N\over 2}-q $ supercharges ${(S^{Bm}_1)}$ to annihilate the vacuum
and their complex conjugates to be creation operators. The resulting
massive
irreducible representation of supersymmetry has $2^{2(N-q)}$
states and it has the same form as a massive representation
of $N-q$  extended supersymmetry. The states will be classified by
$USp(2N-2q)$.
\par
Clearly, a  representation in which some or all of
the central charges are equal to their mass has fewer states
 than the massive representation  formed  when none of the
central charges saturate the mass, or a massive representation for
which  all
the central charges vanish. This is a consequence of the fact that the
latter  Clifford algebra has more of  its
supercharges active
in the  irreducible representation.
In almost all cases,
the representation with some of its central charges saturated
 contains  a smaller range of  spins than the massive  representation
with no central charges. This
feature plays a very important role in discussions  of duality in
supersymmetric theories.
\par
Let us consider the irreducible representations of   $N=4$
supersymmetry which has
both of its two  possible central charges saturated. These
representations are like the corresponding
$N=2$ massive representations. An important example
has  a ${\underline 1}$ of spin one, a ${\underline 4}$ of spin
one-half  and ${\underline 5}$ of spin zero.
The underlined numbers
are their
$USp(4)$ representations. This representation arises when
the $N=4$ Yang-Mills theory is spontaneously broken by
one of its scalars acquiring a vaccum expectation value. The
theory before being spontaneously broken,
 has a  massless
representation with  one spin one, 2 spin 1/2 ,  and six spin 0.
Examining the  massive representations for $N=4$ in the absence of a
central charge one finds that the  representation with the smallest
spins  has all spins from  spin 2 to  spin 0.   Hence the
spontaneously broken theory can only be supersymmetric if the
representation has a central charge.  Another way to get the count in
the above representation is to take the massless representation and
recall that  when the theory is spontaneously broken
one of the scalars
has been eaten by the vector as a result of the Higgs mechanism.

\par
We close this section by answering a question which may have arisen
in the mind of the reader. For the $N$ extended
supersymmetry algebra the supersymmetry algebra in the rest frame of
equation (3) representation has ${N\over 2 }$ possible central charges.
This makes  one central charge for the case of $N=2$. However, this
number conflicts with our understanding that a particle in $N=2$
supersymmetric Yang-Mills theory can have two central charges
corresponding to its electric and magnetic fields. The resolution of
this conundrum is that although one can use a unitary  transformation
to bring the central charge of  a given irreducible  representation
(i.e. particle) to
have only one independent component one can not do this
simultaneously for all
irreducible multiplets or particles.
\par
Some examples of massive
representations with central charges are given
in the table below.
\medskip
\centerline{Table 8.4 Some massive representations with one central
charge $(|Z|=m)$.}
\centerline{All states are complex.}
$$\offinterlineskip\halign{\hfil$# $\hfil&\quad\strut#&\vrule\quad
\hfil$# $\hfil&\quad\hfil$# $\hfil&\quad\strut#&\vrule\quad\hfil$#
$\hfil&
\quad\hfil$# $\hfil&\quad\strut#&\vrule\quad\hfil$#
$\hfil&\quad\hfil$# $\hfil
&\quad\strut#&\vrule\quad# \cr
\noalign{\hrule}
\qquad N&&&&&&&&&&&\cr
{\rm Spin}&&2&&&4&&&6&&&8\cr
\noalign{\hrule}
{\rm Spin}\ 2&&&&&&&&&1&&1\cr
{\rm Spin}\ {3\over2}&&&&&&1&&1&6&&8\cr
{\rm Spin}\ 1&&&1&&1&4&&6&14+1&&27\cr
{\rm Spin}\ {1\over2}&&1&2&&4&5+1&&14&14^\prime+6&&48\cr
{\rm Spin}\ 0&&2&1&&5&4&&14^\prime&14&&42\cr
\noalign{\hrule}}$$
\par
We have seen that when the central charges take  on
 special values
the irreducible representation contain fewer states than the
generic irreducible massive representation. When this occurs
the states in the representation are called BPS states and the
multiplet of states is called a BPS or short multiplet.
Quantum corrections do not, except in very unusual circumstances,
alter the number of states in a theory,  and as such one expects the
number  and type of
BPS multiplets to be the same in the classical and quantum
theories. This follows from the observation that  for a BPS multiplet
to become a generic   massive multiplet,  the
presence of more states would be  required in the theory and one
does not expect to suddenly find new states in the theory as one
smoothly alters the parameters of the theory.  As such, we can expect
this result to hold regardless of the value of the  coupling constant
of the quantum theory.
\par
This property is very useful since one can verify the existence of
certain BPS states when the coupling constant of the theory is small
by conventional  techniques and it then  follows that
these BPS states  will be present in the same theory when the coupling
constant is large.  In fact, the presence of BPS states is one of the
few things that one can reliably establish in the strong coupling
regime of a  supersymmetric theory.    BPS
multiplets play an important role in discussions on duality;
 if one suspects that two theories are dual one of the things one
can reliably check is the correspondence between the BPS states in
each theory.
\par
 The account of
the  massive irreducible representations of supersymmetry given here
is  along similar lines to the review by Ferrara and Savoy given in
 [21].
\bigskip

{\centerline{\bf 4. Three Ways to Construct a Supergravity Theory}}
\bigskip
In this section it is explained how to construct a supergravity
theory from the knowledge of its on-shell states.
\par
There are three main ways. The method
which has been used most often is the Noether method.
It was used to construct the four-dimensional $N=1$ supergravity in
its on-shell [14][15] and off-shell formulations [16][17]. This method
was  also used in the  construction of  the d=11 supergravity theory
[106]
 and a variant of it was used to find many of the properties  of the
IIB theory [111].
Although this method does not make use of any sophistocated
mathematics and can be rather lengthy,  it is very powerful.
Starting from the linearised theory  for the relavent supermultiplet,
  it gives a systematic  way of finding the final non-linear theory. To
illustrate the method clearly  and without undue technical complexity
 we explain,  in section 4.1.1, how to  construct the
Yang-Mills theory from the linearised theory. In section 4.1.2 we
find the linearised supergravity theory in four dimensions and
apply the Noether proceedure to find the full non-linear theory.
In fact, we will only carry out the first steps in this Noether
procedure, but these   clearly illustrate how to find the final
result, most of whose features are  already apparent at an early
stage of the process. We then give the $N=1$
supergravity theory and show it is invariant.
\par
The second method uses the superspace description of supergravity
theories and in section 4.2 we begin by summarising this
approach. Supergravity theories in superspace share a number of
similarities with the usual theory of general relativity.
They are built from a supervierbein and a spin-connection
and are invariant under  superdiffeomorphisms. Using the
supervierbein and spin-connection we can  construct covariant
derivatives and then define torsions and curvatures in the usual way.
The superspace formulation differs from the usual formulation
of general relativity  in the nature   of the tangent space
group.  The tangent  space of the superspace formulation of
supergravity   contains  fermionic and  bosonic subspaces; however,
the tangent space group is only the Lorentz group which rotates the
odd (bosonic) and even (fermionic) subspaces  seperately. In general
relativity the tangent space group  is also the Lorentz group but in
this case it   rotates  all vectors of a given length into each
other.
\par
 The use of a restricted tangent space group in the superspace
formulation allows us to take some of the torsions and
curvatures to be zero since these constraints are respected by the
Lorentz tangent space group. In fact the torsions and curvatures
form a highly reducible representation of the Lorentz group.
Indeed, the imposition of such constraints  is precisely what is
required to find  the
 correct theory of supergravity. Hence the problem of
finding  the superspace formulation of supergravity is to find which
of the torsions and curvatures are zero. We require different sets of
constriants   for the on-shell  and off-shell supergravity theory.
Clearly one gets from the latter to the former by imposing more
constraints.  To find the constraints for the off-shell theory, when
this exists,  can be rather difficult; however it turns out that to
find the  constraints for the on-shell theory is very straightforward
and  remarkably requires only dimensional analysis. Given a knowledge
of the on-shell states in x-space, which are determined in a
straightforward way from the supersymmetry algebra, as explained in
section three,  we can deduce
 the dimensions of all the  gauge-invariant quantities. By introducing
an notion of  geometric dimension, which in effect absorbs all factors
of Newton's constant
$\kappa$ into the fields,
one finds that for sufficently low dimensions
and certain Lorentz character there are no gauge-invariant tensors
in x-space.
The superspace torsions and  curvatures are  gauge-invariant and so
the superspace tensors with  these dimensions and Lorentz character
must then vanish as there is no available x-space tensor that their
lowest  component could equal. We can substitute  these superspace
constraints  into the the Bianchi idenitites satisfied by the torsions
and curvatures to find constraints on the  torsions
and curvatures of higher dimension. From this set of constraints one
can find an  x-space theory and it turns out that in all known cases
  this theory is none other than the corresponding on-shell
supergravity theory. In other words, the constraints deduced from
dimensional analysis and the use of the Bianchi identities are
sufficient to find the on-shell supergravity in superspace and hence
also in x-space. In section 4.2.2 we explicity carry out this
programme for the four-dimensional
$N=1$ supergravity theory and recover the theory we found
by the Noether
method. This procedure was used to find the  full IIB supergravity in
superspace and in x-space [111]. We  refer the reader to reference
[111]  for the details of this construction.
\par
Finally,  we  briefly mention in section 4.3 the third method of
finding  supergravity theories by the gauging certain space-time
groups.
\par
Several parts of this section are taken from reference [0] and we
keep the same equation numbering as in this reference.
\medskip
{\centerline{\bf 4.1 The Noether Method}}
\medskip
{\bf 4.1.1  Yang-Mills Theory and the Noether Technique}
\medskip
Any theory whose nonlinear form is determined by a gauge principle
can be constructed by a Noether procedure [9]. Because of the
importance of the Noether technique in constructing theories of
supergravity we will take this opportunity to illustrate
the technique within the framework of the simpler
supersymmetric Yang-Mills theory [10].

Let us begin by considering the construction of the Yang-Mills theory
itself from the linearised (free) theory. At the free level the
theory  is invariant under two distinct transformations: rigid and
local  Abelian transformations. Rigid transformations belong to a
group  $S$  with generators $R_i$ which satisfy
$$[R_i,R_j]=s_{ij}^{\ \ k} R_k\eqno(7.1)$$
The structure constants $s_{ij}^{\ \ k}$ may be chosen to be totally
antisymmetric and the indices $i,j,k,\ldots$ can be raised and
lowered with the Kronecker delta $\delta _i^j$. Under these rigid
transformations the vector fields
$A_a^i$ transform as
$$\delta A_a^i=s_{jk}^{\ \ i}T^jA_a^k\eqno(7.2)$$
where $T_j$ are the infinitesimal group parameters. The other type of
transformations are local Abelian transformations
$$\delta A_a^i=\partial_a\Lambda^i\eqno(7.3)$$
Clearly both these transformations form a closed
algebra\footnote{$\dagger$}{In particular one finds
$[\delta_\Lambda,\delta_T]A^i_a
=\partial_a(s_{jk}^{\ \ i}T^j\Lambda^k)$.} The
linearised theory which
is invariant under the transformations of Eqs.  (7.2) and (7.3) is
given by
$$
A^{(0)}=\int d^4x \left\{-{1\over4}f^i_{ab}f^i_{ab}\right\}
\eqno(7.4)$$
where $f^i_{ab}=\partial_aA_b^i-\partial_bA_a^i$
The nonlinear theory is found in a series of steps, the first of
which  is to make the rigid transformations local, i.e. $T^j=T^j(x)$.
Now,
$A^{(0)}$ is no longer invariant under
$$\delta A_a^i= s_{jk}^{\ \ i}T^j(x)A_{a}^k\eqno(7.5)$$
but its variation may be written in the form
$$\delta A^{(0)}=\int
d^4x\big\{\big(\partial_aT^k(x)\big)j^{a k}\big\}\eqno(7.6)$$
where
$$j^{a k}=-s_{ij}^{\ \ k}A_b^if^{ab j}\eqno(7.7)$$
Now consider the action $A_1$
$$A_1=A^{(0)}-{1\over2}g\int d^4x(A_{a}^ij^{ai})
\eqno(7.8)$$
where $g$ is the gauge coupling constant; it is invariant {\it to
order $g^0$ provided} we combine the local transformation $T^i(x)$
with  the local transformation $\Lambda^i(x)$ with the identification
$\Lambda^i(x)=(1/g)T^i(x)$. That is, the initially separate local and
rigid transformations of the linearised theory become knitted
together  into a single local transformation given by
$$\delta A_a^i={1\over
g}\partial_aT^i(x)+s^{ijk}T_j(x)A_a^k(x)
\eqno(7.9)$$
The first term in  the transformation of $\delta A_a^i$
yields in   the variation of the last term in $A_1$,  a
term which cancels the unwanted variation  of $A^{(0)}$.

We now continue with this process of amending the Lagrangian and
transformations order-by-order in $g$ until we obtain an invariant
Lagrangian.
The variation of $A_1$ under the second term of the transformation of
Eq. (7.9) is of  order $g$ and is given by
$$\delta A_1=\int
d^4x\{-g(A_a^iA_b^js_{ij}^{\
\ k})(A_b^l\partial_aT^ms_{lm}^{\ \ k})\}\eqno(7.10)$$ An action
invariant to order $g$ is
$$\eqalignno{
A_2&=A_1+\int
d^4x{g^2\over4}(A_a^iA_b^js_{ij}^{\ \ k})(A^{bl}A^{am}s_{lm}^{\ \
k})\cr &={1\over4}\int d^4x(F_{ab}^i)^2&(7.11)}$$
where
$$F_{ab}^i
=\partial_aA_b^i-\partial_bA_a^i-gs_{jk}^{\ \
i}A_a^jA_b^k
\eqno(7.12)$$
In fact the action $A_2$ is invariant under
the transformations of Eq.  (7.9) to
all orders in $g$ and so represents the final answer, and is, of
course, the well-known action of Yang-Mills theory. The commutator of
two transformations on $A^i_a$ is
$$\eqalignno{
[\delta_{T_1},\delta_{T_2}]A_a^i&=s_{ij}^{\ \ k}T_{2}^j\left({1\over
g}\partial_aT_{1}^k+s_{lm}^{\ \ k}T_1^lA_a^m\right)-(1\
\leftrightarrow\ 2)\cr &={1\over
g}\partial_aT_{12}^i+s_{ij}^{\ \ k}T_{12}^jA_{a}^k&(7.13)}$$
where
$$T_{12}^i=s_{jk}^{\ \ i}T_{2}^jT_{1}^k$$
and so the transformations form a closed algebra. In the last step we
used the Jacobi identity in terms of the structure constants.

For supergravity and other local theories the procedure is similar,
although somewhat more complicated. The essential steps are to first
make the rigid transformations local and find invariant Lagrangians
order by order in the appropriate gauge coupling constant. This is
achieved in general not only by adding terms to the action, but also
adding terms to the transformation laws of the field. If the
latter process occurs one must also check the closure order by order
in the gauge coupling constant.

Although one can use a Noether procedure which relies on the
existence  of an action, one can also use a Noether method which uses
the  transformation laws alone. This works, in the Yang-Mills case, as
follows: upon making the rigid transformation local as in Eq. (7.5)
one  finds that the algebra no longer closes, i.e.
$$[\delta_\Lambda,\delta_T]A_a^i
=\partial_a\big(s_{ij}^{\ \ i}(T^j\Lambda^k)\big)-s_{jk}^{\ \
i}(\partial_aT^j)\Lambda^k
\eqno(7.14)$$
The cure for this is to
regard the two transformations as simultaneous  and knit them together
as explained above. Using the new transformation  for $A_a^i$ of
Eq. (7.9) we then test the closure to order $g^0$. In fact, in this
case the  closure works to all orders in $g$ and the process stops
here; in general however, one must close the algebra order by order
in  the coupling constant modifying the transformation laws and the
closure  relations for the algebra. Having the full transformations it
is then  easy to find the full action when that exists.

Let us apply the above Noether proceedure to find the $N=1$
supersymmetric  Yang-Mills theory from its linearised theory. In the
linearised theory the fields
$A_a^i,\ \lambda^i,\ D^i$ have the rigid transformations $T_i$ and
local transformations
$\Lambda^i(x)$ given by
$$\eqalignno{
\delta
A_a^i&=s_{jk}^{\ \ i}T^jA_{a}^k, \
\delta\lambda^i=s_{jk}^{\ \ i}T^j\lambda^k\cr
\delta D^i&=s_{jk}^{\ \ i}T^jD^k&
(7.15)\cr
\noalign{and}
\delta A_a^i&=\partial_a\Lambda^i\quad\delta
D^i=0,\quad\delta\lambda^i=0&
(7.16)}$$
The  supersymmetry transformations of the linearised theory are
given by
$$\eqalignno{
\delta A_a^i&=\bar\varepsilon\gamma_a\lambda^i\cr
\delta\lambda^i&=-{1\over2}\sigma^{cd}f^i_{cd}\varepsilon
+i\gamma_5D^i\varepsilon\cr
\delta D^i&=i\bar\varepsilon\gamma_5\partial\!\!\!/\lambda^i&(7.17)}$$
These transformations form a closed algebra, and leave invariant the
following linearised Lagrangian
$$L=-{1\over4}(f^i_{cd})^2
-{1\over2}\bar\lambda^i\partial\!\!\!/\lambda^i+{1\over
2}(D^i)^2
\eqno(7.18)$$
Let us use the Noether method on the algebra to find the nonlinear
theory. Making the rigid transformation on $A_a^i$ local we must, as
in  the Yang-Mills case, knit the rigid and local transformations
together
\big(i.e.
$\Lambda^i(x)=(1/g) T^i(x)\big)$ to gain closure of gauge
transformations on $A_\mu^i$. Closure of
supersymmetry and gauge transformations implies that the rigid
transformations
on $\lambda^i$ and $D^i$ also become local. This particular closure
also requires that all the supersymmetry transformations are modified
to involve covariant quantities. For example, we find that on $D^i$
$$[\delta_T,\delta_\varepsilon]D^i
=i\bar\varepsilon\gamma_5\gamma^as_{jk}^{\ \ i}(\partial
_aT^j)\lambda^k
\eqno(7.19)$$
and as a result we must replace $\partial_a\lambda^i$ by ${\cal
D}_a\lambda^i=
\partial_a\lambda^i-gs_{jk}^{\ \ i}A_{a}^j\lambda^k$ in the $\delta
D^i$ of  Eq. (7.17)
and then the commutator $[\delta_\Lambda,\delta_\varepsilon]$ is zero
to all orders in $g$. The algebra then takes the form
$$\eqalignno{
\delta A_a^i&=\bar\varepsilon\gamma_a\lambda^i\cr
\delta\lambda^i&=\left(-{1\over2}\sigma^{cd}F_{cd}^i
+i\gamma_5D^i\right)\varepsilon\cr
\delta D^i&=i\bar\varepsilon\gamma_5{\cal
D}\!\!\!/\lambda^i&
(7.20)}$$ where
$$F_{ab}^i=\partial_aA_b^i-\partial_bA_a^i-gs_{jk}^{\ \
i}A_a^jA_b^k$$  We must now verify that the above supersymmetry
transformations close.  For other supersymmetric gauge theories we
must add further terms to  the supersymmetry transformations in order
to regain closure. However,  in this case gauge covariance and
dimensional analysis ensure that  there are no possible terms that one
can  add to these supersymmetry  transformations and so the
transformations of Eqs. (7.20) must be the  complete laws for the full
theory. The reader may verify that  there are no inconsistencies by
showing that the algebra does indeed  close.

The action invariant under these transformations is
$$A=\int d^4x\left\{-{1\over4}(F_{ab}^i)^2
-{1\over2}\bar\lambda^i{\cal
D}\!\!\!/\lambda^i+{1\over2}(D^i)^2\right\}
\eqno(7.21)$$
One could also have used the Noether procedure on the action. Gauge
invariance implies that the action is that given in Eq. (7.21).
Demanding that this gauge invariant action be supersymmetric requires
us to modify the supersymmetry transformations to those of Eq.
(7.20).
\medskip
{\bf 4.1.2 $N=1$ $D=4$ Supergravity}
\medskip
 We will now construct  $N=1$ $D=4$ supergravity using the Noether
method. The starting point is the linearised theory which
we now  construct  along similar lines to the
method used to find the Wess-Zumino model and $N=1$ super QED
in chapters 5 and 6 in reference [0].
\medskip
{\bf {The Linearised Theory}}
\medskip
 We will start with the on-shell
states and construct the   linearized theory, without and then with
auxiliary fields. The $N=1$ irreducible representations of
supersymmetry which include a   spin 2 graviton contain either a
spin 3/2 or a spin 5/2 fermion. The   spin 5/2 particle would seem to
have considerable problems in    coupling to other fields and so we
will choose the spin 3/2 particle.

As in the Yang-Mills case, the linearized theory possesses rigid
supersymmetry and local Abelian gauge invariances. The latter
invariances are required, in order that the fields do describe the
massless on-shell states alone without involving
ghosts. We recall that a rigid symmetry is one
whose parameters are space-time independent while a local symmetry
has  space-time dependent parameters.

These on-shell states are represented by a symmetric second rank
tensor field, $h_{\mu\nu}\ (h_{\mu\nu}=h_{\nu\mu}$) and a Majorana
vector spinor, $\psi_{\mu\alpha}$. For these fields to represent
a spin 2 particle and a spin 3/2 particle they must possess the
infinitesimal gauge transformations
$$\eqalignno{
\delta h_{\mu\nu}&=\partial_\mu\xi_\nu(x)+\partial_\nu\xi_\mu(x)\cr
\partial\psi_{\mu\alpha}&=\partial_\mu\eta_\alpha(x)&(9.1)}$$
The unique ghost-free gauge-invariant, free field equations are
$$E_{\mu\nu}=0,\quad R^\mu=0\eqno(9.2)$$
where $E_{\mu\nu}=R^L_{\ \mu\nu}-{1\over2}\eta_{\mu\nu}R^L,\
R^{Lab}_{\ \ \ \mu\nu}$ is the
linearized Riemann tensor given by
$$\eqalignno{R^{Lab}_{\ \ \ \mu\nu}&=-\partial_a\partial_\mu
h_{b\nu}+\partial_b\partial_\mu h_{a\nu}+\partial_
a\partial_\nu h_{b\mu}-\partial_b\partial_\nu h_{a\mu}\cr
\noalign{and}
R^\mu
&=\varepsilon^{\mu\nu\rho\kappa}i\gamma_5\gamma_\nu\partial_\rho\psi_\kappa\cr
R^{Lb}_{\ \ \mu}&=R^{Lab}_{\ \ \ \mu\nu}\delta^\mu_a\cr
R^L&=R^{La}_{\ \ \mu}\delta^\mu_a&(9.3)}$$
For an explanation of this point see van Nieuwenhuizen. [13].

We must now search for the supersymmetry transformations that form an
invariance of these field equations and represent the supersymmetry
algebra on-shell. On dimensional grounds the most
general transformation is
$$\eqalignno{
\delta
h_{\mu\nu}&={1\over2}(\bar\varepsilon\gamma_\mu\psi_\nu
+\bar\varepsilon\gamma_\nu\psi_\mu)
+\delta_1\eta_{\mu\nu}\bar\varepsilon\gamma^\kappa\psi_\kappa\cr
\delta\psi_\mu&=+\delta_2\sigma^{ab}\partial_ah_{b\mu}
\varepsilon+\delta_3\partial_\nu
h^\nu_{\ \mu}\varepsilon&(9.4)}$$
The parameters $\delta_1,\ \delta_2$ and $\delta_3$
 will be determined by the demanding that the set of
transformations which comprise the supersymmetry transformations of
equation  (9.4) and the gauge transformations of equation  (9.1)
should
form a  closing algebra when the field equations of
equation  (9.2) hold. At
the  linearized level the supersymmetry transformations are linear
rigid transformations, that is, they are {\it first order} in the
fields
$h_{\mu\nu}$ and $\psi_{\mu\alpha}$ and parametrized by {\it
constant} parameters
$\varepsilon^\alpha$.

Carrying out the commutator of a Rarita-Schwinger gauge
transformation, $\eta_\alpha(x)$ of Eq.
(9.1) and a supersymmetry transformation, $\varepsilon$ of Eq. (9.4)
on $h_{\mu\nu}$, we get:
$$[\delta_\eta,\delta_\varepsilon]h_{\mu\nu}={1\over2}(\bar\varepsilon\gamma
_\mu\partial_\nu\eta
+\bar\varepsilon\gamma_\nu\partial_\mu\eta)
+\delta_1\eta_{\mu\nu}\bar\varepsilon\partial\!\!\!/\eta\eqno(9.5)$$
This is a gauge transformation with parameter
${1\over2}\bar\varepsilon\gamma_\mu\eta$ on $h_{\mu\nu}$ provided
$\delta_1=0$. Similarly, calculating the  commutator of a gauge
transformation of $h_{\mu\nu}$ and a  supersymmetry transformation on
$h_{\mu\nu}$  automatically yields the  correct  result, i.e., zero.
However, carrying out the commutator of a supersymmetry
transformation and an Einstein  gauge transformation on
$\psi_{\mu\alpha}$ yields
$$[\delta_{\xi_\mu},\delta_\varepsilon]\psi_\mu=+\delta_2\sigma^{ab}\partial
_a(\partial_\mu\xi_b)
\varepsilon+\delta_3\partial_\nu\partial^\nu\xi_\mu\varepsilon
+\delta_3\partial_\nu\partial_\mu\xi^\mu\varepsilon
\eqno(9.6)$$
which is a Rarita-Schwinger gauge transformation on $\psi_\mu$
provided $\delta_3=0$. Hence we take
$\delta_1=\delta_3=0$.

We must test the commutator of two supersymmetries. On $h_{\mu\nu}$
we  find the commutator of two supersymmetries to give
$$\eqalignno{
[\delta_{\varepsilon_1},\delta_{\varepsilon_2}]h_{\mu\nu}&
=+{1\over2}\{\bar\varepsilon_2\gamma_\mu
\delta_2\sigma^{ab}\partial_ah_{b\nu}\varepsilon_2
+(\mu\leftrightarrow\nu)\}-(1\leftrightarrow2)\cr
&=\delta_2\{\bar\varepsilon_2\gamma^b\varepsilon_1\partial_\mu
h_{b\nu}-\bar\varepsilon_2\gamma^a
\varepsilon_1\partial_a h_{\mu\nu}-(\mu\leftrightarrow\nu)\}&(9.7)}$$
This is a gauge transformation on $h_{\mu\nu}$ with parameter
$\delta_2\bar\varepsilon_2\gamma^b
\varepsilon_1h_{b\nu}$ as well as a space-time translation. The
magnitude of this translation   coincides with that dictated
by the supersymmetry group provided $\delta_2=-1$ which is the value
we now adopt.

It is important to stress that linearized supergravity differs from
the Wess-Zumino model in that  one must take into account the gauge
transformations of Eqs. (9.1) as well as the rigid supersymmetry
transformations of Eq. (9.4) in order to obtain a closed algebra. The
resulting algebra is the $N=1$ supersymmetry algebra when
supplemented  by gauge transformations. This algebra reduces to the
$N=1$  supersymmetry algebra only on gauge-invariant states.

For the commutator of two supersymmetries on $\psi_\mu$ we find
$$\eqalignno{
[\delta_{\varepsilon_1},\delta_{\varepsilon_2}]\psi_\mu&
=-\sigma^{ab}\partial_a\varepsilon_2{1\over2}
(\varepsilon_1\gamma_b\psi_\mu+\bar\varepsilon_1\gamma_\mu\psi_b)
-(1\leftrightarrow2)\cr
&=+{1\over2.4}\sum_Rc_R\bar\varepsilon_1\gamma_R\varepsilon_2\sigma^{ab}
\partial_a\gamma^R
(\gamma_b\psi_\mu+\gamma_\mu\psi_b)-(1\leftrightarrow2)\cr
&={1\over8}\bar\varepsilon_1\gamma_R\varepsilon_2
\sigma^{ab}\gamma^R\left(\gamma_b\psi_{a\mu}+
{1\over2}\gamma_\mu\psi_{ab}\right)\cr
&\quad+\partial_\mu\left({1\over8}\bar\varepsilon_1\gamma_R
\varepsilon_2\sigma^{ab}\gamma^R\gamma
_b\psi_a\right)-(1\leftrightarrow2)&(9.8)}$$
where $\psi_{\mu\nu}=\partial_\mu\psi_\nu-\partial_\nu\psi_\mu$.
Using  the different forms of the Rarita-Schwinger equation of motion,
given  by
$$\eqalignno{
&R^\mu=0\cr
&\Leftrightarrow\gamma^\mu\psi_{\mu\nu}=0\cr
&\Leftrightarrow\psi_{\mu\nu}+{1\over2}i\gamma_5
\varepsilon_{\mu\nu\rho\kappa}\psi^{\rho\kappa}=0&(9.9)}$$
we find the final result to be
$$\eqalignno{
[\delta_{\varepsilon_1},\delta_{\varepsilon_2}]\psi_\mu&
=2\bar\varepsilon_2\gamma^c\varepsilon_1
\partial_c\psi_\mu+\partial_\mu
\big(-\bar\varepsilon_2\gamma^c\varepsilon_1\psi_c\cr
&\quad+\sum_Rc_R{1\over8}\bar\varepsilon_1
\gamma_R\varepsilon_2\sigma^{ab}\gamma^R
\gamma_b\psi_a-(1\leftrightarrow2)\big)&(9.10)}$$
This is the required result: a translation and a gauge transformation
on $\psi_\mu$.

The reader can verify that the transformations of Eq. (9.4) with the
values of $\delta_1=\delta_3=0,\ \delta_2=-1$ do indeed leave the
equations of motion of
$h_{\mu\nu}$ and $\psi_{\mu\alpha}$ invariant.

Having obtained an irreducible representation of supersymmetry
carried  by the fields $h_{\mu\nu}$ and $\psi_{\mu\alpha}$ when
subject to their  field equations we can now find
the algebraically on-shell Lagrangian. The action (Freedman, van
Nieuwenhuizen and Ferrara [14], Deser and Zumino [15]) from which the
field equations of Eq. (9.2) follow, is
$$A=\int
d^4x\left\{-{1\over2}h^{\mu\nu}E_{\mu\nu}-{1\over2}\bar\psi_\mu
R^\mu\right\}\eqno(9.11)$$
It is invariant under the transformations of Eq. (9.4) provided we
adopt the values for the parameters $\delta_1,\ \delta_2$ and
$\delta_3$ found above. This invariance holds without use of the
field  equations, as it did in the Wess-Zumino case.

We now wish to find a linearized formulation which is built from
fields which carry a representation of supersymmetry without imposing
any restrictions (i.e., equations of motion), namely, we find the
auxiliary fields. As a guide to their number we can
apply our Fermi-Bose counting rule which, since the algebra contains
gauge transformations, applies only to the gauge-invariant states.
On-shell, $h_{\mu\nu}$ has two helicities and so does
$\psi_{\mu\alpha}$; however off-shell, $h_{\mu\nu}$ contributes
$(5\times4)/2=10$
degrees of freedom minus 4 gauge degrees of freedom giving 6 bosonic
degrees of freedom.
On the other hand, off-shell $\psi_{\mu\alpha}$ contributes
$4\times4=16$ degrees of freedom minus 4 gauge degrees of freedom,
giving 12 fermionic degrees of freedom. Hence the auxiliary fields
must  contribute 6 bosonic degrees of freedom. If there are $n$
auxiliary    fermions there  must be $4n+6$ bosonic auxiliary fields.

Let us assume that a minimal formulation exists, that is, there are
no  auxiliary spinors. Let us also assume that the bosonic auxiliary
fields  occur in the Lagrangian as squares without derivatives (like
$F$ and
$G$) and so are of dimension two. Hence we have 6 bosonic auxiliary
fields; it only remains to find their Lorentz character and
transformations. We will assume that they consist of a scalar $M$, a
pseudoscalar $N$  and a pseudovector $b_\mu$, rather than an
antisymmetric tensor or 6 spin-0 fields. We will give the motivating
arguments for this later.

Another possibility is the two fields $A_\mu$ and $a_{\kappa\lambda}$
which possess the gauge transformations $\delta
A_\mu=\partial_\mu\Lambda;\ \delta a_{\kappa\lambda}=\partial_\kappa
\Lambda_\lambda-\partial_\lambda\Lambda_\kappa$. A contribution
$\varepsilon_{\mu\nu\rho\kappa}
A^\mu\partial^\nu a^{\rho\kappa}$ to the action would not lead to
propagating degrees of freedom.

The transformation of the fields $h_{\mu\nu},\ \psi_{\mu\alpha},\ M,\
N$ and $b_\mu$ must reduce on-shell to the on-shell transformations
found above. This restriction, dimensional arguments and the
fact that if the auxiliary fields are to vanish on-shell they must
vary into field equations gives
the transformations to be [16,17]
$$\eqalignno{
\delta
h_{\mu\nu}&={1\over2}(\bar\varepsilon\gamma_\mu\phi_\nu
+\bar\varepsilon\gamma_\nu\psi_\mu)\cr
\delta\psi_{\mu\alpha}&=-\sigma^{ab}\partial_ah_{b\mu}
\varepsilon-{1\over3}\
\gamma_\mu(M+i\gamma_5N)
\varepsilon+b_\mu
i\gamma_5\varepsilon+\delta_6\gamma_\mu/\!\!\!bi\gamma_5\varepsilon\cr
\delta M&=\delta_4\bar\varepsilon\gamma\cdot R\cr
\delta N&=\delta_5i\bar\varepsilon\gamma_5\gamma\cdot R\cr
\delta
b_\mu&=+\delta_7i\bar\varepsilon\gamma_5R_\mu+\delta_8i
\bar\varepsilon\gamma_5\gamma_\mu\gamma\cdot
R&(9.12)}$$ The parameters $\delta_4,\ \delta_5,\ \delta_6,\ \delta_7$
and
$\delta_8$ are determined by the restriction
that the above transformations of Eq. (9.12) and the gauge
transformations of Eq. (9.1) should form a closed
algebra. For example, the commutator of two supersymmetries on $M$
gives
$$[\delta_{\varepsilon_1},\delta_{\varepsilon_2}]M
=\delta_4\{-\bar\varepsilon_2\gamma^\mu\varepsilon_1
\partial_\mu
M+16\bar\varepsilon_2\sigma^{\mu\nu}i\gamma_5
\varepsilon_1(1+3\delta_6)\partial_\mu
b_\nu\}\eqno(9.13)$$
which is the required result provided $\delta_4=-{1\over2}$ and
$\delta_6=-{1\over3}$.
Carrying out the commutator of two supersymmetries on all fields we
find a closing algebra provided
$$\delta_4=-{1\over2},\quad\delta_5=-{1\over2},\quad\delta_6
=-{1\over3},\quad\delta_7={3\over2}\quad{\rm
 and}\quad\delta_8=-{1\over2}\eqno(9.14)$$
We henceforth adopt these values for the parameters. An action which
is constructed from the fields
$h_{\mu\nu},\ \psi_{\mu\alpha},\ M,\ N$ and $b_\mu$ and is invariant
under the transformations of Eq.
(9.12) with the above values of the parameters is [16],[17]
$$A=\int
d^4x\left\{-{1\over2}h_{\mu\nu}E^{\mu\nu}-{1\over2}\bar\psi_\mu
R^\mu-{1\over3}(M^2+N^2-b^\mu b_\mu)\right\}\eqno(9.15)$$
This is the action of linearized $N=1$ supergravity and upon
elimination of the auxiliary field $M,\ N$
and $b_\mu$ it reduces to the algebraically on-shell Lagrangian of
Eq.  (9.11).
\medskip
{\bf The Nonlinear Theory}
\medskip
The full nonlinear theory of supergravity can be found from the
linearized theory discussed above by applying
the Noether technique discussed at the beginning of this section.
Just as in the case of Yang-Mills  the reader will observe that the
linearized  theory possesses the local Abelian invariances of Eq.
(9.1) as well as  the rigid (i.e., constant parameter) supersymmetry
transformations of Eq. (9.4).

We proceed just as in the case of the Yang-Mills theory and make the
parameter of rigid transformations space-time
dependent, i.e., set $\varepsilon=\varepsilon(x)$ in Eq. (9.4). The
linearized action of Eq. (9.15) is then no longer
invariant, but its variation must be of the form
$$\delta A_0=\int d^4x\partial_\mu\bar\varepsilon^\alpha j^\mu_{\
\alpha}\eqno(9.16)$$
since it is invariant when $\bar\varepsilon^\alpha$ is a constant.
The  object $j^\mu_{\ \alpha}$ is proportional to
$\psi_{\mu\alpha}$ and linear in the bosonic fields $h_{\mu\nu},\ M$
or $N$ and $b_\mu$. As such, on dimensional
grounds, it must be of the form
$$j_{\mu\alpha}\propto\partial_\tau h_{\rho\mu}\psi_{\nu\beta}+\dots
$$
Consider now the action, $A$, given by
$$A_1=A_0-{\kappa\over4}\int d^4x\bar\psi^\mu j_\mu
\eqno(9.17)$$
where $\kappa$ is the gravitational constant. The action $A$ is
invariant to order $\kappa^0$ {\it provided} we
combine the now local supersymmetry transformation of Eq. (9.12) with
a local Abelian Rarita-Schwinger gauge transformation
of Eq. (9.1) with parameter $\eta(x)=(2/\kappa)\varepsilon(x)$. That
is, we make a transformation
$$\eqalignno{
\delta\psi_\mu&
={2\over\kappa}\partial_\mu\varepsilon(x)
-\partial_ah_{b\mu}\sigma^{ab}\varepsilon(x)-{1\over3}\gamma
_\mu(M+i\gamma_5N)\varepsilon(x)\cr
&\quad+i\gamma_5\left(b_\mu-{1\over3}\gamma_\mu \gamma^\nu
b_\nu\right)
\varepsilon(x)&(9.18)}$$
the remaining fields transforming as before except that $\varepsilon$
is now space-time dependent.

As in the Yang-Mills case the two invariances of the linearized
action  become knitted together to form one transformation, the role
of gauge  coupling being played by the gravitational constant,
$\kappa$. The addition of the term $(-\kappa/4)\bar\psi^\mu j_\mu$ to
$A_0$ does the required job; its variation is
$$-{\kappa\over4}\cdot2\cdot\left({2\over\kappa}\right)(\partial_\mu
\bar\varepsilon)j^\mu+{\rm
terms\ of\ order\ }\kappa^1\eqno(9.19)$$
The order $\kappa^1$ terms do not concern us at the moment. We note
that $j_{\mu\alpha}$ is linear in $\psi_{\mu\alpha}$
and so we get a factor of 2 from $\delta\psi_{\mu\alpha}$.

In fact, one can carry out the Noether procedure  in the context of
pure gravity where one finds at the linearized level the rigid
translation
$$\delta h_{\mu\nu}=\zeta^\lambda\partial_\lambda
h_{\mu\nu}\eqno(9.20)$$
 and the local gauge transformation
$$\delta
h_{\mu\nu}=\partial_\mu\xi_\nu+\partial_\nu\xi_\mu\eqno(9.21)$$
These become knitted together at the first stage of the Noether
procedure to give
$$\delta
h_{\mu\nu}={1\over\kappa}\partial_\mu\zeta_\nu
+{1\over\kappa}\partial_\nu\zeta_\mu+\zeta^\lambda\partial_\lambda
h_{\mu\nu}\eqno(9.22)$$
since $\xi_\nu=(1/\kappa)\zeta_\nu$. This variation of $h_{\mu\nu}$
contains the first few terms of an Einstein general
coordinate transformation of the vierbein which is given in terms of
$h_{\mu\nu}$ by
$$e^{\ a}_\mu=\eta^{\ a}_\mu+\kappa h^{\ a}_\mu\eqno(9.23)$$
We proceed in a similar way to the Yang-Mills case. We obtain order
by  order in $\kappa$ an invariant Lagrangian by adding
terms to the Lagrangian and in this case also adding terms to the
transformations of the fields. For example, if we added a
term to $\delta\psi_\mu$ say,
$\delta\bar\psi_\mu=\dots+\bar\varepsilon X_\mu\kappa$, then from the
linearized action we receive
a contribution $-\kappa\bar\varepsilon X_\mu R^\mu$ upon variation of
$\psi_\mu$. It is necessary at each step (order of
$\kappa$) to check that the transformations of the fields form a
closed algebra. In fact, any ambiguities that arise in the
procedure are resolved by demanding that the algebra closes.

The final set of transformations [16,17] is
$$\eqalignno{\delta e^{\ a}_\mu&
=\kappa\bar\varepsilon\gamma^a\psi_\mu\cr
\delta\psi_\mu&=2\kappa^{-1}D_\mu\big(w(e,\psi)\big)\varepsilon
+i\gamma_5\left(b_\mu-{1\over3}\gamma_\mu/\!\!\!b\right)\varepsilon
-{1\over3}\gamma_\mu(M+i\gamma_5N)\varepsilon\cr
\delta M&=-{1\over2}e^{-1}\bar\varepsilon\gamma_\mu
R^\mu-{\kappa\over2}i\bar\varepsilon\gamma_5\psi_\nu
b^\nu-\kappa\bar\varepsilon
\gamma^\nu\psi_\nu
M+{\kappa\over2}\bar\varepsilon(M+i\gamma_5N)\gamma^\mu\psi_\mu\cr
\delta N&=-{e^{-1}\over2}i\bar\varepsilon\gamma_5\gamma_\mu
R^\mu+{\kappa\over2}\bar\varepsilon\psi_\nu b^\nu-\kappa\bar\varepsilon
\gamma^\nu\psi_\nu
N-{\kappa\over2}i\bar\varepsilon\gamma_5(M+i\gamma_5N)
\gamma^\mu\psi_\mu\cr
\delta
b_\mu&={3i\over2}e^{-1}\bar\varepsilon\gamma_5\left(g_{\mu\nu}-{1\over3}
\gamma_\mu\gamma_\nu\right)R^\nu+\kappa\bar\varepsilon
\gamma^\nu
b_\nu\psi_\mu-{\kappa\over2}\bar\varepsilon\gamma^\nu\psi_\nu b_\mu\cr
&\quad-{\kappa\over2}i\bar\psi_\mu\gamma_5(M+i\gamma_5N)
\varepsilon-{i\kappa\over4}\varepsilon_\mu^{\ bcd}b_b\bar\varepsilon\gamma_5
\gamma_c\psi_d&(9.24)}$$
where
$$\eqalignno{
R^\mu&=\varepsilon^{\mu\nu\rho\kappa}i\gamma_5\gamma_\nu
D_\rho\big(w(e,\psi)\big)\psi_\kappa\cr
D_\mu\big(w(e,\psi)\big)&=\partial_\mu+w_{\mu
ab}{\sigma^{ab}\over4}}$$
 and
$$\eqalignno{
w_{\mu ab}&={1\over2}e^\nu_{\ a}(\partial_\mu e_{b\nu}-\partial_\nu
e_{b\mu})-{1\over2}e_b^{\ \nu}(\partial_\mu e_{a\nu}-\partial_\nu
e_{a\mu})\cr
&\quad-{1\over2}e_a^{\ \rho}e_b^{\ \sigma}(\partial_\rho e_{\sigma
c}-\partial_\sigma a_{\rho c})e_\mu^{\ c}\cr
&\quad+{\kappa^2\over4}(\bar\psi_\mu\gamma_a\psi_b
+\bar\psi_a\gamma_\mu\psi_b-\bar\psi_\mu\gamma_b\psi_a)&(9.25)}$$
They form a closed algebra, the commutator of two supersymmetries on
any field being
$$\eqalignno{
[\delta_{\varepsilon_1},\delta_{\varepsilon_2}]&=\delta_{\rm
supersymmetry}(-\kappa\xi^\nu\psi_\nu)+\delta_{\rm general\
coordinate}(2\xi_\mu)\cr
&\quad+\delta_{\rm Local\
Lorentz}\left(-{2\kappa\over3}\varepsilon_{ab\lambda\rho}
b^\lambda\xi^\rho\right.\cr
&\quad\left.-{2\kappa\over3}\bar\varepsilon_2\sigma_{ab}
(M+i\gamma_5N)\varepsilon_1+2\xi^dw_d^{\
ab}\right)&(9.26)}$$
where
$$\xi_\mu=\bar\varepsilon_2\gamma_\mu\varepsilon_1$$
The transformations of Eq. (9.24) leave invariant the action
[16], [17]
$$A=\int d^4x\left\{{e\over2\kappa^2}R-{1\over2}\bar\psi_\mu
R^\mu-{1\over3}e(M^2+N^2-b_\mu b^\mu)\right\}\eqno(9.27)$$
where
$$\eqalignno{
R&=R_{\mu\nu}^{\ \ ab}e_a^{\ \mu}e_b^{\ \nu}\cr
\noalign{and}
R_{\mu\nu}^{\ \ ab}{\sigma_{ab}\over4}&=[D_\mu,D_\nu]}$$
The auxiliary fields $M,\ N$ and $b_\mu$ may be eliminated to obtain
the nonlinear algebraically on-shell Lagrangian which was the form in
which supergravity was originally found in Refs. 14 and 15.

As discussed  at the beginning of this section  one could also build
up the non-local theory  by working with the algebra of field
transformations alone.
\medskip
{\bf { Invariance of $N=1$ Supergravity}}
\medskip
We refer the reader to chapter 10 of reference [0] for a
demonstration of the invariance of the supergravity under
local supersymmetry transformations. This proof [18,19] uses the
1.5 order formalism.
\medskip
{\centerline{\bf 4.2 On-Shell $N=1$ $D=4$ Superspace}}
\medskip
In this section we will construct the on-shell superspace
 formulation of $N=1$ $D=$ supergravity, from which we recover the
equations of motion in $x$-space.
\medskip
{\bf { 4.2.1 Geometry of Local Superspace}}
\medskip
The geometrical framework [69] of superspace
supergravity has many of
the constructions of general relativity, but also requires additional
input. A useful guide in the  construction of local superspace is that
it should admit rigid  superspace as a limit.

We begin with an eight-dimensional manifold
$z^\Pi=(x^u,\theta^{\underline\alpha})$ where $x^u$ is a commuting
coordinate and  $\theta^{\underline\alpha}$ is a Grassmann odd
coordinate. On this manifold a  super-general coordinate
reparametrization has the form
$$z^\Pi\rightarrow z^{\prime\Pi}=z^\Pi+\xi^\Pi\eqno(16.1)$$
where $\xi^\Pi=(\xi^u,\xi^{\underline\alpha})$ are arbitrary
functions  of $z^\Pi$.

Just as in general relativity we can consider scalar superfields,
that  is, fields for which
$$\phi^\prime(z^\prime)=\phi(z)\eqno(16.2)$$  and superfields with
superspace world indices $\varphi_\Lambda$; for  example
$$\varphi_\Lambda={\partial\phi\over\partial z^\Lambda}\eqno(16.3)$$
The latter transform as
$$\varphi^\prime_\Lambda(z^\prime)={\partial z^\Pi\over\partial
z^{\prime\Lambda}}\varphi_\Pi(z)\eqno(16.4)$$
The transformation properties of higher order tensors is obvious.

We must now specify the geometrical structure of the manifold. For
reasons that will become apparent, the superspace formulation is
essentially a vierbien formulation. We introduce supervierbiens
$E_\Pi^{\ N}$ which transform under the supergeneral coordinate
transformations as
$$\delta E_\Pi^{\ N}=\xi^\Lambda\partial_\Lambda
E_\Pi^{\ N}+\partial_\Pi\xi^\Lambda E_\Lambda^{\ N}\eqno(16.5)$$
The $N$-index transforms under the tangent space group which is taken
to be just the Lorentz group;
and so $\delta E_\Pi^{\ N}=E_\Pi^{\ M}\Lambda_M^{\ N}$ where
$$\Lambda_M^{\ N}=\left(\matrix{
\lambda_m^{\ \ n}&0&0\cr 0&-{1\over4}(\sigma_{mn})_A^{\
B}\Lambda^{mn}&0\cr
0&0&+{1\over4}(\sigma_{mn})_{\dot A}^{\ \dot
B}\Lambda^{mn}\cr}\right)\eqno(16.6)$$
The matrix $\Lambda_m^{\ n}$ is an arbitrary function on superspace
and it governs not only the rotation of the vector index, but also
the  rotation of the spinorial indices. Since we are dealing
with an eight-dimensional manifold one could choose a much larger
tangent space group. For example,
$\Lambda_M^{\ N}$ could be an arbitrary matrix that preserves the
metric
$$g_{NM}={\rm
diag}(a_1\eta_{mn},a_2\varepsilon_{AB},a_3\varepsilon_{\dot A\dot
B})\eqno(16.7)$$
where $a_1,\ a_2$ and $a_3$ are non-zero arbitrary constants.
Demanding reality of the metric implies
$a_2^*=a_3$ and we may scale away one factor. This corresponds to
taking the tangent space group to be $OSp(4,1)$. In such a
formulation
 one could introduce a metric $g_{\Pi\Lambda}=
E_\Pi^{\ N}g_{NM}E_\Lambda^{\ M}$ and one would have a formulation
which mimicked Einstein's
general relativity at every step [70].

Such a formulation, however, would not lead to the $x$-space
component
$N=1$ supergravity given
earlier. One way to see this is to observe that the above tangent
space group does not coincide
with that of rigid superspace (super Poincar\'e/Lorentz), which has
the Lorentz group, as given
in Eq. (16.6) with $\Lambda_m^{\ n}$ a constant matrix, as its
tangent  space group. As linearized
superspace supergravity must admit a rigid superspace formulation,
any  formulation based on an
$OSp(4,1)$ tangent group will not coincide with linearized
supergravity. In fact the $OSp(4,1)$
formulation has a higher derivative action.

An important consequence of this restricted tangent space group is
that tangent supervectors $V^N=V^\Pi E_\Pi^{\ N}$ belong to a
reducible representation of the Lorentz group. This allows one
to write down many more invariants. The objects $V^mV_m,\
V^AV^B\varepsilon_{AB},\ V^{\dot A}V^{\dot B}\varepsilon_{\dot B\dot
A}$ are all separately invariant.

In other words, in the choice of metric in Eq. (16.7) the constants
$a_1,\ a_2,\ a_3$ can have any value including zero.

We define a Lorentz valued spin connection
$$\Omega_{\Lambda M}^{\ \
N}=\left(\matrix{
\Omega_{\Lambda m}^{\ \ \  n}&0&0\cr 0&-{1\over4}\Omega_\Lambda^{\
mn}(\sigma_{mn})_A^{\ B}&0\cr
0&0&{1\over4}\Omega_\Lambda^{\ mn}(\bar\sigma_{mn})_{\dot A}^{\ \dot
B}\cr}\right)\eqno(16.8)$$
This object transforms under super general coordinate transformations
as
$$\delta\Omega_{\Lambda M}^{\ \ \
N}=\xi^\Pi\partial_\Pi\Omega_{\Lambda M}^{\ \ \ N}+\partial_\Lambda
\xi^\Pi\Omega_{\Pi M}^{\ \ \ N}\eqno(16.9)$$
and under tangent space rotation as
$$\delta\Omega_{\Lambda M}^{\ \ \ N}=-\partial_\Lambda\Omega_M^{\
\ N}+\Omega_{\Lambda M}^{\ \ \ S}
\Lambda_S^{\ \ N}+\Omega_{\Lambda R}^{\ \ N}\Lambda_M^{\
\ R}(-1)^{(M+R)(N+R)}\eqno(16.10)$$
The covariant derivatives are then defined by
$$D_\Lambda=\partial_\Lambda+{1\over2}\Omega_\Lambda^{\
mn}J_{mn}\eqno(16.11)$$
where $J_{mn}$ are the appropriate Lorentz generators (see Appendix
A of reference [0]). The covariant derivative with tangent indices is
$$D_N=E_N^{\ \Lambda}D_\Lambda\eqno(16.12)$$
where $E_N^{\ \Lambda}$ is the inverse vierbien defined by
$$\eqalignno{
E_N^{\ \Lambda}E_\Lambda^{\ M}&=\delta_N^{\ M}&(16.13)\cr
\noalign{or}
E_\Lambda^{\ M}E_M^{\ \Pi}&=\delta^\Pi_\Lambda&(16.14)}$$
Equipped with super-vierbien and spin-connection we define the
torsion  and curvature tensors as usual
$$[D_N,D_M\}=T_{NM}^{\ \ R}D_R+{1\over2}R_{NM}^{\ \
\ \ mn}J_{mn}\eqno(16.15)$$
Using Eqs. (16.11) and (16.12) we find that
$$\eqalignno{
T_{NM}^{\ \ \ \ R}&=E_M^{\ \Lambda}\partial_\Lambda E_N^{\
\Pi}E_\Pi^{\ R}+\Omega_{MN}^{\ \ \ R}-(-1)^{MN}(M\leftrightarrow
N)&(16.16)\cr
R_{MN}^{\ \ \ \ rs}&=E_M^{\ \Lambda}E_N^{\
\Pi}(-1)^{\Lambda(N+\Pi)}\{\partial_\Lambda\Omega_\Pi^{\
rs}+\Omega_\Lambda^{\ \ rk}\Omega_{\Pi k}^{\ \
s}-(-1)^{\Lambda\Pi}(\Lambda\leftrightarrow\Pi)\}&(16.17)}$$
The super-general coordinate transformations can be rewritten using
these tensors
$$\eqalignno{
\delta E_\Lambda^{\ M}&=-E_\Lambda^{\ R}\xi^NT_{NR}^{\ \
M}+D_\Lambda\xi^M&(16.18)\cr
\delta\Omega_{\Lambda M}^{\ \ \ N}&=E_\Lambda^{\ R}\xi^PR_{PRM}^{\ \ \
\ \ N}&(16.19)}$$
where $\xi^N=\xi^\Lambda E_\Lambda^{\ N}$ and we have discarded a
Lorentz transformation.

The torsion and curvature tensors satisfy Bianchi identities which
follow from the identity
$$\big[D_M,[D_N,D_R\}\big\}-\big[[D_M,D_N\},D_R\big\}
+(-1)^{RN}\big[[D_M,D_R\},D_N\big\}=0\eqno(16.20)$$
They read
$$\eqalignno{
0&=I^{(1)\ \ \ \ \ F}_{\ RMN}=[-(-1)^{(M+N)R}D_RT_{MN}^{\ \
\ \ F}+T_{MN}^{\
\ \ \ S}T_{SR}^{\ \ \ \ F}+R_{MNR}^{\ \ \ \ \ \ F}]\cr
&\quad+[+(-1)^{MN}D_NT_{MR}^{\ \ \ \ F}-(-1)^{NR}T_{MR}^{\ \
\ \ S}T_{SN}^{\
\ \ \ F}-(-1)^{NR}R_{MRN}^{\ \ \ \ \ \ F}]\cr
&\quad+[-D_MT_{NR}^{\ \ \ \ F}\cr
&+(-1)^{(N+R)M}T_{NR}^{\ \ \
\ S}T_{SM}^{\ \
\ \ F}+(-1)^{(N+R)M}R_{NRM}^{\ \ \ \ \ \ F}]&(16.21)\cr
\noalign{and}
I^{(2)\ \ \ \ mn}_{\ RMN}&=[(-1)^{(M+N)R}D_RR_{MN}^{\ \
\ \ mn}+T_{MN}^{\ \ \ S}R_{SR}^{\ \ \ \ mn}]\cr
&-(-1)^{NR}(R\rightarrow N,\ N\rightarrow R,\ M\rightarrow M\
{\rm in\ the\ first\ bracket)}\cr
&+(-1)^{(N+R)M}(M\rightarrow N,\ N\rightarrow R,\ R\rightarrow M\
{\rm in\ the\ first\ bracket})
=0&(16.22)}$$
It can be shown that if $I^{(1)\ \ \ \ F}_{\ MNR}$ holds then
$I^{(2)\ \ \ \ \ mn}_{\ RMN}$ is automatically
satisfied. This result holds in the presence of constraints on
$T_{MN}^{\ \ \ R}$ and $R_{MN}^{\ \ \ \ mn}$
and is a consequence of the restricted tangent space choice. We refer
to this as Dragon's theorem [71].
For all fermionic indices we find that
$$\eqalignno{
I_{ABC}^{\ \ \ \ N}&=-D_AT_{BC}^{\ \ \ N}+T_{AB}^{\ \ \ S}T_{SC}^{\ \
\ N}+R_{ABC}^{\ \ \ \ \ N}-D_CT_{AB}^{\ \ \ N}+T_{CA}^{\ \
\ S}T_{SB}^{\
\ \ N}\cr &\quad+R_{CAB}^{\ \ \ \ \ N}-D_BT_{CA}^{\ \ \ N}+T_{BC}^{\ \
\ S}T_{SA}^{\ \ \ N}+R_{BCA}^{\ \ \ \ \ N}=0&(16.23)}$$
and for fermionic indices with one bosonic index
$$\eqalignno{I_{ABr}^{\ \ \ \ N}&=-D_AT_{Br}^{\ \ \ N}+T_{AB}^{\ \
\ S}T_{Sr}^{\ \ \ N}+R_{ABr}^{\ \ \ \ \ N}
-D_rT_{AB}^{\ \ \ N}+T_{rA}^{\ \ \ S}T_{SB}^{\ \ \ N}\cr
&\quad+R_{rAB}^{\ \ \ \ \ N}+D_BT_{rA}^{\ \ \ N}-T_{Br}^{\ \
\ S}T_{SA}^{\
\ \ N}-R_{BrA}^{\ \ \ \ \ N}=0&(16.24)}$$
while
$$\eqalignno{
I_{Anr}^{\ \ \ \ N}&=-D_AT_{nr}^{\ \ \ N}+T_{An}^{\ \ \ s}T_{sr}^{\ \
\ N}+R_{Anr}^{\ \ \ \ \ N}-D_rT_{An}^{\ \ \ N}
+T_{rA}^{\ \ \ s}T_{sn}^{\ \ \ N}\cr
&\quad+R_{rAn}^{\ \ \ \ \ N}-D_nT_{rA}^{\ \ \ N}+T_{nr}^{\ \
\ s}T_{sA}^{\
\ \ N}+R_{nrA}^{\ \ \ \ \ N}=0&(16.25)}$$
Clearly one can replace any undotted index by a dotted index and the
signs remain the same.
We recall that for rigid superspace all the torsions and curvatures
vanish except for $T_{A\dot B}^{\ \ \ n}=-2i(\sigma^n)_{A\dot B}$.
Clearly this is inconsistent with an $OSp(4,1)$ tangent space group.

The dimensions of the torsions and curvature can be deduced from the
dimensions of $D_N$. If $F$
and $B$ denote fermionic and bosonic indices respectively then
$$[D_F]={1\over2};\quad[D_B]=1\eqno(16.26)$$
and
$$\eqalignno{
[T_{FF}^{\ \ B}]&=0;\quad[T_{FF}^{\ \ F}]=[T_{FB}^{\ \ B}]
={1\over2}\cr
[T_{FB}^{\ \ F}]&=[T_{BB}^{\ \ B}]=1;\quad[T_{BB}^{\ \
F}]={3\over2}&(16.27)}$$
while
$$[R_{FF}^{\ \ \ \ mn}]=1;\quad[R_{FB}^{\ \
\ \ mn}]={3\over2};\quad[R_{BB}^{\
\ \ \ mn}]=2\eqno(16.28)$$
It is useful to consider the notion of the geometric dimension of
fields. This is the  dimension of the field as it appears in the
torsions and curvature. Such expressions never
involve $\kappa$ and as they are nonlinear in certain bosonic fields,
such as the vierbien
$e_\mu^{\ n}$, these fields must have zero geometric dimensions. The
dimensions of the other
fields are determined in relation to $e_\mu^{\ n}$ to be given by
$$[e_\mu^{\ n}]=0\quad[\psi_\mu^{\
\alpha}]={1\over2}\quad[M]=[N]=[b_\mu]=1\eqno(16.29)$$
These dimensions can,for example,  be read of from  the  supersymmetry
transformations.  These dimensions differ from the
canonical assignment of  dimension by one unit. The difference comes
about as we have absorbed factors of
$\kappa$ into the fields.
\medskip
{\bf 4.2.2 On-shell Derivation of $N=1$ $D=4$ Superspace Supergravity}
\medskip
Having set up the appropriate geometry of superspace we are now in
a position to derive on-shell $N=1$ $D=4$ supergravity using its
superspace setting and solely from a knowledge  of
the  on-shell  states of a given spin in the irreducible
representation.  The  result is derived
by using dimensional analysis and the Bianchi identities in
superspace discussed above.
 We now illustrate this procedure for $N=1$
supergravity. It was this method that was used [111] to find the full
equations of motion in superspace and in $x$-space of IIB
supergravity.

The on-shell states are represented by $h_{\mu\nu}\
(h_{\mu\nu}=h_{\nu\mu})$ and $\psi_\mu^{\ \alpha}$ which have the
gauge transformations
$$\eqalignno{
\delta h_{\mu\nu}&=\partial_\mu\xi_\nu+\partial_\nu\xi_\mu\cr
\delta\psi_{\mu\alpha}&=\partial_\mu\eta_\alpha&(16.115)}$$
We have omitted the nonlinear terms, as only the general form is
important. The geometric
dimension of $h_{\mu\nu}$ is zero while that of $\psi_\mu^{\ \alpha}$
is one half. The lowest
dimension gauge covariant objects are of the form
$$\partial\psi\quad{\rm and}\quad\partial\partial h\eqno(16.116)$$
which have dimensions 3/2 and 2 respectively.

Consider now the super torsion and curvature; these objects at
$\theta=0$ must correspond
to covariant $x$-space objects. If there is no such object then the
corresponding tensor must
vanish at $\theta=0$ and hence to all orders in $\theta$. The only
dimension-0 tensors
are $T_{AB}^{\ \ n}$ and $T_{A\dot B}^{\ \ n}$. There are no
dimension-0 covariant objects
except the numerically invariant tensor $(\sigma^n)_{A\dot B}$.
Hence,  we must conclude that
$$T_{AB}^{\ \ \ n}=0\quad T_{A\dot B}^{\ \ \ n}=c(\sigma^n)_{A\dot
B}\eqno(16.117)$$
where $c$ is a constant. We choose $c\ne0$ in order to agree with
rigid superspace. The
reality properties of $T_{A\dot B}^{\ \ \ n}$ imply that $c$ is
imaginary and we can normalize
it to take the value $c=-2i$.

There are no dimension-${1\over2}$ covariant tensors in $x$-space and
so
$$T_{A\dot B}^{\ \ \ \dot C}=T_{AB}^{\ \ \ C}=T_{Am}^{\ \
\ n}=0\eqno(16.118)$$
There are no dimension-1 covariant objects in $x$-space. This would
not be the case if
one had an independent spin connection, $w_\mu^{\ rs}$, for $\partial
e+w+\dots$ would
be a covariant quantity. When $w_\mu^{\ rs}$ is not an independent
quantity it must be
given in terms of $e_\mu^{\ n}$ and $\psi_\mu^{\ \alpha}$ in such a
way as to render the
above dimension-1 covariant quantity zero. Hence, for a dependent
spin  connection, i.e.,
in second-order formalism, we have
$$T_{nA}^{\ \ \ \dot B}=T_{nA}^{\ \ \ B}=R_{AB}^{\ \ \ \
mn}=0=R_{A\dot B}^{\
\ \ \ mn}\eqno(16.119)$$
In other words, every dimension 0-, ${1\over2}-$, 1- torsion and
curvature vanishes with the
exception of $T_{A\dot B}^{\ \ \ n}=-2i(\sigma^n)_{A\dot B}$.

The reader who is familiar with the off-shell constraints for $N=1$
supergravity can compare them with the on-shell constraints found
here.  The set of constraints of off-shell
supergravity is given in  Section 16.2 of reference [0]. We find that
that the extra constraints are
$T_{nA}^{\ \ \ B}=T_{nA}^{\
\ \ \dot B}=0=R_{AB}^{\ \ \ \ AB}$. In terms of the superfields $R,\
W_{(ABG)}$ and $G_{A\dot B}$
this is equivalent to
$$R=G_{A\dot B}=0$$
\par
Returning to the on-shell theory.
The dimension 3/2 tensors can involve
at
$\theta=0$ the spin 3/2  object $\partial\psi$ and
so these will not all be zero. The only remaining non-zero tensors
are
$T_{mn}^{\ \ \ A},\
R_{Ar}^{\ \ \ \ mn}$ and $R_{st}^{\ \ \ \ mn}$ and of course $T_{A\dot
B}^{\
\ \ n}=-2i(\sigma^n)_{A\dot B}$. However, the previous constraints of
Eqs.   (16.117)-(16.119) are sufficient to
specify the entire theory, as we will now demonstrate. The first
nontrivial Bianchi identity
has dimension 3/2 and is
$$\eqalignno{
I_{nB\dot D}^{\ \ \ \ \ \dot C}&=-D_nT_{B\dot D}^{\ \ \ \dot
C}+T_{nB}^{\
\ \ F}T_{F\dot D}^{\ \ \ \dot C}
+R_{nB\dot D}^{\ \ \ \ \ \dot C}+D_{\dot D}T_{nB}^{\ \ \ \dot
C}-T_{\dot Dn}^{\ \ \ F}T_{FB}^{\ \ \ \dot C}\cr
&\quad-R_{\dot DnB}^{\ \ \ \ \ \dot C}-D_BT_{\dot Dn}^{\ \ \ \dot
C}+T_{B\dot D}^{\ \ \ \dot F}T_{Fn}^{\ \ \ \dot C}+R_{B\dot Dn}^{\ \ \
\ \ \dot C}=0&(16.120)}$$
Using the above constraints this reduces
$$-2i(\sigma^m)_{B\dot D}T_{mn}^{\ \ \ \dot C}-R_{nB\dot D}^{\ \
\ \ \dot C}=0\eqno(16.121)$$
Tracing on $\dot D$ and $\dot C$ then yields
$$(\sigma^m)_{B\dot D}T_{mn}^{\ \ \ \dot D}=0\eqno(16.122)$$
This is the Rarita-Schwinger equation as we will demonstrate shortly.

The spin 2 equation must have dimension two and is contained in the
$I_{Bmn}^{\ \ \ A}$
Bianchi identity.
$$\eqalignno{
I_{Bmn}^{\ \ \ A}&=-D_BT_{mn}^{\ \ \ A}+T_{Bm}^{\ \ \ F}T_{Fn}^{\ \
A}+R_{Bmn}^{\ \ \ \ A}-D_n
T_{Bm}^{\ \ \ A}+T_{nB}^{\ \ \ F}T_{Fm}^{\ \ \ A}\cr
&\quad+R_{nBm}^{\ \ \ \ \ A}-D_mT_{nB}^{\ \ \ A}+T_{mn}^{\ \
\ F}T_{FB}^{\
\ \ A}+R_{mnB}^{\ \ \ \ \ A}=0&(16.123)}$$
Application of the constraints gives
$$-D_BT_{mn}^{\ \ \ A}+T_{mnB}^{\ \ \ \ A}=0\eqno(16.124)$$
On contracting with $(\sigma^m)_{\dot BA}$ we find
$$(\sigma^m)_{\dot BA}D_BT_{mn}^{\ \ \ A}=0=R_{mnB}^{\ \ \
\ \ A}(\sigma^m)_{\dot BA}\eqno(16.125)$$
Using the fact that $R_{mnB}^{\ \ \ \ \ A}=-{1\over4}R_{mn}^{\ \
\ \ pq}(\sigma_{pq})_B^{\ A}$
yields the result
$$R_{mn}-{1\over2}\eta_{mn}R=0$$
or
$$R_{mn}=0\quad{\rm where}\quad R_{mn}=R_{msn}^{\ \ \
\ \ s}\eqno(16.126)$$ We now wish to demonstrate that these are the
spin 3/2 and spin 2  equations. The $\theta=0$  components of
$E_\mu^{\ n}$ and $E_\mu^{\ A}$ are denoted as follows:
$$E_\mu^{\ n}(\theta=0)=e_\mu^{\ n},\quad E_\mu^{\
A}(\theta=0)={1\over2}\psi_\mu^{\ A}\eqno(16.127)$$
At this stage the above equation is simply a definition of the fields
$e_\mu^{\ n}$ and
$\psi_\mu^{\ A}$. The $\theta=0$ components of $E_A^{\ n}$ may be
gauged away by an
appropriate super general coordinate transformation. As
$$\delta E_A^{\ n}(\theta=0)=\xi^\Pi\partial_\Pi E_A^{\
n}|_{\theta=0}+\partial_A\xi^\Pi
E_\Pi^{\ n}|_{\theta=0}=\dots+\partial_A\xi^\mu e_\mu^{\
n}+\dots\eqno(16.128)$$
we may clearly choose $\partial_A\xi^\mu$ so that $E_A^{\ n}=0$.
Similarly we may choose
$$E_{\underline{\dot A}}^{\ \dot B}=\delta_{\dot A}^{\ \dot B},\quad
E_{\underline A}^{\ B}=\delta_A^{\ B},\quad E_{\underline A}^{\ \dot
B}=0\eqno(16.129)$$
To summarize
$$E_\Pi^{\ m}(\theta=0)=\left(\matrix{
e_\mu^{\ n}&{1\over2}\psi_\mu^{\ A}&{1\over2}\psi_\mu^{\ \dot A}\cr
0&\delta_B^{\ A}&0\cr 0&0&\delta_{\dot B}^{\ \dot
A}\cr}\right)\eqno(16.130)$$
For the spin connection $\Omega_\Pi^{\ m}$ we define
$$\Omega_\mu^{\ mn}(\theta=0)=w_\mu^{\ mn}
\eqno(16.131)$$
and we use a Lorentz transformation to gauge
$$\Omega_\alpha^{\ mn}(\theta=0)=0\eqno(16.132)$$
At $\theta=0$ we then find
$$T_{\mu\nu}^{\ \ \dot A}=-{1\over2}\partial_\mu\psi_\nu^{\ \dot
A}+\Omega_{\mu\nu}^
{\ \ \dot A}-(\mu\leftrightarrow\nu)=-{1\over2}\psi_{\mu\nu}^{\ \ \dot
A}\eqno(16.133)$$
where
$$\psi_{\mu\nu}^{\ \ \dot A}\equiv D_\mu\psi_\nu^{\ \dot
A}-(\mu\leftrightarrow\nu)$$
and
$$D_\mu\psi_\nu^{\ \dot A}=\partial_\mu\psi_\nu^{\ \dot A}-\psi_\nu^{\
\dot B}w_{\mu\dot B}^{\ \ \dot A}\eqno(16.134)$$
Here we have used the results
$$\Omega_{\mu\nu}^{\ \ \dot A}=E_\nu^{\ N}w_{\mu N}^{\ \ \dot
A}={1\over2}\psi_\nu^{\ \dot B}
w_{\mu\dot B}^{\ \ \dot A}\eqno(16.135)$$
The torsion with all tangent indices is given in terms of
$T_{\mu\nu}^{\ \ \dot A}$ by the relation
$$\eqalignno{
T_{\mu\nu}^{\ \ \dot A}(\theta=0)&=E_\mu^{\ N}(\theta=0)E_\nu^{\
M}(\theta=0)T_{NM}^{\ \ \dot A} (\theta=0)(-1)^{NM}\cr
&=e_\mu^{\ n}e_\nu^{\ m}T_{nm}^{\ \ \dot A}(\theta=0)&(16.136)}$$
where we have used the constraints
$T_{Bn}^{\ \ \dot A}=T_{\dot BC}^{\
\ \dot A}=0$. Consequently
$$0=(\sigma^m)_{A\dot B}T_{mn}^{\ \ \dot
B}(\theta=0)=-{1\over2}(\sigma^m)_{A\dot B}e_m^{\ \mu}e_n^{\
\nu}\psi_{\mu\nu}^{\ \ \dot B}\eqno(16.137)$$
and we recognize the Rarita-Schwinger equation on the right-hand
side.

Actually to be strictly rigorous we must also show that $w_\mu^{\
mn}$  is the spin connection
given in terms of $e_\mu^{\ n}$ and $\psi_\mu^{\ A}$. In fact, this
follows from the constraint
$T_{nm}^{\ \ r}=0$. We note that
$$T_{\mu\nu}^{\ \ r}(\theta=0)=-\partial_\mu e_\nu^{\ r}+w_{\mu\nu}^{\
\ r}-(\mu\leftrightarrow\nu)\eqno(16.138)$$
However,
$$\eqalignno{
T_{\mu\nu}^{\ \ r}(\theta=0)&=E_\mu^{\ N}(\theta=0)E_\nu^{\
M}(\theta=0)T_{NM}^{\ \ r}(\theta
=0)(-1)^{MN}\cr
&\quad-{1\over4}\psi_\mu^{\ A}\psi_\nu^{\ \dot B}T_{A\dot B}^{\ \
r}(\theta=0)-\psi_\mu^{\ \dot B}
\psi_\nu^{\ A}T_{\dot BA}^{\ \ r}(\theta=0)\cr
&=+{1\over2}i\psi_\nu^{\ \dot B}(\sigma^r)_{A\dot B}\psi_\mu^{\
A}-(\mu\leftrightarrow\nu)&(16.139)}$$
Consequently we find that
$$w_{\mu n}^{\ \ m}e_\nu^{\ n}-\partial_\nu e_\mu^{\
m}-(\mu\leftrightarrow\nu)=+{i\over2}\psi_\nu
^{\ \dot B}(\sigma^m)_{A\dot B}\psi_\mu^{\
A}-(\mu\leftrightarrow\nu)\eqno(16.140)$$
which can be solved in the usual way to yield the correct expression
for $w_{\mu n}^{\ \ m}$.

The spin 2 equation is handled in the same way:
$$R_{\mu\nu}^{\ \ mn}(\theta=0)=\partial_\mu w_\nu^{\ mn}+w_\mu^{\
mr}w_{\nu r}^{\ \ \ n}-(\mu\leftrightarrow\nu)\eqno(16.141)$$
However
$$\eqalignno{
R_{\mu\nu}^{\ \ nm}(\theta=0)&=E_\mu^{\ N}(\theta=0)E_\nu^{\
M}(\theta=0)R_{NM}^{\ \ \ \ mn}(\theta=0)(-1)^{mN}\cr
&=e_\mu^{\ p}e_\nu^{\ q}R_{pq}^{\ \
\ nm}(\theta=0)+{1\over2}\big(\psi_\mu^{\ \dot A}e_\nu^{\ p}R_{\dot
Ap}^{\ \ \ nm}\cr
&\quad+\psi_\mu^{\ A}e_\nu^{\ p}R_{Ap}^{\ \
\ nm}(\theta=0)-(\mu\leftrightarrow\nu)\big)&(16.142)}$$
The object $R_{Ap}^{\ \ \ \ nm}$ can be found from the Bianchi
identity
$I_{Anr}^{\ \ \ s}$
$$\eqalignno{
0&=I_{Anr}^{\ \ \ \ s}=-D_AT_{nr}^{\ \ \ s}+T_{An}^{\ \ \ F}T_{Fr}^{\ \
\ s}+R_{Anr}^{\ \ \ \ \ s}-D_rT_{An}^{\ \ \ s}
+T_{rA}^{\ \ \ F}T_{Fn}^{\ \ \ s}\cr
&\quad+R_{rAn}^{\ \ \ \ \ s}-D_nT_{rA}^{\ \ \ s}+T_{nr}^{\ \
\ F}T_{FA}^{\
\ \ s}+R_{nrA}^{\ \ \ \ \ s}&
(16.143)}$$
Using the constraints we find that
$$R_{Anr}^{\ \ \ \ \ s}+R_{rAn}^{\ \ \ \ \ s}=+2iT_{nr}^{\ \ \dot
B}(\sigma^s)_{A\dot B}\eqno(16.144)$$
From Eq. (16.137) we find that
$$R_{Anr}^{\ \ \ \ \ s}+R_{rAn}^{\ \ \ \ \ s}=-i(\sigma^s)_{A\dot
B}e_n^{\
\mu}e_r^{\ \nu}\psi_{\mu\nu}^{\ \ \dot B}\eqno(16.145)$$
Contracting Eq. (16.142) with $e^\nu_{\ m}$ we find
$$e^{\ \nu}_mR_{\mu\nu}^{\ \ nm}(\theta=0)\equiv R_\mu^{\ n}=e_\mu^{\
p}R_{pm}^{\ \ nm}+{1\over2}
(\psi_\mu^{\ A}R_{Am}^{\ \ \ \ nm}+\psi_\mu^{\ \dot A}R_{\dot Am}^{\ \
\ \ nm})\eqno(16.146)$$
Equation (16.145) then gives
$$e_\mu^{\ p}R_{pm}^{\ \ \ \ nm}=R_\mu^{\ n}
-\left({i\over2}\psi_\mu^{\
A}(\sigma^m)_{A\dot B}e_m^{\ \lambda}
e^{n\tau}\psi_{\lambda\tau}^{\ \ \dot B}+{\rm
h.c.}\right)\eqno(16.147)$$
Equation (16.126) $(R_{mn}=0)$ then yields the spin 2 equation of
$N=1$ supergravity, which is the
left-hand side of the above equation.

At first sight it appears that the task is not finished; one should
also analyze all the remaining
Bianchi identities and show that they do not lead to any
inconsistencies. However, it can be shown
that the other Bianchi identities are now automatically
satisfied [71,82].
\medskip
{\centerline{\bf 4.3 Gauging of Space-time Groups}}
\medskip
It has been know for many years  that Einstein's theory of
general relativity contains a local Lorentz symmetry.
When the action is expressed  in
firstorder formalism the spin-connection is the gauge field
 and the Riemann tensor the field
strength   for  Lorentz group [114]. It was only in reference [18]
that  a space-time group was gauged and  general
relativity theory was deduced from a gauge theory viewpoint. In fact,
 reference [18] gauged  the super Poinca\'e group and deduced the
supergravity theory from this view point.  It is
straightforward to restrict the calculation to that for the Poinca\'e
group and  deduce just general relativity. Of course at that time
supergravity had been constructed [14],[15], but the gauging proceedure
provided the first analytical proof [18] of its invariance under
local supersymmetry. The theory of
supergravity with a cosmological constant by gauging the
the super de Sitter group was independently found in reference [113].
\par
Let us consider gauging the $N=1$ super Poinca\'e group; corresponding
to the generators $J_{ab},P_a$ and $Q^\alpha$ we introduce  the gauge
fields  $w_\mu^{ab},e_\mu^a$ and $Q_{\mu \alpha}$ which will  become
the  spin-connection, the vierbein and gravitino. It is
straightforward  to calculate the field strengths $R_{\mu
\nu}^{ab},C_{\mu \nu}^a$ and
$\phi_{\mu \nu \alpha}$ and the gauge transformations of the
gauge fields. Supergravity is not the gauge theory of the
super Poinca\'e group in an obvious way and we must
proceed by setting the  field strength $C_{\mu \nu}^a$
associated to translations to zero. We now construct an action
to the appropriate order  in derivatives that is invariant under the
gauge transformations of the super Poinca\'e group subject to the
constraint $C_{\mu \nu}^a$. In particular, we start  from the most
general action which is first-order in the field strengths i.e.
$$ \int d^4 x\ \epsilon ^{\mu\nu\rho\lambda }(\epsilon_{abcd }e_\mu^a
e_\nu^b R_{\rho\lambda}^{cd} + if\bar \psi_\mu \gamma_5\gamma_\nu
\phi_{\rho\lambda} ).
$$
The   constant $f$  is readily fixed by demanding the
invariance of this action. Hence,
we rapidly arrive at the  supergravity action. In fact
the constraint $C_{\mu \nu}^a=0$ is just that required to
 correctly express the spin-connection in terms of the
vierbein and gravitino, that is to go from first- to second-order
formalism. The  constraint is also
just that required to  convert gauge
transformations associated with translations  into general coordinate
trasformations.
\par
When carrying out the variation of the action subject to
the constraint the variation of the spin-connection is
irrelevant, since its variation  multiplies
its equation  of motion which vanishes due to the constraint $C_{\mu
\nu}^a=0$. However, this constriant  is none other than the condition
for expressing the spin-connection in terms of the vierbein, that is
the transition to second-order formalism.
Thus the invarinace of the above action subject to the constraint
provides    an analytical proof of the invariance of supergravity
under local supersymmetry [18],[19]. This way of proceeding became
known as the 1.5-order formalism and it is reviewed in chapter 10 of
reference [0].
\par
The construction of the theories of conformal supergravity
were carried out using this gauging method [118]. The key to getting
the gauge method  to work is to guess  the appropriate
constraints. However, since these constraints break the original gauge
transformations  they are not always easy to find. Much effort  has
been devoted to  developing  the method discussed into  a systematic
procedure. One such work was that of reference [115] where the full
gauge symmetry was realised, but was spontaneously broken.
\par
The gauge techninque has not been used to construct the theories
of supergravity in ten and eleven dimensions and it may be
instructive to derive them using this method.  It cannot be a
coincidence that gravity and supergravity admit such
 simple formulations as a gauge theories and this connection
suggests  that there is something to be understood at  a deeper
level.

\bigskip

\centerline{\bf 5.  Eleven-dimensional  Supergravity}
\medskip
  In this section,  we give the eleven-dimensional
supergravity  theory and describe its properties. Eleven
dimensional supergravity  is thought
to be the low energy effective action of a new kind of  theory
called M theory which is believed to underlie string theory.
Little is known about M theory apart from its relation  to eleven
dimensional supergravity.
\par
The non-trivial representation of the   Clifford
algebra in eleven dimensions  is inherited from that in ten dimensions
and so has dimension
$2^{10\over2}=32$. We also inherit the properties
$\epsilon=1$ and so $ B^T=B$ and $ C=-C^T$ which were discussed in
section one. The resulting  properties of
the $\gamma$ matrices are given in equations (1.4.6) and (1.4.8).
\par
Eleven-dimensional supergravity  is based on the  $D=11$
supersymmetry algebra with Majorana spinor
$Q_\alpha$ which
has 32  real components. As we discussed in chapter two the algebra
takes the form [117]
$$\{Q_\alpha,Q_\beta\}
=(\gamma^mC^{-1})_{\alpha\beta}P_m+(\Gamma^{mn}C^{-1})_{\alpha\beta}
Z_{mn}+(\Gamma^{mnpqr}C^{-1})_{\alpha\beta}Z_{mnpqr}
\eqno(5.1)$$
where $P_m$ is the translation operator and
 $Z_{mn}$ and $Z_{mnpqr}$ are
central charges. Although these play little apparent role in
the construction of the supergravity theory they are very
important  in M theory.
\par
Eleven dimensions is the maximal dimension in which one can have a
supergravity theory [105]. By a supergravity theory we mean a theory
with spins two and less. This observation follows from the study of
the massless irreducible representations of supersymmetry, which  can
be deduced in a straightforward way from the relevant supersymmetry
algebras.  The irreducible representations of the four-dimensional
supersymmetry algebras were  given in
section three. We found that the maximal supergravity theory
in four dimensions corresponded to  $N=8$ supersymmetry. This
is  the algebra  with eight Majorana supercharges, each of which has
four real components. In fact, this theory can be obtained the eleven
dimensional supergravity theory by dimensional reduction.
\par
We now explain  this result and show that it implies
that  a supergravity theory can live in at most
eleven dimensions. The four
dimensional result follows in an obvious way from the  fact that in
the massless case from each four component  supercharge only two of
the components  act non-trivially on
the physical states. Further, these  two components form a Clifford
algebra,  one of which raises the helicity by  1/2 and one of which
lowers the
helicity by 1/2. Choosing the supercharges
 that raises the helicity to
annihilate the vacuum, the physical states are given  by the action of
the remaining  supercharges. If the supersymmetry algebra has $N$
Majorana supercharges,
 the physical states are given by the action of
N creation operators each of which lowers the helicity by 1/2.
Consequently, if we take the vacuum to have helicity two the lowest
helicity state in the representation will be
$2-{N\over 2}$. To have a supergravity theory we cannot have less than
helicity $-2$  and hence the limit $N\le 8$. Given a supergravity
theory in a dimension greater than four we can reduce it in a trivial
way by taking all the fields to be independent of the
extra dimensions, to  find a
supergravity theory in four dimensions. However,  the number of
supercharges is unchanged in the reduction and so the maximal number
is $4\times 8=32$. Hence the supergravity in the higher dimension must
arise in a dimension whose spinor representation has  dimension 32 or
less. Thus we find the desired result;  eleven dimensions
is the highest dimension in which a supergravity theory can exist.
It also follows that any supergravity theory  must have 32, or fewer
supercharges and that  the maximal,
or largest, supergravity  theory in a given dimension  has 32
supercharges.
\par
The irreducible representation, or particle content, of eleven
dimension supergravity was found in reference [105] by analysing
the irreducible representations of the supersymmetry algebra
of equation (5.1). One
could also deduce it by requiring that it reduce to the
irreducible representation of the four-dimensional $N=8$ supergravity
theory given in section three. We now give a more intuitive argument
for the particle content.
\par
Eleven-dimensional supergravity must be invariant under general
coordinate transformations (i.e. local translations)  and local
supersymmetry  transformations. To achieve these symmetries it must
possess the  equivalent "local gauge"
fields, the vielbein $e_\mu^a$ and
the  gravitino $\psi_{\mu \alpha}$. The latter  transforms as
$\delta \psi_{\mu \alpha}= \partial _\mu \eta_\alpha +\ldots $
and so must be the same type of spinor as the supersymmetry
parameter which in this case is a Majorana spinor.
\par
For future use we now give the on-shell count of degrees of
freedom  of the graviton and gravitino in $D$ dimensions. The
relevant bosonic  part of the
 little group which classifies  the irreducible
representation is $SO(D-2)$. The graviton encoded  in $e_\mu^a$ is a
second-rank symmetric traceless tensor of
$SO(D-2)$ and as such has ${1\over 2}(D-2)(D-1)-1$ degrees of freedom
on-shell.  The gravitino has
$(D-3)cr$ real components. Here  $c$ is the dimension of the Clifford
algebra in dimension $D-2$ and so is given by
$c=2^{{D\over 2}-1}$  if $D$ is even and
$c= 2^{{(D-1)\over 2}-1}$ if
$D$ is odd. The quantity  r is 2,1 or 1/2 if $\psi_{\mu \alpha}$ is a
Dirac, Majorana or  Majorana-Weyl spinor respectively. In terms of
little group representations, the gravitino
is a vector spinor $\phi_{i},i=1,\dots , D-2$ which is
$\gamma$-traceless $\gamma^i\phi_{i}=0,$. A general vector
spinor in $D-2$ dimensions has
$(D-2){c}r$  components, but the $\gamma$-trace subtracts
another spinor's worth of components (i.e. ${c}r$).
\par
For eleven dimensions we find that the graviton  and gravitino have
44 and 128 degrees of freedom on-shell. However, in any supermultiplet
of on-shell physical states  the fermionic and bosonic degrees of
freedom must be equal. Assuming that there are no further fermionic
degrees of freedom we require another 84 bosonic on-shell degrees of
freedom. If we take these to belong to an irreducible  representation
of $SO(9)$ then the unique solution would be a third rank
anti-symmetric tensor. This can only arise from a third rank gauge
field $A_{\mu_1\mu_2\mu_3}$ whose fourth rank gauge field
$F_{\mu_1 \mu_2\mu_3\mu_4}
\equiv 4\partial _{[\mu_1} A_{\mu_2\mu_3\mu_4]}$, the anti-symmetry
being with weight one.  We note that in $D$ dimensions a  rank $p$
anti-symmetric gauge field belongs to the anti-symmetric   rank $p$
tensor representation of $SO(D-2)$ and so has
${(D-2)\dots (D-p-1)\over p!}$ degrees of freedom on-shell.
\par
The eleven-dimensional supergravity Lagrangian  was constructed in
reference [106] and is given by
$$
L=-{e\over4\kappa^2}R\big(\Omega(e,\psi)\big)-
{e\over 48}F_{\mu_1\dots\mu_4}F^{\mu
_1\dots\mu_4}-{ie\over2}
\bar\psi_\mu\Gamma^{\mu\nu\varrho}D_\nu\left({1\over2}
(\Omega+\hat\Omega)\right)\psi_\varrho
$$
$$
+{1\over192}e\kappa
(\bar\psi_{\mu_1}\Gamma^{\mu_1\dots\mu_6}
\psi_{\mu_2}
+12\bar\psi^{\mu_3}
\Gamma^{\mu_4\mu_5}\psi^{\mu_6})(F_{\mu_3\dots\mu_6}+\hat
F_{\mu_3\dots\mu_6} )
$$
$$+{2\kappa\over(12)^4}
\epsilon^{\mu_1\dots\mu_{11}}F_{\mu_1\dots\mu_4}F_{\mu_5
\dots\mu_8}A_{\mu_9\mu_{10}\mu_{11}}
\eqno(5.2)$$
where
$$ F_{\mu_1\dots\mu_4}=4\partial_{[\mu_1}A_{\mu_2\mu_3\mu_4]}
$$
$$
\hat F_{\mu_1\dots\mu_4}=F_{\mu_1\dots\mu_4}-3\bar\psi_{[\mu_1}
\Gamma_{\mu_2\mu_3}\psi_{\mu_4]},
\eqno(5.3)$$
and
$$\Omega _{\mu mn}= \hat \Omega _{\mu mn}-{i\over 4}
\bar \psi_\nu\Gamma_{\mu m n}^{\ \ \ \ \ \nu\lambda}\psi_{\lambda},
$$
$$\hat  \Omega _{\mu mn}= \Omega _{\mu mn}^0 (e)
+{i\over 2}
(\bar \psi_\nu\Gamma_{ n}\psi_{m}-\bar \psi_\nu\Gamma_{ m}\psi_{n}
+\bar \psi_n\Gamma_{ \mu}\psi_{m})
\eqno(5.4)$$
The symbol $\Omega _{\mu mn}^0 (e) $ is the usual expression for the
spin-connection in terms of the vielbein $e_\mu^n$ which can be found
in section four.
\par
It is invariant under the local supersymmetry transformations
$$\eqalign{\delta e_\mu^m&=-i\kappa\bar\epsilon\Gamma^m\psi_\mu\cr
\delta\psi_\mu&={1\over
\kappa}D_\mu(\hat\Omega)\epsilon+{i\over12^2}(\Gamma_\mu^{\
\nu_1\dots\nu_4}-8\delta_\mu^{\nu_1}\Gamma^{\nu_2\nu_3\nu_4})
F_{\nu_1\dots\nu_4}\epsilon\cr
\delta
A_{\mu_1\mu_2\mu_3}
&={3\over2}\bar\epsilon\Gamma_{[\mu_1\mu_2}\psi_{\mu
_3]}}
\eqno(5.5)$$
\par
Although the result may at first sight look complicated most of the
terms can be understood if one were to consider constructing the
action using the Noether method. In this method, which we discussed
in chapter four, we  start  from the linearized theory for the
graviton, the gravitino and the  gauge field $A_{\mu_1\mu_2\mu_3}$.
The
linearized action is bilinear in the fields and is invariant under a
set of rigid supersymmetry transformations which are linear in the
fields as well as the local Abelian transformations
$\delta \psi_{\mu \alpha}= \partial _\mu \eta_\alpha$,
$A_{\mu_1\mu_2\mu_3}= \partial _{[\mu_1}\Lambda_{\mu_2\mu_3]}$ as
well as  an appropriate analogous transformation for the  linearised
vielbein. The linearised  supersymmetry transformations  are found by
using dimensional analysis and closure of the linearised
supersymmetry algebra. We now let the rigid supersymmetry parameter
$\epsilon$ depend on space-time  and
identify the two spinor parameters by
$\eta_\alpha= {1\over \kappa}\epsilon_\alpha$.  We know that the final
result will be invariant under general coordinate transformations and
so we may at each step in the Noether procedure insert the vielbein
in all terms so as to ensure this invariance.
Even at this stage
 in the Noether procedure we recover all the terms
in the transformations in the fields given above in equation (5.5)
except for some of the terms in the spin-connection $\hat \Omega$. For
the action, we find all the terms except the last two terms and again
some terms in the spin-connection of the gravitino. The action at this
stage is not invariant under the now local supersymmetry
transformations
and  as explained in chapter four  we can gain invariance at order
$\kappa^0$ by adding a term of the form
$\bar \psi ^{\mu \alpha}j_{\mu \alpha}$ where $j_{\mu \alpha}$ is the
Noether current for the supersymmetry of the linearised theory. This
term is none other than the second to
last term in the above action. To
gain invariance to order $\kappa^1$ we must cancel the variations of
this second to
last term under the supersymmetry transformation. This
is achieved if we add the last term to the action. Hence even
at this stage in the procedure we have accounted for essentially all
the  terms in the action and transformation laws.
While one can pursue the Noether procedure to the end, to find the
final form of the action and transformations laws, it is perhaps best
to guess the final form of the connection and  verify that the action
is invariant and the local supersymmetry transformations close.
\par
The eleven-dimensional action  contains only one coupling constant,
Newton's constant $\kappa$, which has the dimensions of
$mass^{-{9\over 2}}$ and so  defines a Planck mass $m_p$ by
$\kappa= m_p^{-{9\over 2}}$.   We note that  there are
 no scalars in the
action whose expectation value could be used to define another
coupling constant.  If we scale the fields by
$\psi_\mu\to \kappa^{-1}\psi_\mu$ and
$A_{\mu_1\mu_2\mu_3} \to \kappa^{-1}A_{\mu_1\mu_2\mu_3}$
we find that all factors of $\kappa$ drop out of the
action except for a prefactor of $\kappa^{-2}$ and all
factors of $\kappa$ drop out of the supersymmetry transformation laws.
As such, when expressed in terms of these variables,
the classical field
equations do not contain $\kappa$.  In fact,  the value of the
coupling constant
$\kappa$  has no physical meaning. One way to see this fact is to
observe that if, after carrying out  the above redefinitions,
we
Weyl-scale  the fields by
$$e_\mu^m\to e^{-\alpha} e_\mu^m, \ \psi_\mu\to
e^{-{\alpha\over 2}}\psi_\mu, \  {\rm {and}}
,\ A_{\mu_1\mu_2\mu_3} \to e^{-3\alpha}A_{\mu_1\mu_2\mu_3}
\eqno(5.6)$$
as well as scale  the coupling constant
by  $\kappa\to
e^{-{9\alpha\over 2}}\kappa$,  we then find that the  action is
invariant.  In deriving this result we used the equation
$$ R(e^\tau e_\mu^m)=e^{-2\tau}(R(e_\mu^m) +2(D-1)D^\mu D_\mu \tau
+(D-1)(D-2)g^{\mu\nu}\partial_\mu\tau\partial_\nu \tau)
\eqno(5.7)$$
where $R$ is the Ricci tensor in dimension $D$.
Of course this is not a symmetry of the
action in the usual sense
 as we have rescaled   the coupling constant.  However, as the
coupling constant only occurs as a prefactor multiplying the entire
action,   it   is a symmetry of the classical equations of motion.
Hence, we can only  specify the value of the constant
$\kappa$ with  respect to a  particular metric.
  These transformations
are   not a symmetry of  the
quantum theory where, in the path integral, the prefactor of
$\kappa^{-2}$ which  multiplies all terms in the action becomes
$\kappa^{-2}\hbar$  where $\hbar$ is Planck's constant. In this case
we can absorb the rescaling either in  $\kappa$ or $\hbar$.
The above Weyl scaling of the vielbein  implies that the
proper distance $d^2s$ scales  as $d^2s\to e^{-\alpha}d^2s$. Taking
the scaling to be absorbed by $\hbar$ we find that $\hbar \to
e^{-9\alpha}\hbar$ and so small proper distance corresponds to
small $\hbar$. Put another  small $\kappa$, or
equivalently $\hbar$, is the same as working at small proper
distance.
\par
Although we have constructed the supergravity theories in ten and
eleven dimensions in this section we have omitted many of the
significant formulae, such as the supersymmetry transformations.
Since these are contained in the original papers
[105],[106],[107][108][109] on the subject, we have used the same
metric as in these papers, that is  the tangent space  metric
$\eta_{mn} = diag(+1,-1,\ldots ,-1)$.
Since many practitioners nowadays prefer the
other signature we now give the
rules to change to the  tangent space metric which is mainly plus i.e.
$\eta_{nm} =
diag(-1,1,\ldots ,1)$. To go to the latter metric we must take
$\eta_{nm} \to -\eta_{nm},\
\gamma^a\to i\gamma^a,\  e^n_\mu\to e_\mu^n$. Using this rule it is
easy to carry out the change. One finds, for example, that
$g^{\mu\nu} \partial _\mu\sigma \partial_\nu \sigma
\to - g^{\mu\nu} \partial _\mu\sigma \partial_\nu \sigma$ and
$R\to -R$.

\bigskip

\centerline{\bf 6. IIA and IIB Supergravity}
\medskip
In this section we give  the supergravity theories in ten
dimensions which have the maximal supersymmetry. There are two such
theories, called IIA and IIB, and they are the effective
low energy actions
for the IIA and IIB string theories respectively. We describe the
properties of these supergravity theories that are relavent for the
 string theories and play an important role in string duality.
\medskip
{\bf  {6.1 Supergravity Theories in Ten Dimensions}}
\medskip
In ten dimensions the non-trivial representation of the
Clifford algebra has dimension
$2^{10\over2}=32$. As we found in section one the matrices $B$ and $C$
associated with the complex conjugation and transpose of the
$\gamma$-matrices obey the properties
$ B^T=B$ and $ C=-C^T$ (i.e. $\epsilon=1$). The properties of
the $\gamma$ matrices are given in equations (1.2.16) and (1.2.17).
In ten dimensions a Majorana spinor has 32 real components,
however, we can also have Majorana-Weyl spinors and these only have 16
real components.
\par
The supersymmetry algebra for  a single
Majorana-Weyl
supercharge which has 16 real components is given in equation
(1.5.4).  There are two supersymmetric theories that are based on this
algebra, the  so called $N=1$ Yang-Mills theory [119] and the $N=1$
supergravity  theory [120] which is more often called  type I
supergravity. The  coupling between the two theories was given in
references [121].
\par
In the discussion at the beginning of section  five we found that
 a supergravity theory can be based on a supersymmetry algebra
with   32 or fewer supercharges. If  we  consider the
supersymmetry algebra with a 32 component Majorana spinor we find
 the IIA supergravity theory which was constructed in
references [107],[108] and [109].  Clearly, when we decomposed the
Majorana  spinor into Majorana-Weyl spinors we get two such spinors:
one of each chirality.
The other ten-dimensional  supersymmetry algebra with 32
supercharges has two
Majorana-Weyl spinors of the same chirality and  the corresponding
supergravity is  IIB supergravity. This theory  was constructed in
references [110],[112] and [111]. Unlike the other supergravity
theories  in ten dimensions IIB supergravity has an internal symmetry
which is  the group $SL(2,{\bf R})$ [110].
\par
Upon reduction of eleven-dimensional supergravity to ten dimensions
by taking the eleventh dimension to be a circle we will obtain
a ten-dimensional theory that possesses a
supersymmetry algebra based on a 32 component Majorana supercharge
 which  decomposes into two Majorana-Weyl spinors of
opposite  Weyl chiralities. Thus the resulting theory
 can only be  IIA supergravity. Indeed, this was how IIA
supergravity was found  [107],[108] [109].
\par
The importance of the IIA and IIB
supergravity theories, which was the main motivation for their
 construction,   is that they are the low energy effective
theories of the  corresponding IIA and IIB closed string theories  in ten
dimensions. Type I supergravity coupled to $N=1$ Yang-Mills theory  is
the effective action for the low energy limit of the $E_8\otimes E_8$ or
$SO(32)$  heterotic string.  All these supergravity theories were
constructed at a time when string theory was deeply
unpopular and when it did become fashionable little interest
was taken in supergravity theories. However,
they now  form the basis for many of the  discussions of
duality in string theories.

\medskip
{\bf {6.2 IIA Supergravity}}
\medskip
As we have mentioned, this theory is based on a
supersymmetry algebra which contains one Majorana spinor
$Q_\alpha\,\  \alpha=1,\dots,32$. Following
the discussion of section 2.5, we conclude that the anti-commutator
of two supersymmetry generators can have central charges of rank
$p$ where  $p=1,2$; mod 4 and so the corresponding
anti-commutator  is given by [117]
$$
\{ Q_\alpha,\ Q_\beta\}
=(\gamma^mC^{-1})_{\alpha\beta}P_m+(\gamma^{mn}C^{-1})_{\alpha\beta
}Z_{mn}+(\gamma^{m_1\dots m_5}C^{-1})_{\alpha\beta}Z_{m_1\dots
m_5}
$$
$$+(\gamma^{m_1\dots m_4}\gamma_{11}C^{-1})_{\alpha\beta}Z_{m_1\dots
m_4}
+(\gamma^m\gamma_{11}C^{-1})_{\alpha\beta}Z_m
+(\gamma_{11}C^{-1})_{\alpha\beta} Z
\eqno(6.2.1)$$
We need only take $p\le 5$ since we may use the equation
$$
\gamma^{m_1\dots m_s}\gamma_{11} ={1\over (10-s-1)!}\eta
\epsilon^{m_1\dots m_s m_{s+1}\dots m_{10}}\gamma_{m_{s+1}\dots
m_{10}},
\eqno(6.2.2)$$
where $\eta=\pm 1$, to eliminate terms with $p\ge 6$.
\par
The IIA theory was obtained in references [107],[108] and [109]
from the $D=11$ theory
by compactification  and this is the method of construction
we now follow. We consider the eleven-dimensional supergravity of
equation (5.2) and take the eleventh dimension to be a circle $S^1$
of radius $R$. To be precise, we take the eleventh
coordinate
$x^{10}$ to  be such that $x^{10 \prime}\sim x^{10}$ if
$x^{10 \prime}= x^{10}+ 2\pi n R,\ n\ \in\ {\bf Z}$ where
$2\pi R$ parameterizes the range of $x^{10}$.
 We will also write $x^{10}$ as $x^{10}= \theta R$
for $0 \le \theta < 2\pi$.
We adopt the convention that hatted indices run over all eleven
dimensional indices, but unhatted indices only run over the ten
dimensional indices,  for example $\hat \mu,\hat \nu=0,1,\ldots ,10$
while
$\mu,\nu=0,1,\ldots ,9$.
Given any of the fields in the eleven-dimensional supergravity
 we can
take its Fourier transform on
 $x^{10}$. In particular,  if $\phi$ represents any of these
fields   whose Lorentz and possible spinor  indices are
suppressed, we find that
$$\phi(x^\mu,x^{10})=\phi(x^\mu)+\sum_{n,
n\ne0}e^{in\theta}\phi_n
(x^\mu)
\eqno(6.2.3)$$
Thus from each particle in ten dimensions we find an infinite number
of particles in ten dimensions. The non-zero modes (i.e. $n\not=0$),
however, will lead  to  massive particles whose masses are given by
their momentum in the eleventh direction.
 Such massive particles are called Kaluza-Klein
particles. In the limit when the radius of the circle is large  these
particles become infinitely massive and  can be neglected, whereupon one
is  left with a finite set of massless particles which form a
supergravity theory. Discarding the  massive particles can be achieved
by  taking  all the eleven-dimensional fields to be independent of
$x^{10}$ and this we now do.
\par
This reduction proceeds in the following generic manner
$$\matrix{
D=11&&e_{\hat\mu}^{\ \hat m}&
A_{\hat\mu_1\hat\mu_2\hat\mu_3}&\psi_{\hat\mu\alpha}\cr
&&\downarrow&\downarrow&\downarrow\cr
D=10&&e_\mu^{\ m},B_\mu,\phi& A_{\mu_1\mu_2\mu_3},A_{\mu_1\mu_2}&
\psi_{\mu \alpha},\lambda_\alpha\cr}
\eqno(6.2.4)$$
While one can reduce the fields in many ways only some
definitions of the ten-dimensional fields will lead to a final result
which is in  the  generic form  in which a supergravity theory is
usually written. The three-form gauge field reduces in an obvious way
$A_{\mu_1\mu_2\mu_3}= A_{\mu_1\mu_2\mu_3},
\  A_{\mu_1\mu_2}=
A_{\mu_1\mu_2 10}$ where $\mu_1,\mu_2\mu_3=0,1\ldots
9$. However, the useful reductions  for the
other fields are more subtle; the vielbein takes the  form
$$e_{\hat\mu}^{\ \hat m}=
\left(\matrix{ e^{-{1\over 12}\sigma }e_\mu^{\ m} &2e^{{2\over 3}\sigma}
B_\mu\cr
  0                  & e^{{2\over 3}\sigma} \cr}\right),\
(e^{-1})^{\ \hat\mu}_{ \hat m}=
\left(\matrix{
 e^{{1\over 12}\sigma }e^{\ \mu}_{\ m} &-2e^{{1\over 12}\sigma}
B_\nu (e^{-1})^{\ \nu}_{  m}\cr
  0                  & e^{-{2\over 3}\sigma} \cr}\right)
\eqno(6.2.5)$$
while the eleven-dimensional gravitino becomes
$$
\psi_{\hat m }
=(e^{-{1\over24}\sigma}e_m^\mu\psi_\mu^\prime,\ {2\over 3}\sqrt 2
e^{{17\over24}\sigma}\lambda)
\eqno(6.2.6)$$
where
$\psi_\mu^\prime=e^{-{1\over 24}\sigma}(\psi_\mu - \sqrt {{1\over
72}}\Gamma_\mu\Gamma^{11}\lambda) -\sqrt {{32\over 9}}
e^{{3\over 4}\sigma} B_\mu\lambda$ and
$\Gamma^{11}= i\Gamma^{1}\dots \Gamma^{10}$.
The above formulae and those
below are related to those of  reference [107] by carrying out the
transformation
$\sigma \to {2\over 3}\sigma$ on the latter.
\par
The field $B_\mu$ is a gauge field whose gauge transformation
has a parameter that came from   the  general coordinate
transformations with parameter
$\xi^{10}$ in the  eleven-dimensional theory.
 The component
$e_{\hat\mu=10}^{\ m}$  of the vielbein
can be chosen to be zero as a
result of a local Lorentz transformation $w^m_{\ 10}$.
The strange factors involving $e^\sigma$ and other redefinitions  are
required in order to get the usual  Einstein and spinor kinetic energy
terms. The ten-dimensional Newtonian coupling constant $\kappa$ is given
in terms of the eleven-dimensional Newtonian constant $\kappa_{11}$ by
$\kappa^2= {(\kappa_{11})^2\over 2\pi R}$ and has the dimensions of
$(mass)^{-8}$. This equation follows from examining the $\kappa$ that
results from the-dimensional reduction. We have  set
$\kappa=1$ in this section.
\par
The resulting ten-dimensional IIA  supergravity theory is given by
$$ L=L^B+L^F
\eqno(6.2.7)$$
where the first term contains all the terms which are independent of
the fermions and the second term is the remainder. The bosonic part
is given by  [107],[108],[109]
$$\eqalign{
L^B&=-eR\big(w(e)\big)-{1\over12}ee^{\sigma\over2}
F^\prime_{\mu_1\dots\mu_4}F^
{\prime\mu_1\dots\mu_4}+{1\over3}ee^{-\sigma}F_{\mu_1\dots\mu_3}
F^{\mu_1\dots
\mu_3}\cr
&\quad-ee^{{3\over2}\sigma}F_{\mu_1\mu_2}F^{\mu_1\mu_2}
+{1\over2}\partial_\mu
\sigma\partial^\mu\sigma\cr
&\quad+{1\over2\cdot(12)^2}\epsilon^{\mu_1\dots\mu_{10}}
F_{\mu_1\dots\mu_4}F_{
\mu_5\dots\mu_8}A_{\mu_9\mu_{10}}\cr}
\eqno(6.2.8)$$
where
$$F_{\mu_1\mu_2}=2\partial_{[\mu_1}B_{\mu_2]}
\eqno(6.2.9)$$
$$
F_{\mu_1\mu_2\mu_3}=3\partial_{[\mu_1}A_{\mu_2\mu_3]}
\eqno(6.2.10)$$
$$
F^\prime_{\mu_1\dots\mu_4}=4\partial_{[\mu_1}A^\prime_{\dots\mu_4]}
+12A_{[\mu_1\mu_2}G_{\mu_3\mu_4]}
\eqno(6.2.11)$$
In the last definition we have  used the
field
$A^\prime_{\mu_1\mu-2\mu_3}=
A_{\mu_1\mu-2\mu_3}-6B_{[\mu_1}A_{\mu_2\mu_3]}
$ which is invariant under the   gauge transformation  with
parameter
$\xi^{10}$. The fermionic part of the Lagrangian is much more
complicated and the first two terms are
$$L^F=-{i\over2}e\psi_{\mu_1}\Gamma^{\mu_1\mu_2\mu_3}
D_{\mu_2}\psi_{\mu_3}+{i
\over2}e\bar\lambda\Gamma^\mu D_\mu\lambda+\dots
\eqno(6.2.12)$$
\par
The transformations of the fields can be deduced in a similar fashion
from the  transformation of eleven-dimensional fields of equation
(5.1). For example, the veilbein and dilaton transform as
$$\delta e_\mu^{\ m}= -i\bar \epsilon  \Gamma ^m \psi _\mu,\
\delta \sigma =\sqrt {2} i\bar \lambda \Gamma^{11}\epsilon
\eqno(6.2.13)$$
where $\epsilon$ is the suitably defined parameter of local
supersymmetry transformations.
We refer the reader to references [107],[108]
and [109] for the  transformations  of the other fields
and the terms in the fermionic part of the action.
\par
The IIA action has an
$SO(1,1)$ invariance with  parameter c that transforms the
fields as
$$\sigma \to \sigma +c,\ B_\mu \to e^{-{3\over 4}c}B_\mu,\
A_{\mu\nu}\to e^{{1\over 2}c}A_{\mu\nu},\
A_{\mu\nu\rho}\to e^{-{1\over 4}c}A_{\mu\nu\rho}
\eqno(6.2.14)$$
while the  vielbein in ten dimensions is inert.
This symmetry has its origin in the eleven-dimensional theory
and in particular the Weyl scalings of equation
 (5.6) given in the previous section.
Although these are not a symmetry
of the  action in eleven dimensions
we can convert them into a symmetry
of the action in ten dimensions provided we combine
them with a diffeomorphism on $x^{10}$, in particular the
diffeomorphism
$x^{10} \to
e^{-9\alpha } x^{10}$. To keep the range of
$x^{10}$ the same we also scale $R$ by
$R\to e^{-9\alpha } R$. This diffeomorphism
 is a symmetry of the  eleven
dimensional theory and  from the active viewpoint
transforms  $\int dx^{10}\to
 e^{-9\alpha }\int dx^{10}$ and the Lagrangian $L$
in eleven-dimensions
Lagrangian
to  $L\to  e^{9\alpha }L$. The theory in ten dimensions
is obtained by substituting the field expansion of equation (6.2.3)
into the action in eleven-dimensions. However, the effect of
the
$\int d x^{10}$ is just to extract  the part of the
Lagrangian $L$ in eleven dimensions that
is independent of
$x^{10}$, which then becomes the Lagrangain in ten dimensions
 and gives a factor of
$2\pi R$. The latter  is then combined with the factor
$\kappa_{11}^{-2}$ to define the ten-dimensional
Newtonian coupling constant
$\kappa_{10}^{-2}=2\pi R \kappa_{11}^{-2}$.
Clearly, the  coupling constant $\kappa_{10}$
in ten dimensions is inert
under the combined transformation as $R\to e^{-9\alpha } R$ under the
diffeomorphism and $\kappa_{11}^{-2}\to e^{9\alpha }\kappa_{11}^{-2}$
under the Weyl scaling. Similarly the action in ten dimensions
is inert as it scales by $e^{9\alpha }$ under the diffeomorphism and
$e^{-9\alpha }$  under  the Weyl scaling.
Thus we have found that  dimensional  reduction has
transformed a symmetry of the equations of motion  into a symmetry of
the action.
\par
An alternative way of writing the IIA Lagrangian is to use the
so-called string metric $g_{\mu\nu}^s$. This metric is the one that
occurs in the  sigma model approach to string theory which starts with
the  sigma model action in two dimensions:
$$-{1\over 4\pi\alpha'}\int d^2\xi (\sqrt {-g} g^{\alpha \beta }\partial
_\alpha x^\mu
\partial _\beta x^\nu g_{\mu\nu}^s+ \epsilon ^{\alpha \beta }
\partial _\alpha x^\mu\partial _\beta x^\nu A_{\mu\nu})
+{1\over 4\pi}\int d^2\xi\sigma R^{(2)}
\eqno(6.2.15)$$
where $R^{(2)}$ is the  curvature scalar
in two dimensions . The fields
$A_{\mu\nu}$  and $\sigma$ are the anti-symmetric tensor gauge field and
dilaton  which appear in the massless string spectrum and occur in the
IIA action given above. The constant $\alpha'$ has the dimensions
of $(mass)^{-2}$ and  defines the mass scale $m_s$ of the string
by $m_s^2= {1\over 4\pi {\alpha'}}$.  The combination in front of the
first two terms of the string action is often called the
string tension $T$ (i.e. $T= {1\over 4\pi\alpha'}$).
 One recovers the
tree-level string equations and thus at lowest order in $\alpha'$
we find the supergravity equations of motion,
by demanding  conformal invariance.
The corresponding string vielbein
$e^{s \ m}_\mu$ is related to the above vielbein by
$e^{s \ m}_\mu=e^{{1\over 4}\sigma}e^{ \ m}_\mu$.
\par
The last term in the above sigma model action, for constant $\sigma$,
takes the form
$<\sigma>\chi$ where $\chi$ is the Euler number and is given by
$\chi= {1\over 4\pi}\int d^2\xi R^{(2)}$. For a closed Riemann
surface of genus g it is given by $\chi=2-2g$.

\par
Making this change
in the bosonic part of the IIA Lagrangian of equation (6.2.8)
and dropping
the "s" superscript on the string vielbein  we find that
the Lagrangian  becomes
$$
L_B=ee^{-2\sigma}\{-R+4\partial_\mu\sigma\partial^\mu\sigma
-{1\over3}F_{\mu_1\mu_2\mu_3}F^{\mu_1\mu_2\mu_3}\}
$$
$$+\left\{-{1\over12}eF^\prime_{\mu_1\dots\mu_4}
F^{\prime\mu_1\dots\mu_4}-
eF_{\mu_1\mu_2}F^{\mu_1\mu_2}\right\}
+{2\over12^2}\epsilon^{\mu_1\dots\mu_{10}}
F_{\mu_1\dots\mu_4}F_{\mu_5\dots
\mu_8}A_{\mu_9\mu_{10}}
\eqno(6.2.16)
$$
\par
As we have mentioned, the IIA action is the lower energy limit of the
IIA string theory, which is obtained as the string tension goes to
zero (i.e. $\alpha ^\prime \to \infty$).
 In this limit one is left with
only the massless  particles of the IIA supergravity theory. It will be
very useful to know how these particles  arise  in the
IIA string. This closed string theory in its formulation
with manifest world surface supersymmetry, that is the
Neveu-Schwarz-Rammond formulation [7], has four
sectors;  the
$NS\otimes NS$, the $R\otimes R$, $R\otimes NS$ and the
$NS\otimes R$
corresponding to the different boundary conditions
that can be adopted for the two
dimensional spinor in the theory. Clearly, the
$NS\otimes R$ and $R\otimes NS$ sectors contain
the fermions while the $NS\otimes NS$ and  the $R\otimes R$ sectors
contain the bosons. It is straightforward to solve the physical state
conditions in these sectors and one finds that the bosonic fields of
the  IIA supergravity arise as
$$\underbrace{e_\mu^a, A_{\mu\nu},\sigma}_{NS\otimes
NS};\quad\underbrace{A_{\mu \nu\varrho}, B_\mu}_{R\otimes R};
\quad\underbrace{\psi_{\mu\alpha},\lambda_\alpha}_{NS\otimes R\  and
\ R\otimes NS}
\eqno(6.2.17)$$
\par
Looking at the IIA Lagrangian in the string frame of equation (6.2.16)
we find that all the fields that arise in the $NS\otimes NS$
sector occur in a different way from those in the
$R\otimes R$ sector; While the former have a factor of $e^{-2
\sigma }$ the latter do not have such a factor. We will find
 the same phenomenon for the IIB string.
\par
The IIA supergravity  has two parameters. It has the Newtonian coupling
constant $\kappa$, which we have suppressed, and  the
parameter $<e^\sigma>$. The IIA string also has two parameters
the string tension $T$, or equivalently the string mass scale $m_s$, and
the string coupling constant $g_s$. Since the low energy effective
action of the string is  IIA supergravity these two sets of
parameters must be  related.
\par
We first consider how the parameters arise
in the string theory.
In a second
quantized formulation of string theory, one finds that the action
can be written in a way where the string coupling only occurs as a
prefactor of $g_s^{-2}$. Examples of such formulations are
the    light-cone gauge action or the gauge covariant action. The
parameter $\alpha'$ only occurs in these formulations through the
masses of the the particles or equivalently the $L_n$ operators that
occur in these formulations. In the path integral formulation, the
action becomes multiplied by $\hbar^{-1}$ and so we find that
$\hbar$ and $g_s$  only occur in the   combination $\hbar g_s^2$. This
situation is identical to the
 way in which  the gauge coupling  occurs in Yang-Mills theory.
The
 power of Planck's constant measures the number of loops. Indeed if
we have any Feynman graph with $n$ loops, $I$ propagators and
$V$ vertices, the power of $\hbar$ associated with an $n$-loop graph
 is given by
$\hbar^{(I-V)}$, since each vertex carries a power of  $\hbar$ and
each propagator an inverse power of $\hbar$.
Using the topological relation
$n=I-V+1$, the power of $\hbar$ for a $n$-loop diagram becomes
$\hbar^{(n-1)}$.  Our previous discussion then implies that each
$n$-loop diagram has a power $g_s^{(2n-2)}$ associated with it.
\par
Now let us examine  how the parameters arise in the effective action.
The   first quantized string action
in two dimensions of equation (6.2.15) contains  the dilaton  $\sigma$
multiplied by  the
Euler number $\chi$ of the Riemann  surface. A surface of genus n
corresponds to a n-loop string amplitude and so in the
path integral of this action one finds a factor of
$e^{(2n-2)<\sigma>}$. As a result, we conclude that
  the IIA string coupling and the expectation value of the dilaton
are related by $g_s=<e^\sigma>$  Since $\kappa$ has the
dimension of $(mass)^{-4}$ it must be proportional to
$(\alpha')^{2}$ and is given by the relation
$\kappa=
(\alpha')^2 e^{<\sigma>}$.
\par
We now finish our discussion of the IIA supergravity by reiterating some
of the above features  that will be useful for
discussions of string duality.  The IIA supergravity theory
has the gauge fields
$\sigma, B_\mu,\ A_{\mu\nu}$, and
$A_{\mu\nu\varrho}$.
This means the IIA theory has  gauge fields of rank $q$ where
$q=1,2,3$ and these have corresponding field strengths of rank
$q+1=2,3,4$.  Given a field strength $F_{\mu_1\ldots \mu_q}$ of rank
$q+1$
we can define a dual field strength by
$F_{\mu_1\ldots \mu_(D-1-q)}={1\over (q+1)!}\epsilon _{\mu_1\ldots
\mu_{10}}F^{\mu_{D-q}\ldots\mu_{10}}$. Hence the  duals of the above
field strengths are forms of rank
$q=6,7,8$. When the original field strengths are on-shell  we can, at
least at the linearized level  write the dual field strengths  in terms
of dual gauge fields  of ranks
5,6 and 7. Hence the IIA theory has gauge fields of ranks
p=1,2,3,5,6,7 if we include the dual gauge fields as well as
the original ones. It is instructive to list the above gauge fields
according to the string sector in which they arise. In the $NS\otimes
NS$ we find gauge fields of ranks
2 and 6
while in the $R\otimes R$ sector we find gauge
fields of ranks
1,3,5 and 7. We observe that classifying the gauge fields according
to the different sectors splits them into fields of  odd and even
rank.
\par
From equation (6.2.5) we read off the component of the  vielbein
associated with the circle to be  $e_{10}^{\ 10}=e^{{2\over3}<\sigma>}$
with corresponding  metric   $g_{10 \ 10}=e^{{4\over3}<\sigma>}$. The
parameter $R$ introduced into the defining range of the variable
$x^{10}$ has no physical meaning as it only parametrizes the range of
$x^{10}$. However, from the metric  we can compute the radius $R_{11}$ of
the circle in the eleventh dimension. We  find that  $R_{11}=
Re^{{2\over3}<\sigma>}$. We recall that the string coupling is given  by
$g_s= e^{<\sigma>}$ and as a result we find that
$$g_s= {\left( {R_{11}\over R}\right)}^{3\over 2}
\eqno(6.2.18)$$
The above relationship between the radius $R_{11}$ of  compactification
of the eleven-dimensional theory and  the IIA string coupling
constant implies  in particular that as $R_{11}\to \infty$ we find
that $g_s\to \infty$. Thus in the strong coupling limit of the
IIA string the radius of the circle of compactification becomes
infinite suggesting that the theory decompactifies [124].
\par
We now consider some properties of the Kaluza-Klein modes
which we have so far ignored in the reduction from eleven dimensions.
Their massess are given by the action of
$(e^{-1})_{11}^{\ \hat \mu}\partial _{\hat \mu}= e^{-{2\over
3}\sigma}\partial _{11}$ where we have used equation (6.2.5).
Examining the expansion of equation (6.2.3)  we find that
 the masses of Kaluza-Klein particles are  given by
$ n{e^{-{2\over3}<\sigma>}\over R}$ for integer $n$. However, using the
relationship between $R$ and $R_{11}$ given just above  we find
that  the masses of the Kaluza-Klein particles are  given by
${ n\over R_{11}}$. The
 gauge field $B_\mu$ which originated from the eleven-dimensional
metric couples to the   Kaluza-Klein particles in a way which is
governed by the derivative
$$(e^{-1})_{m}^{\ \hat \mu}\partial _{\hat \mu}=
e^{{1\over 12}\sigma}(e^{-1})_{m}^{\  \mu}(\partial _{\mu}
-2B_\mu\partial _{11})
$$
where we have again used equation (6.2.5). As such, we find that the
Kaluza-Klein particles have charges given by
${2n\over R}$ for integer $n$. From the IIA string perspective,
the $B_\mu$ field is in the $R\otimes R$ sector and so
the Kaluza-Klein particles couple with these charges to the
$R\otimes R$ sector. It turns out that the IIA supergravity
possess solitonic particle solutions that have precisely the
masses and charges of the Kaluza-Klein particles [123].
Thus the  IIA supergravity and so in
effect the IIA string knows   about all the particle content of
the  theory in eleven dimensions and  not only the
massless modes that arise
after the  compactification.
\par
It is these
observations that underlie the conjecture [123],[124] that the strong
coupling limit of the IIA string theory is an eleven-dimensional
theory, called M-theory, whose low energy limit is eleven-dimensional
supergravity.
\medskip
{\bf 6.3 IIB Supergravity}
\medskip
The IIB supergravity was found in references [110],[111] and [112] using
two  different methods. In reference [110], a variant, [126],  of the
Noether method  was used: rather than working with an action and
transformation laws,  one can  just use the transformation laws. One
begins with the rigid supersymmetry transformations and local Abelian
transformations of the linearised theory. Letting the supersymmetry
parameter become space-time dependent,
  the transformations laws no longer close; however  we may still close
the supersymmetry algebra  order-by-order in
$\kappa$ by adding terms to
the transformation laws provided we also identify the now local spinor
parameter of the supersymmetry transformation with the spinor
parameter that occurs in the local Abelian transformation of the
gravitino. In this way the supersymmetry transformations laws of the
IIB  theory and the fact that the scalars belong to the coset
$SU(1,1)/U(1)$ were found [110]. Using the fact that  the transformations
laws  only close on-shell this work was extended in reference [112]
to find the equations of motion in the absence of fermions. In the
independent work of reference [111] the full  equations of motion in
superspace and $x$-space were found using the on-shell superspace
techniques of section 4.2. The third-order terms of the IIB theory
were constructed for the  light-cone gauge Hamiltonian in reference
[125].
\par
 The strategy behind these calculations   is explained in
chapter 4 and although the ideas are straightforward the
calculations themselves are technically complicated to the extend
that they will not be reproduced here. Nonetheless we will describe
the essential features of the IIB theory so that the reader will
grasp some of the ideas involved and gain a feel for the IIB theory
itself.
\medskip
{\bf 6.3.1 The Algebra}
\medskip
The IIB supergravity is based on a supersymmetry algebra
whose two supercharges $Q_\alpha^i,\ i=1,2;\ \alpha=1,\dots,32$ are
Majorana-Weyl spinors of the same chirality.
They therefore obey the conditions
$${(Q_\alpha^i)}^*=Q_\alpha^i,\quad\Gamma_{11}Q^i=Q^i
\eqno(6.3.1)$$
The supersymmetry
algebra is given by
$$\{Q_\alpha^i,Q_\beta^j\}= (\gamma^\mu C^{-1})_{\alpha \beta}\delta
^{ij}P_\mu +\dots
\eqno(6.3.2)$$
where $+\dots$ denote terms with  central charges whose form the
reader may readily find by following the discussion at the end of
section 1.5.
\par
It is more useful to  work instead with the complex Weyl
supercharges
$Q_\alpha=Q_\alpha^1+iQ_\alpha^2$,
$\bar Q_\alpha=Q_\alpha^1-iQ_\alpha^2=(Q_\alpha)^*$
The supersymmetry algebra also
contains a  $U(1)$ generator denoted  $R$
($R^\dagger = - R$) which acts on the supercharges as
$$[Q_\alpha,\ R]=iQ_\alpha,\ [\bar Q_\alpha,\ R]=-i\bar Q_\alpha
\eqno(6.3.3)$$
\medskip
{\bf 6.3.2 The Particle Content}
\medskip
The field content of the IIB theory is
$$ e_\mu^{\ m},\ A_{\mu\nu},\ a,\ B_{\mu\nu\rho\kappa},\
\psi_{\mu \alpha},\ \lambda_\alpha
\eqno(6.3.4)$$
The fields  $A_{\mu\nu}$ and $a$ are complex while the gauge field
$B_{\mu\nu\rho\kappa}$ is real. The spinors are complex
Weyl spinors;  the graviton   $\psi_{\mu \alpha}$ is of the
opposite chirality
 to $\lambda_\alpha$, but  has the same  chirality
as  the supersymmetry parameter $\epsilon_\alpha$.
Recalling our discussion above equation (5.2) we find that these
fields  lead to 35, 56, 2, 35 ,112 and 16 on-shell degrees of freedom
respectively. The gauge field $B_{\mu_1\mu_2\mu_3\mu_4}$ defines
the linearised five-rank field strength $g_{\mu_1\mu_2\mu_3\mu_4\mu_5}
\equiv 5\partial
_{[\mu_1}  B_{\mu_2\mu_3\mu_4\mu_5]}$ which satisfies a self-duality
condition. At the linearized level this self-duality condition is
given by
$$ g_{\mu_1\mu_2\mu_3\mu_4\mu_5}={1\over 5 !}
\epsilon_{\mu_1\mu_2\mu_3\mu_4\mu_5\nu_1\nu_2\nu_3\nu_4\nu_5}
g^{\nu_1\nu_2\nu_3\nu_4\nu_5}\equiv {}^* g_{\mu_1\mu_2\mu_3\mu_4\mu_5}
\eqno(6.3.5)$$
Without the self-duality condition this gauge field
corresponds to a particle that  belongs to the fourth-rank
totally anti-symmetric representation of the little group
$SO(8)$. The self-duality condition above corresponds to the
constraint that this representation  is self-dual and hence the
35 degrees of freedom given above. Thus the supermultiplet of IIB
supergravity has 128 bosonic degrees of freedom and 128
fermionic degrees of freedom on-shell.
\par
Most of the fields of equation (6.3.4) transform under the $U(1)$
transformations; their $R$ weights are
0,2,4,0,1 and 3 respectively. Clearly, real fields must have
$R$ weight zero and the gravitino must have the opposite $R$ weight as
the supercharge $Q_\alpha$ since $\delta \psi_\mu= \partial _\mu
\epsilon$.
\par
Following the  pattern of the standard action for gauge fields one may
be tempted to use the linearised action
$$\int d^{10} x\
\ g_{\nu_1\nu_2\nu_3\nu_4\nu_5}g^{\nu_1\nu_2\nu_3\nu_4\nu_5}
\eqno(6.3.6)$$
for the fourth-rank gauge field. However,
if $g$ is the five-form
associated with the gauge field, the above
action is given by $\int g\wedge {}^* g=
\int g\wedge g =0$. This  discussion
can be rephrased without using forms as follows: using the
self-duality condition  we can rewrite one of the field strengths in
terms of ${}^* g$;  swopping the indices on the $\epsilon$ symbol such
that the last five indices are at the beginning, we incur a minus
sign;  using the  self-duality condition once more we again find that
the above
action vanishes  as its negative.
\par
Clearly
this result holds for  any rank
$2n+1$ self-dual gauge field strength  in a space-time of dimension
$4n+2,\ n\in\ {\bf Z}$. Indeed there is no simple action  for the
fourth-rank gauge field and so for the IIB theory itself.  Although
some actions have been suggested there are reasons to  believe that
they do not correctly  capture all the physics of the theory. As a
result we will content ourselves with  deriving the  equations of
motion.
\par
The IIB supergravity is the theory that describes the
effective action of the low energy
limit of the IIB string. The massless fields in the IIB string being
those that occur in the IIB supergravity. To find the massless fields
we must examine the physical state conditions for the IIB string
which, being a closed superstring, has in the Neveu-Schwarz- Rammond
formulation, the usual
 $NS\otimes NS$,  $R\otimes R$, $R\otimes NS$ and  $NS\otimes R$
corresponding to the possible boundary conditions for the two
dimensional spinor in the theory. Clearly, the last two sectors contains
the fermions while the  $NS\otimes NS$ and  the
$R\otimes R$ sectors contain
the bosons. The bosons arise as
$$\underbrace{ e_{\mu}^a,\ A^1_{\mu\nu},\
\sigma}_{NS\otimes NS}\quad\underbrace
{A^2_{\mu\nu},\ l,\ B_{\mu\nu\rho\tau}}_{R\otimes R}
\eqno(6.3.7)$$
where $A^1_{\mu\nu}$ and  $A^2_{\mu\nu}$, are real
fields which make up the complex fields
$A_{\mu\nu}$, and we
replace the complex scalar $a$ by two real fields
$l$ and $\sigma $. The precise   way in  which
these decompositions are defined will be specified later.
 The fact
that the two rank-two  gauge fields are split between the $NS\otimes
NS$ and  $R\otimes R$ sectors has important consequences for
discussions of string duality. Since the physical state condition
in the $NS\otimes NS$ sector are exactly the same as in the
IIA string we should not be surprised to find that this sector
contains exactly the same bosonic field content.
\medskip
{\bf 6.3.3 The Scalars}
\medskip
The gauge fields, the graviton and the gravitini possess gauge
transformations
as a result of which they can only occur in gauge invariant
quantities, which in the sense of section 4,  have geometric dimensions
greater than zero. At first sight, this is not the case for the
scalars
of the theory since they have geometric dimension zero.
As we explained in section,  dimensional analysis plays
an important role in the construction of supergravity theories
and as such it might seem that the role of the scalars in the theory
is difficult to determine.  Fortunately, however, the
scalars   belong [110] to
the coset space
 ${SU(1,1)\over U(1)}$  and as a consequence the  way they can
  can occur in the theory is strongly constrained.
\par
The use of coset spaces to describe
scalar fields was described in reference [127]. Since  it is just as
simple to describe the general theory [127] we will give the
construction for a  general coset space.  Let
 $G$ be  any group,  $H$ one of its subgroups and denote
the coset space by
$G/H$. Let us
consider any space-time dependent
${\cal {V}}\in G$ which is taken to transform as
$${\cal {V}}\ \rightarrow\ g{\cal {V}} h
\eqno(6.3.8)$$
where $ h\in H$ is a local (i.e. space-time dependent) transformation
and
$g\in G$ is a  rigid transformation. We may use the local $H$
transformations to gauge away dim$H$ scalar fields
leaving ${\rm {dim}} G- {\rm {dim}}H$ scalar fields in ${\cal {V}}$.
The object
$\Omega_\mu \equiv {\cal{V}}^{-1}\partial_\mu
{\cal {V}}$ belongs to the Lie algebra of $G$ and so
can be written in the form
$$\Omega_\mu={\cal{V}}^{-1}\partial_\mu
{\cal {V}}
=f_\mu^a T_a+w_\mu^iH_i
\eqno(6.3.9)$$
where $H_i$ are generators of $H$ and $T_a$ are the remaining
generators of $G$. The object
$\omega _\mu$ is invariant under the rigid transformations
$g\in\ G$, but transforms under local $H$ transformations as
$$\omega_\mu\ \rightarrow\ h^{-1}\partial_\mu h+h^{-1}\omega_\mu h
\eqno(6.3.10)$$
The theory simplifies if we restrict ourselves to
reductive cosets which are those for which the commutator
$[T_a,\ H_i]$ can be written in terms of only the coset generators
$T_a$.  In this case the above transformation rule implies that
$f_\mu\equiv f_\mu^a T_a \to h^{-1}f_\mu h$ and
$w_\mu\equiv w_\mu^iH_i \to h^{-1}w_\mu h+ h^{-1}\partial_\mu h$.  We
can think of
$f_\mu^a$ as a vielbein on
${G/ H}$ defining a set of preferred frames and $w_\mu^i$ as the
connection  associated with local $H$ transformations.
\par
An invariant Lagrangian  is given by
$$\eta ^{\mu\nu} Tr (f_\mu f_\nu)
\eqno(6.3.11)$$
The corresponding equation of motion is given by
$$D_\mu f^\mu_a=0
\eqno(6.3.12)$$
where we have introduced the covariant derivative $D_\mu f_\nu
\equiv \partial_\mu f_\nu+
[w_\mu,\ f_\nu]$. It is straightforward to verify that
the above Lagrangian and  equation of motion are invariant under
both local $H$ transformations and rigid $G$ transformations.
\par
Let us work out the above expressions for the
case of interest, namely for $G=SU(1,1)$ and
$H=U(1)$. The group $G=SU(1,1)$ is the set of two by two matrices $g$
of determinant one which acts on the column vector
$\left(\matrix{z_1\cr z_2\cr}\right) $ by $\left(\matrix{z_1\cr
z_2\cr}\right) \to g\left(\matrix{z_1\cr z_2\cr}\right) $
in such a way  as to preserve
$|z_1|^2-|z_2|^2$. The most general element
of
$SU(1,1)$ can be written in the form
$$U= \left(\matrix{u&v\cr
                   v^*&u^*\cr}\right)
\eqno(6.3.13)$$
subject to $uu^*-vv^*=1$. An  infinitesimal element of
$G=SU(1,1)$  can therefore be written in the form
$g=I+A$ where $A$ is given by
$$A= -2i\hat a \sigma _3 + b_1\sigma _1 -b_2 \sigma _2
= \left(\matrix{-2i\hat a& b\cr
                   b^*&+2i\hat a\cr}\right)
\eqno(6.3.14)$$
where $\hat a ,\ b_1$ and $b_2$ are real, $b=b_1+i b_2$,
and $\sigma _i,\ i=1,2,3$, are the Pauli matrices.
An alternative parameterisation of elements of $G=SU(1,1)$
is given by exponentiating the above infinitesimal element;
$$U=e^A=
\left( \matrix{ cosh \rho -2i\hat a{sinh\rho \over \rho}&b{sinh\rho
\over
\rho}\cr
        b^* {sinh\rho \over \rho}&cosh \rho +2i\hat a{sinh\rho
\over \rho}\cr}\right)
\eqno(6.3.15)$$
where $\rho^2= b^*b-4 {\hat a}^2$.
The $U(1)$ subgroup is generated by
$i\sigma_3$ and its elements take the form
$$ h=\left(\matrix{e^{-2i\hat a}&0\cr 0&e^{+2i\hat a}\cr}\right)
\eqno(6.3.16)$$
\par
Taking ${\cal{V}}$ to be a general elements of $SU(1,1)$ we find that
$\omega_\mu$ takes the form
$$\omega_\mu=\left(\matrix{2iQ_\mu& P_\mu\cr \bar
P_\mu&-2iQ_\mu\cr}\right)
\eqno(6.3.17)$$
where  $Q_\mu=( Q_\mu)^*$.
Under an infinitesimal  local $U(1)$ transformation of equation
(6.3.16) we find from equation (6.3.10) that the vielbein and
$U(1)$-connection transform as
$$\delta Q_\mu=-\partial_\mu \hat a,\quad\delta P_\mu=4i\hat a P_\mu
\eqno(6.3.18)$$
The last  equation corresponds to the $U(1)$ weight 4 assignment
 given earlier to
the  scalars. If
we write
${\cal{V}}=(C_+,C_-)$, where
$C_+$ and
$C_-$ are two  column vectors,  then $C_{\pm}$ transforms as
$C_\pm\to g C_\pm e^{\mp 2i\hat a}$ under local and rigid
transformations. If we further take the ratio of the top and
 bottom component of the column vectors
 $C_{\pm}$ and denoted it
by
$c_{\pm}$, then it follows that $c_{\pm}$ is inert under local $H$
transformations but transforms as
$$ c_{\pm} \to {uc_{\pm}+v\over v^*c_{\pm}+u^*}
\eqno(6.3.19)$$
under a rigid transformation of the form of equation (6.3.13).
\par
We can use a local
$U(1)$ transformation to bring
${\cal{V}}$  to be of the form
$$ {\cal{V}} =
{1\over\sqrt{1-\phi\phi^*}}
\left( \matrix{1&\phi\cr
\phi^*&1\cr}\right)
\eqno(6.3.20)$$
This
choice is most easily achieved using the form of
$SU(1,1)$ elements given in  equation (6.3.15).
Examining the second column vector in ${\cal{V}}$ we find
that $c_-=\phi$ and so $\phi$ transforms as under
$G$ as
$$ \phi  \to {u\phi+v\over v^*\phi+u^*}
\eqno(6.3.21)$$
For the choice of ${\cal{V}}$ of equation (6.3.20) we find that
$$Q_\mu=-{i \over 4} {(-\phi \partial_\mu \phi^*
+\phi^*\partial_\mu \phi)\over (1-\phi\phi^*)},\
P_\mu={\partial_\mu\phi\over(1-\phi\phi^*)}
\eqno(6.3.22)$$
\par
The corresponding equation of motion is found by substituting
equation (6.3.17) into equation (6.3.12) to find
$$D_\mu P^\mu\equiv\partial_\mu P^\mu+4iQ_\mu P^\mu=0
\eqno(6.3.23)$$
The actual IIB equations of
motion of the scalars must  be of this form, but will also  include
other terms containing  the superpartners of the scalars. We
observe that the scalars only occur through
$P_\mu$ or, for derivatives of other fields with non-zero $U(1)$
weight, through $Q_\mu$. Both these fields are given by equation
 (6.3.9) which contains the  space-time derivatives
acting on the group
element and so they have geometric  dimension one. This fact plays a
crucial role in the way the scalars are encoded into the IIB
theory as it allows the use of dimensional analysis to restrict
the way the scalars can occur in the theory.
\par
It has been found that the scalar fields that occur in supergravity
theories always belong to a coset space [169],[170],[171]. Some of the
other interesting cases are the $N=4$ and $N=8$ supergravity theories in
four dimensions  where the coset spaces are $SU(1,1)/U(1)$[169] and
$E_7/ SU(8)$ [170]. In table 6.1 we give the coset spaces
associated with the maximal supergravities. this  table was
taken from references [167] and  [181]. All these
theories  except for the IIB theory arise from the dimensional
reduction of the  eleven-dimensional supergravity theory.
\bigskip
\bigskip
\bigskip
\vskip3cm
\centerline {\bf 6.1 Coset spaces of the Maximal
Supergravities}
\bigskip
$$\vbox{\tabskip=0pt \offinterlineskip

\halign to238pt{\strut#& \tabskip=1em& \ #\hfil& \ # \hfil& \ # \hfil&\ #
\tabskip=0pt\cr \noalign{\hrule}

&D&G&H&\cr \noalign{\hrule}

&11&1&1&\cr

&10, \ IIB&$SL(2)$&SO(2)&\cr

&10,\ IIA&$SO(1,1)/Z_2$&1&\cr

&9&$GL(2)$&$SO(2)$&\cr

&8&$E_3\sim SL(3)\times SL(2)$&$U(2)$&\cr

&7&$E_4\sim SL(5)$&$USp(4)$&\cr

&6&$E_5\sim SO(5,5)$&$USp(4)\times USp(4)$&\cr

&5&$E_6$&$USp(8)$&\cr

&4&$E_7$&$SU(8)$&\cr

&3&$E_8$&$S0(16)$&\cr
\noalign{\hrule}}}$$

\medskip
{\bf 6.3.4 The Gauge Fields}
\medskip
Let us now examine how the gauge fields associated with the
rank-two tensor gauge field $A_{\mu\nu}$ can occur in the theory.
We define the field strength of
 $A_{\mu\nu}$ to be
$\Im_{\mu_1\mu_2\mu_3}=3\partial_{[\mu_1}A_{\mu_2\mu_3]}$.
This gauge field must be inert under the local $U(1)$ transformations
associated with the  coset ${SU(1,1)\over U(1)}$. If it were to have a
non-trivial $U(1)$  transformation then it could  only  transform
covariantly, but in this case   the
field strength would not transform covariantly. One could attempt to
avoid this latter conclusion by including the  $U(1)$ connection
$Q_\mu$ in the  definition of the field strength, however then
the corresponding  field strength would not be invariant under the
standard
$U(1)$  gauge transformations of the gauge field.
Under rigid
$g\in SU(1,1)$ it transforms as
$$(\bar\Im,\Im)\ \rightarrow\ (\bar\Im,\Im)g^{-1}
\eqno(6.3.24)$$
We can define a $SU(1,1)$ invariant field strength by
$$(\bar F,F) =(\bar\Im,\Im){\cal {V}}
\eqno(6.3.25)$$
Using equations (6.3.8) and (6.3.16) we find that under a local $U(1)$
transformation the new fields $(\bar F,F)$ have  weights
 $( -2,2)$
and  so  transform as
$$
(\bar F,F)\ \rightarrow\ (\bar F,F)h
\eqno(6.3.26)$$
Using the form of ${\cal{V}}$
 given in  equation (6.3.20) we can readily find
$(\bar\Im,\Im)$ in terms of $(\bar F,F)$.
\par
We note that the coset space description of the scalar fields
plays an important part in the formulation of the gauge field
which will occur in  the equations of motion. It will follow that  the
equations of motion will admit duality transformations. Since almost
all maximal   supergravity have gauge fields and coset space scalars
this is a general feature of supergravity theories. We refer the
reader to the lectures of M-K. Gaillard and B. Zumino in this volume for
a complete discussion of this topic.
\medskip
{\bf {6.3.5 The  Equations of Motion}}
\medskip
As we have discussed, the derivation of the equations of motion
was carried out using a variant of the Noether method (see section 4.1)
and  the on-shell superspace technique discussed in section 4.2.
These calculations are too involved to reproduce here;
however, many features of the equations can be deduced using
the features of the IIB theory discussed above. For example,
the equation of motion  must reduce to the correct linearised equations,
obey the requirements of dimensional analysis and contain terms  of
the  same $U(1)$ weight. In addition the gauge fields can only occur in
terms of their field strengths $F^{\mu_1\mu_2\mu_3}$ and $\bar
F^{\mu_1\mu_2\mu_3}$  as well as a five  rank field strength
$G_{\mu_1\dots\mu_5}$. These field strengths all have geometric dimension
one. The vielbein must occur through the  usual Ricci tensor $R_{\mu\nu}$
which has geometric dimension two and  as we discussed above  the scalars
belong to the coset $SU(1,1)/U(1)$ and hence are only contained in the
geometric dimension one objects
$P_\mu$ and $Q_\mu$. The latter can only occur as part of a
covariant derivative.
\par
Let us begin with the scalars;  their equation of motion must
generalise equation (6.3.23) and hence the terms in the equation must
have $U()$ weight 4 and geometric dimension two. The only possible
candidate is
$F_{\mu_1\mu_2\mu_3}F^{\mu_1\mu_2\mu_3}$.
The equation for
$A_{\mu\nu}$ must contain $D^{\mu_3}F_{\mu_1\mu_2\mu_3}$  and so it
has geometric dimension two and $U(1)$ weight 2. The only such terms
we can add are  $F_{\mu_1\mu_2\mu_3}P^{\mu_3}$ and
$G_{\mu_1\dots\mu_5}F^{\mu_3\dots\mu_5}$. The
equation of motion for the fourth rank gauge field is just the
self-duality condition on the five-rank field strength and so
this equation has geometric dimension one and $U(1)$ weight zero.
There are no terms one can add to this equation other than the duality
condition on he field strength itself. Analysing the  veilbein equation
in the same way we  essentially determine the  field equations up to
constants.
\par
The equations of motion of IIB supergravity in the absence of fermions is
given by [111],[112]
$$D^\mu P_\mu={1\over 6}F_{\mu_1\mu_2\mu_3}F^{\mu_1\mu_2\mu_3}
\eqno(6.3.27)$$
$$
D^{\mu_3}F_{\mu_1\mu_2\mu_3}=\bar
F_{\mu_1\mu_2\mu_3}P^{\mu_3}-{i\over6}
G_{\mu_1\dots\mu_5}F^{\mu_3\dots\mu_5}
\eqno(6.3.28)$$
$$R_{\mu\nu}=-2\bar
P_{(\mu}P_{\nu)}-F_{(\mu}^{\ \ \nu_1\nu_2}F_{\nu)\nu_1\nu_2}
+{1\over12}g_{\mu\nu}\bar
F_{\mu_1\mu_2\mu_3}F^{\mu_1\mu_2\mu_3}
-{1\over96}G_\mu^{\ \mu_1\dots\mu_4}G_{\nu\mu_1\dots\mu_4}
\eqno(6.3.29)$$
$$
G_{\mu_1\dots\mu_5}={}^*G_{\mu_1\dots\mu_5}
\eqno(6.3.30)$$
where
$$G_{\mu_1\dots\mu_5}=5\partial_{[\mu_1}A_{\mu_2\dots\mu_5]}
+20i(A_{[\mu_1\mu_2} \Im^*_{\mu_3\dots\mu_5]}-
A_{[\mu_1\mu_2}^* \Im _{\mu_3\dots\mu_5]})
\eqno(6.3.31)$$
\par
The reader is referred to reference [111] for the fermionic
contribution.  They are invariant [111,112] under the local
supersymmetry  and $U(1)$ transformations of reference [110].
\medskip
{\bf {6.3.6 The $SL(2,{\bf R})$ Version}}
\medskip
The group $SU(1,1)$ is isomorphic to the group $SL(2,{\bf R})$.
 For some purposes it is better to formulate the theory
in a manner where the $SL(2,{\bf R})$ form of the invariance is
manifest  rather than as above
where the $SU(1,1)$ symmetry is apparent. As we explained above,
 $g\ \in\ SU(1,1)$  acts on the column vector
$\left(\matrix{z_1\cr z_2\cr}\right) $ by $\left(\matrix{z_1\cr
z_2\cr}\right) \to g\left(\matrix{z_1\cr z_2\cr}\right) $. If we
denote the ratios of the column vector by $z={z_1\over z_2}$ then
 the action of  $SU(1,1)$ becomes
$$z\to {uz+v\over v^*z+ u^*}
\eqno(6.3.32)$$
This action is such that it takes the unit disc $|z|\le 1$ to itself.
We can  map the unit desk to the upper
half plane $H=\{ w: Im w\ge 0\}$
by the transformation
$$z \ \rightarrow\
w=i\left({1-z\over1+z}\right)
\eqno(6.3.33)$$
The action induced by the transformation  of equation (6.3.32)
on $H$
is given by
$$w\ \rightarrow\ {aw+b\over cw+d}
\eqno(6.3.34)$$
where $ ad-bc=1$ and  $a,b,c,d$ are real. In this last
transformation we recognise the action of the group $SL(2,{\bf R})$,
corresponding to the element
$$\hat g=\left(\matrix{a&b\cr c&d\cr
}\right)\in SL(2,{\bf R}).
\eqno(6.3.35)$$
It is  well known that $SL(2,{\bf R})$ is the largest
group which maps
the upper half plane to itself and so we should not be surprised
that  in  mapping from the unit disc onto the upper half plane
the action of
$SU(1,1)$ becomes that of  $SL(2,{\bf R})$.
The precise
relationship between the parameters of the two groups is given by
$$
a= {1\over 2}(u+u^*-v-v^*),\ b={i\over 2}(-u+u^*-v+v^*),$$
$$
c=\ -{i\over 2}(-u+u^*+v-v^*),\ d=\ {1\over 2}(u+u^*+v+v^*)
\eqno(6.3.36)$$
\par
For the $SU(1,1)$ formulation of the IIB theory given above we found
that  the scalar field $\phi$ of equation (6.3.20) transformed under
$SU(1,1)$  in  the same way as the variable $z$ of equation (6.3.21).
 Hence  if we make the transformation from
$\phi$ to the variable
$\varphi$ by
$$\phi\ \to \varphi=i\left({1-\phi\over1+\phi}\right)
\eqno(6.3.37)$$
then $\varphi$ transforms under $SL(2,{\bf R})$ just like $w$, that is
$$\varphi\to {a\varphi+b\over c\varphi+d}
\eqno(6.3.38)$$
\par
It remains to find new variables for the rank-two gauge field that
transform in a recognisable way under $SL(2,{\bf R})$.
Let us write the field strength $\Im_{\mu\nu\rho}$ as
$\Im_{\mu\nu\rho}=\Im_{\mu\nu\rho}^2+i\Im_{\mu\nu\rho}^1 $
then an
explicit calculation shows that the transformation law of equation
(6.3.24) for $\Im_{\mu\nu\rho}$ becomes
$$\left(\matrix{\Im_{\mu\nu\rho}^1\cr \Im_{\mu\nu\rho}^2\cr}\right)\
\rightarrow\ \hat g\left(\matrix{\Im_{\mu\nu\rho}^1\cr
\Im_{\mu\nu\rho}^2\cr}\right)
\eqno(6.3.39)$$
It is straightforward to substitute for the new variables
into the equations
of motion  (6.3.27) to (6.3.31)
to find a formulation that
is manifestly $SL(2,{\bf R})$ invariant. Carrying out this
transformation  and  also making the substitution
$\varphi=l+ie^{\sigma}$,
one finds that the $NS\otimes NS$ fields,
(i.e. $e^{\ a}_\mu,\ A^1_{\mu\nu}$ and $\sigma$) have identical
equations  of motion as the
$NS\otimes NS$ sector of the IIA supergravity. In fact, since these
fields do not include the rank four gauge field  we can formulate the
dynamics of the $NS\otimes NS$ sector of the theory in terms of an
action and this action  will have a Lagrangian which is  none other
than the first term of equation (6.2.16),  if we choose to work with
the string metric.
\par
We close with some comments on some of the features of
the IIB theory discussed that are most relevant to  our
discussion on string duality.
 Like the IIA theory,  the IIB theory has two coupling
constants;  the Newtonian constant $\kappa$, whose dependence we have
suppressed,  and the expectation value of $<e^\sigma>$. As we
have explained in the previous section the latter
plays the role  of the IIB string coupling constant, i.e.
$g_s=<e^\sigma> $. In general, an $SL(2,{\bf R})$ transformation
changes  from weak to strong string coupling. For example, the
transformation
 $\varphi=l+ie^\sigma\ \rightarrow\ \varphi'=
l'+ie^{\sigma'}=-{1\over \varphi}$ implies
that
$$ g^\prime_s =\langle e^{\sigma'}\rangle
={1\over <e^\sigma>}={1\over g_s}
\eqno(6.3.40)$$
maps from the weak to the strong regime.
\par
The $SL(2,{\bf R})$ transformation also mixes the two real  rank-three
field strengths $\Im_{\mu\nu\rho}^i,\ i=1,2 $ one of which arises from
the
$NS\otimes NS$ sector and one from the
$R\otimes R$ sector of the string theory. Hence a
generic  $SL(2,{\bf R})$
transformation   transforms fields in  the $NS\otimes NS$
sector into those in
 the $R\otimes R$ sector and vice versa.
\par
The IIB theory cannot be obtained from the eleven-dimensional
supergravity by a reduction; however if we were to reduce the IIB
theory on a circle to nine dimensions then the resulting
supergravity theory would have an underlying supersymmetry algebra
with 32 supercharges which, being  an
odd dimension, would be of   no fixed chirality.
This is precisely the
same supersymmetry algebra that would result if we were to reduce the
IIA theory on a circle to nine dimensions. In particular, they
form two 16 component Majorana spinors. Since this algebra uniquely
determines the maximal nine-dimensional  supergravity theory we must
conclude that reducing the IIA and IIB  supergravity theories to nine
dimensions leads to the same supergravity theory. [107].

\medskip
{\bf 6.4 Type I Supergravity}
\medskip
The type I Supergravity was in fact the first supergravity
theory in ten
dimensions to be constructed [120]. Its underlying algebra
contains a Majorana-Weyl spinor supercharge whose anti-commutator
was given in equation (1.5.4). The field content is given by
$$\underbrace{e_\mu^a,\phi}_{NS\otimes
NS};\quad\underbrace{A_{\mu \nu}}_{R\otimes R};
\quad{\rm
plus}\quad\underbrace{\psi_{\mu\alpha},\lambda_\alpha}_{NS\otimes R}
\eqno(6.4.1)$$
where we have also indicated the sectors in the type I string from
which they come. The gravitino $\psi_{\mu \alpha}$ and the spinor
$\lambda_\alpha$ are Majorana-Weyl spinors of opposite chirality
and all the bosonic fields are real.
\par
This theory can be obtained form the either the IIA or the IIB
theory by truncation. From the IIA theory we impose the obvious
Weyl conditions on the gravitino and spinor in the IIA theory,
and to be consistent with supersymmetry, we must also set
$A_{\mu\nu\rho}=0=B_\mu$. It is straightforward to find the
Lagrangian for the bosonic fields by truncating the corresponding
action for the IIA theory of
equation (6.2.8).
\par
We can also obtain the type I theory from the IIB theory by a
truncation. In the IIB theory we consider the operator
$\Omega$ which  changes the sign of $\Im_{\mu\nu\rho}^2$, $l$ and
$B_{\mu\nu\rho\sigma}$, which are  in the
$NS\otimes NS$, $R\otimes R$ and
$R\otimes R$ sectors respectively, but leaves inert $e_\mu^a$,
$\Im_{\mu\nu\rho}^1$ and
$\sigma$ which are in  $NS\otimes NS$, $R\otimes R$ and $NS\otimes
NS$ sectors respectively. To recover the type I
supergravity  we keep only fields which are left inert by
$\Omega$ and so we find only the latter fields.
 On the
spinors we impose a Majorana condition which leads to a
gravitini and another fermion which are both Majorana-Weyl although of
opposite chirality. In fact $\Omega$ corresponds in string theory
to world sheet parity, that is it exchanges left and right moving
modes.
\par
The $N=1$ Yang-Mills theory is also based on the supersymmetry
algebra with one Majorana-Weyl supercharge. It consists
of a gauge field $A_\mu$ and one Majorana-Weyl spinor
$\lambda_\alpha$. This theory [119]  is easily derived. We first
write down the linearised transformation  laws that are determined
up to two constants by dimensional analysis.    The
constants are then fixed by demanding that the supersymmetry
transformations and the linearised gauge transformations form a closed
algebra.  The full theory is uniquely found by  demanding that the
algebra  closes and that the action be invariant under the usual
non-Abelian gauge transformation of the gauge field. The result is
given by
$$\int d^{10}x(-{1\over4}F_{\mu\nu}^iF^{\mu\nu i}
-{i\over2}\bar \lambda^i \gamma^\mu\mu (D_\mu\lambda)^i)
\eqno(6.4.2)$$
which is invariant under
$$
\delta A_\mu^i= i\bar \epsilon \gamma_\mu \lambda^i,\
\delta \lambda^i = -{1\over2}F_{\mu\nu}^i\gamma^{\mu\nu}\epsilon
\eqno(6.4.3)$$
where
$$F_{\mu\nu}^i=\partial _\mu A^i_\nu-
\partial _\nu A^i_\mu -gf_{jk}^{\ \ i}A^j_\mu A^k_\nu
\eqno(6.4.4)$$
is the Yang-Mills field strength
and
$$D_\mu\lambda^i=\partial _\mu \lambda^i-
gf_{jk}^{\ \ i}A^j_\mu\lambda^k
\eqno(6.4.5)$$
\par
We can also consider the coupling between the $N=1$ Yang-Mills theory
and the type I supergravity. This was found in reference
[121]. If we take the gauge group to be $SO(32)$ or $E_8\otimes E_8$
we find the theory that results from the low energy limit of
the corresponding heterotic string theory or the type I string theory.

\bigskip

\centerline{\bf { 7. Brane Dynamics}}
\bigskip
{\bf 7.1 Bosonic Branes}
\medskip
Super p-branes are extended objects that sweep out a
$p+1$-dimensional  space-time manifold  in a background
superspace-time.  A 0-brane is just a
 particle and a 1-brane is a string. However p-branes for
$p\ge2$ also occur in string theory as solitons and are
thus non-perturbative objects.  In this section
we find what   possible
superbranes
 can exist  and give their dynamics. Although p-branes with
$p\ge2$ are intrinsically
non-perturbative objects,  they are related by duality
symmetries  to perturbative particle and string states. As such,
they play an important role in discussions of string duality.
\par
 We first consider a brane that has no supersymmetry.
A $p$-brane
sweeps out a $p+1$-dimensional world surface  $M$, with coordinates
$\xi^m,\ m= 0,1,\dots,p$ in a $D$-dimensional target space
$\underline M$ with coordinates $X^{\underline n},\ \underline n=
0,1,\dots,D-1$.  As the symbols imply we use $m,n,p, \ldots $
for the embedded  surface  world indices and $\underline m,\underline
n,\underline p, \ldots $
for target  space world indices. The corresponding tangent space
indices are $a,b,c, \ldots $
for world surface  indices and $\underline a,\underline b,
\underline c, \ldots $ for target    space world indices. This
notation is used extensively in the literature in this subject. The
reader should  have no difficulty making the transition from the
$\mu,\nu\ldots$ and $m,n,p\ldots $ used in the
previous section for the
world and tangent target space indices respectively.  The surface
$M$ swept out by the $p$-brane  in the target space
$\underline M$  is specified by the functions
$X^{\underline n}(\xi^n)$ which extremise the action
$$-T \int d^{p+1}\xi \sqrt{-det g_{mn}}
\eqno(7.1.1)$$
where
$$g_{mn}=\partial _nX^{\underline n}\partial _m X^{\underline m}
g_{\underline n \underline m }
\eqno(7.1.2)$$
and $g_{\underline m \underline m }$ is the metric of the target
space-time often referred to as the background metric.
The constant $T$ is the brane tension and has the dimensions
of $(mass)^{p+1}$. The
action  of equation (7.1.1) is invariant under reparameterisations of
both the target space $\underline {M} $ and the world surface
${M}$. The
metric
$g_{mn}$ is the metric induced on the world surface $M$ by the
background metric of the target space. As such,
 we recognize the action in equation (7.1.1) as the area swept out
by the p-brane.  Hence, like the string and point particle, a p-brane
moves so as to extremise the volume of the surface it sweeps out.  If
the target  space is flat the background metric is just the Minkowski
metric
$g_{\underline m \underline m }= \eta_{\underline m \underline m }$.
A 0-brane is just a point particle and if it has mass $m$ then
$T=m$. A 1-brane is just the usual bosonic string and the action of
equation  (7.1.1) is the Nambu action for the string if we take the
background metric to be flat. In this case we often write
$T={1\over 2\pi\alpha'}$ where $\alpha'$ is the string Regge slope
parameter.
\par
The bosonic brane does not have enough symmetry to  determine
its couplings to the fields in the target space. However, a
$p$-brane naturally couples to a
$p+1$ gauge field $A_{\underline
m_1\dots\underline m_{p+1}}$ of  the target space by a term
 of the form
$$\int d^{p+1} \xi\epsilon^{n_1\dots n_{p+1}}\partial_{n_1}
X^{\underline m_1}\dots
\partial_{n_{p+1}}X^{\underline m_{p+1}}A_{\underline
m_1\dots\underline m_{p+1}}
\eqno(7.1.3)$$
So for example, the motion of  a 0-brane, that is a point particle, is
described  by the  functions
$X^{\underline n}(\tau)$, where $\xi^0=\tau$, and it naturally
couples to a vector
field $A_{\underline n}$ in  the form
$$\int d\tau {dX^{\underline n}\over d\tau} A_{\underline n}
\eqno(7.1.4)$$
If we couple this expression to that in equation (7.1.1) for a flat
target space then the equations of motion for $X^{\underline n}$
are nothing but the Lorentz force law for a charged particle in an
electromagnetic field.  A 1-brane, i.e. string, couples to a two-form
$A_{\underline n
\underline m}$ in
the manner
$$\int d^2\xi\epsilon^{mn}\partial_m
X^{\underline m_1}\partial_n X^{
\underline m_2}A_{\underline m_1\underline m_2}
\eqno(7.1.5)$$
\par
We can split the  target space indices $\underline n,
\underline m$,  into  those associated with the directions
longitudinal to the  brane and those  with
directions which are  transverse to the brane. We
denote the  former by $n,m\ldots =0,\ldots ,p$, and the latter by
$n',m'\ldots=p+1,
\ldots ,D-1$.  A useful gauge  is the static gauge in
which we use the reparameterisation transformations of the world
surface to  identify  the $p+1$ longitudinal
coordinates $X^{n}(\xi), \ { n} =0,1,\dots ,p$,  with the
coordinates
$\xi^n, \  n=0,1,\dots ,p$ of the
$p$ brane; in other words
$$X^{ n}(\xi)=\xi^n, \
 n=0,1,\dots ,p
\eqno(7.1.6)$$
This leaves the transverse coordinates $X^{ n'}(\xi), \
{ n'}=p+1,\dots ,D-1$, to describe the dynamics of the brane.
We can think of the $D-p-1$ transverse coordinates as the Goldstone
bosons or zero modes of the broken translations due to the presence
of the $p$-brane.
\par
As we have explained in section 6, the low energy effective action of
a  string theory is a supergravity theory. It has been found that
$p$-brane solutions arise from this supergravity theory. For such a
static solution the supergravity  fields do not depend on  $p$ of the
spatial coordinates which are the spatial coordinates of the $p$-brane
world surface, but do depend on the $D-p-1$ coordinates
trnasverse to the brane.
It turns out that the supergravity fields usually depend on functions
which are harmonic functions of the transverse coordinates.
\par
It will be instructive to consider a simpler and better
understood example of  solitons, namely the
 monopoles (i.e. a  0-brane) that occur in four
dimensions. Monopoles  arise as static solutions in the  $N=2$
supersymmetric Yang-Mills field theory. The space of parameters
required to specify the solution is called the moduli space. If there
are $K$ monopoles then the moduli space has dimension
$4K$. For one monopole  this four-dimensional space is ${\bf
R}^3\otimes S^1$. It is made up of the three spatial coordinates ${\bf
R}^3$ which  specify the position of the monopole and a further
parameter associated with gauge transformations which have a
non-trivial behaviour at infinity. For
$K\ge 2$ the moduli space is more complicated, but an explicit  metric
is known for the case of $K=2$ [128].  Since there exist static
monopoles   solutions, it follows that monopoles do not
experience any forces when at rest.  However,  when they are set in
motion they do experience velocity dependent forces.  In general the
behaviour of the monopoles is described by the quantum field theory in
which they arise.  However, if the monopoles have only a small amount
of energy above their rest masses then we can approximate their
motion   in terms of the coordinates of their
moduli space [129].  The motion is then described by an action whose
fields are the coordinates of the moduli space  which are given
a  time dependence corresponding to the monopole motion.
 If we had only one monopole then three of
the moduli
$X^i,\ i=1,2,3$, could
  be interpreted as the  Goldstone bosons resulting from
 the breaking
of the  spatial translations due to the presence of the monopole
and the fourth moduli $\eta$ would be  related to the existence of
non-trivial gauge transformations at infinity. As such,
 the motion is described by
$X^i(\tau),\ i=1,2,3,\ \eta(\tau)$ where $\xi^0=\tau$.  A
much more detailed account of these ideas can be found in the
lectures of N. Manton in this volume.
\par
In a similar spirit we can consider the moduli space and action which
describes the low energy behaviour of
 p-brane solitons. The moduli space of the $p$-brane solitons
contains  the positions of the $p$-branes and to describe their low
energy motion we let the moduli   depend on the world-volume
coordinates of the
$p$-brane. Indeed,
we can think of  the action of equation (7.1.1) when we take only its
terms to lowest order in derivatives as the effective action  which
describes the low energy behaviour of  a  single p-brane soliton.
 \par
There is an alternative interpretation of the action of equation
(7.1.1). Were we believe that a certain p-brane was a fundamental
object then we might take the action of equation (7.1.1) to describe
its dynamics completely.  This was precisely the viewpoint of
for the string for many years.
 \medskip
{\bf 7.2  Types of Superbranes}
\medskip
A super $p$-brane can be viewed as a $p+1$-dimensional {\bf bosonic}
sub-manifold
$M$, with coordinates $\xi^n ,\ n=0,1,\dots ,p$, that moves  through
a target superspace
$\underline M$  with coordinates
$$Z^{\underline N}= (X^{\underline n},
\Theta ^{\underline \alpha} )
\eqno(7.2.1)$$
We use the superspace index convention that $\underline N,
\underline M, \ldots $ and $\underline A,\underline B,\ldots $
 represent the world and tangent space  indices of the
target space.
Later, in subsection 7.5, we use the same symbols without the
underlining to represent the corresponding indices of the
world surface superspace.
\par
In this section we wish to find which types of branes can exist in
which space-time dimensions with particular emphasis on ten and
eleven dimensions. We will also discuss the features of brane
dynamics which are generic to all branes, leaving to the next three
subsections a more detailed discussion  of the dynamics of the
specific types of branes.
\par
Superbranes come in various types. The
simplest  are those  whose dynamics  can be described
entirely by specifying the superworld surface
$Z^{\underline N}(\xi^n)$ that the
$p$-brane sweeps out in the target space. We refer to these branes
as {\bf simple superbranes}; they were also previously called type I
branes' not to be confused with type I strings. As we shall see there
are  other types of branes that   have   higher spin fields living on
their world surfaces. In particular, we will discuss  branes that have
vectors and second rank anti-symmetric tensor gauge   fields on their
world surface.
\par
A super brane has an  action of the form
$$A=A_1+A_2
\eqno(7.2.2)$$
The first term is given by
$$A_1= -T \int d^{p+1}\xi \sqrt{-det g_{mn}}+\dots
\eqno(7.2.3)$$
where
$$g_{mn}=\partial _nZ^{\underline N}\partial _m Z^{\underline M}
g_{\underline N \underline M },
\eqno(7.2.4)$$
$$ g_{\underline N \underline M }= E_{\underline N}^{\ \ \underline a}
E_{\underline M}^{\ \ \underline b}\eta _{\underline a \underline
b}
\eqno(7.2.5)$$
and $E_{\underline N}^{\ \ \underline a}$ is the supervielbein
 on the
target superspace. By abuse of notation we use the same
symbol for the world surface metric as for the bosonic case; the
reader will be able to distinguish between the two as a result of
the context.  The $+\dots $ denotes terms involving possible world
surface  fields as well as those that depend on the other background
fields.  The constant $T$ is the p-brane tension and has the
dimensions of $(mass)^{p+1}$ since the action is dimensionless and
$X^{\underline n}$ has the dimension of $(mass)^{-1}$. If we consider
a brane that is static we can think of the tension $T$ as the mass
per unit spatial volume of the brane.
\par
The symbol $g_{\underline N \underline M }$ is not really
a background metric in the usual sense since the sum on the
tangent space indices is restricted to be only over the bosonic
part. Such a restricted summation is possible as a consequence of
the fact that the superspace tangent space group is just the Lorentz
group. In fact,  as we discussed in section 4, the theory of local
superspace is formulated in terms of the supervielbein since the
metric is not uniquely defined.
\par
The second part of the action  $A_2$ of equation (7.2.2) contains,
in addition to others,
the term
$$\int d^{p+1}
\xi\epsilon^{ n_1\dots n_{p+1}}\partial_{n_1}
X^{\underline m_1}\dots
\partial_{n_{p+1}}X^{\underline m_{p+1}}A_{\underline
m_1\dots\underline m_{p+1}}
\eqno(7.2.6)$$
where $A_{\underline m_1\dots\underline m_{p+1}}$ is a background
gauge field.
\par
The background space-time fields can only belong to
 a supermultiplet that exists in the target superspace. In this
review, we will take these supermultiplets to be  the supergravity
theory that
has the background supersymmetry algebra possessed by  the
$p$-brane. The field content of the possible  supergravity theories
can be deduced from the supersymmetry algebra  using the methods given
in section three. However, as we discussed  in section five, the
supergravity theory is essentially unique if the   supersymmetry
algebra has 32 supercharges and the type of spinors contained in the
supersymmetry algebra are specified. Thus, unlike   bosonic branes,
the background fields are  specified
 by the background supersymmetry of the brane, if that supersymmetry
has 32 supercharges. If the brane has 16 supercharges then
although one cannot generally specify the background fields
uniquely,  the possible supergravity theories are very limited. The
general coupling of a
$p$-brane to the background fields is complicated, but it
 always  has a coupling to a $(p+1)$-gauge field in the form of
equation (7.1.3). However, since this $(p+1)$-gauge field must be one
of the  background fields of the supergravity theory, this places an
important restriction on which superbranes can arise for a given
dimension and target space supersymmetry algebra.
\par
If we consider the case of a super 1-brane, the action of
equation (7.2.2) is just the Green-Schwarz action [130] for the
superstring. The case  $p=2$ is often referred to as the
membrane and the  action for the supermembrane  in eleven
dimensions was found in [151,157].
\par
The action of equation (7.2.2) is invariant under super
reparameterisations of
 the target superspace  $\underline {M} $, but
 only  bosonic
reparameterisations of the world surface ${M}$ since the
embedded manifold ${M}$ is a bosonic manifold. It is also
invariant under a  Fermi-Bose symmetry called $\kappa$-supersymmetry.
From the viewpoint  adopted  here this is a complicated
symmetry that ensures that the fermions have the correct number of
degrees of freedom on-shell and we will discuss it in more
detail in the next section. This symmetry relates the terms in
$A_1$ to those in $A_2$ and vice versa, and in fact fixes uniquely
the form of $A_2$ given the form of $A_1$.
\par
Even if the target space is flat superspace the
supervielbein has a non-trivial dependence on the coordinates
 given by
$$\partial _mZ^{\underline N}
E_{\underline N}^{\ \ \underline a}
=\partial _mX^{\underline a}-{i\over 2}
\bar \Theta \gamma^{\underline a}\partial _m \Theta,\
 \partial _mZ^{\underline N}E_{\underline N}^{\ \ \underline \alpha }
= \partial_m \Theta ^{\underline \alpha }
\eqno(7.2.7)$$
In this case, the target space super reparameterisation invariance
reduces to just  rigid supersymmetry'
$$\delta X^{\underline a}={i\over 2}\bar \epsilon \gamma^{\underline a}
\Theta ,
\ \delta \theta ^{\underline \alpha}= \epsilon ^{\underline \alpha}
\eqno(7.2.8)$$
\par
Note that the action of equation (7.2.2)
does not appear to possess world-surface supersymmetry. We recall
for the case of a 1-brane (that is a string) there are two
formulations, the Green-Schwarz formulation given here
and the original Neveu-Schwarz-Ramond  formulation. The latter
is formulated in terms of $X^{\underline n}$ and a spinor which,
 in contrast to
 above, possess  a target space vector index and is a spinor with
respect to  the two-dimensional world sheet.  This formulation is
manifestly invariant under world surface reparameterisations, but not
under target space super reparameterisations. If one goes to
light-cone gauge (i.e. in effect static gauge) then the  two
formulations become the same and so the Green-Schwarz  formulation has
a hidden world sheet supersymmetry and  the Neveu-Schwarz-Ramond [172]
a hidden target space supersymmetry provided we carry out the GSO
projection.
\par
It is thought [173] that all branes which in their
Green-Schwarz formulation admit $\kappa$-supersymmetry actually
have a hidden world surface supersymmetry. Although
the analogue of a  Neveu-Schwarz-Ramond  formulation is not known for
p-branes when $p>1$, there does exist a superembedding formalism
[131,132,133,134,137,138,139]. In this formulation the p-brane
sweeps out a {\bf supermanifold}  which is embedded in the  target
superspace.   Although this approach   leads to equations of motions
and not an action, it has the  advantage that it possesses
super reparmeterisation invariance in both  the world
surface  and the target superspaces. The $\kappa$-symmetry
is then just part of the super reparameterisations of the
world surface. Its particular form is a result of the gauge fixing
required to get from the super embedding formalism to the
so-called Green-Schwarz formulation.  This origin  of
$\kappa$-symmetry  was first found in reference [178] within the
context of the point particle. We will comment further on the
superembedding  formalism  in section 7.5.
\par
The superbranes also
possesses a static gauge for which the bosonic coordinates take the
form of equation (7.1.6) while
$\kappa$-supersymmetry can be used to set half of the fermions
$\Theta^{\underline \alpha}=(\Theta ^\alpha ,\ \Theta ^{\alpha
^\prime})$ to vanish i.e. $\Theta ^{\alpha}=0$. While the
remaining $D-p-1$ bosonic coordinates
$X^{n'}$ correspond to the Goldstone bosons associated with the
breaking of translations by the
$p$-brane, the remaining $\Theta^{\alpha ^\prime } $ Goldstone fermions
correspond to the  breaking of half of the supersymmetries by the
brane.
\par
The fields of the $p$-brane belong to a supermultiplet of the
world surface and so  must have
equal numbers of fermionic and bosonic degrees of freedom on-shell.
Let us first consider a
$p$-brane that arises in a theory that has  maximal
supersymmetry. This  would be the case if the brane describes the
low-energy motion of a soliton
  of a maximal supergravity theory which breaks half the
supersymmetry. In this case, if it breaks half of the 32
supersymmetries of the target space, it will,  in static gauge, have
only 16
$\Theta^{\alpha^\prime}$ which will lead
to 8 fermionic degrees of freedom on-shell. If we are dealing with a
simple superbrane these must be matched by the coordinates
$X^{n^\prime}$ in static
gauge which must therefore be eight in number. Thus if we are in
eleven dimensions the only simple superbrane is a 2-brane while if
we are in ten dimensions the only simple superbrane is a 1-brane.
\par
There also exist D-branes whose
 dynamics requires  a vector field
$A_n,\
n=0,1,\dots,p$,  living on the brane in addition to the coordinates
 $X^{\underline n},\ \Theta ^{\underline \alpha }$ which describe the
embedding of the brane in the target superspace. In this case,
 if we
have a  brane that arises in  the background of a maximal supergravity
theory and which breaks half of the this supersymmetry then we again
have 8  fermionic degrees of freedom on-shell.
This must be balanced by
the  vector which has
$p-1$ degrees of freedom on-shell and the  remaining $D-p-1$ transverse
coordinates from which we deduce that
$D=10$. Hence such $D$-branes can only exist in ten dimensions.
For D-branes this simple counting argument allows branes for
all $p$, however, as we shall see, not all values of $p$ occur for a
given target space   supersymmetry. In addition,  we will discuss
 other branes with  higher rank gauge fields living on
their world surface.
\par
To find further restrictions on
 which branes actually exist we can use the argument given above.
A super p-brane  couples to a $p+1$-gauge field which must
belong to the background supergravity theory. Hence to see which
branes can exist  we need only  see which gauge fields are present in
the corresponding supergravity  theory.
When doing this we must bear in
mind that if we have a  rank $p+1$
gauge field which  has  a  rank
$p+2$-field strength $F_{(p+2)}$ we can
take its dual to produce a rank $D-p-2$ field strength
${}^* F_{(D-p-2)}$
 which, if the original field strength is on-shell,  has  a
corresponding rank-$D-p-3$ dual gauge field ${}^* A_{D-p-3}$. The dual
field strength is defined by
$$
{}^* F_{\underline n_1\ldots \underline n_{(D-p-2)}}= {1\over (p+2)!}
\epsilon _{\underline n_1\ldots \underline n_{(D-p-2)}
\underline n_{(D-p-1)}\ldots \underline n_D}
F^{\underline n_{(D-p-1)}\ldots \underline n_D}
\eqno(7.2.9)$$
Hence, if
the original rank $p+1$ gauge field couples to a
$p$-brane, the dual gauge field is  rank $D-p-3$  and  couples to a
$D-p-4$ brane.  In fact $D$-branes can only couple to the $R\otimes R$
sector of a string theory [135].
\par
Let us begin with branes in eleven dimensions.
This theory was described in section 5
and possesses a third-rank gauge
field
$A_{(3)}$ which should couple to a 2-brane. The  dual potential
has rank six i.e. ${}^* A_6$ and this couples  to a 5-brane. Thus in
eleven dimensions we expect only a 2-brane  and a 5-brane. The
2-brane is just the simple superbrane we discussed above. The 5-brane
has 16 Goldstone fermions and these lead to the same 8 degrees of
freedom on-shell; however it has only five transverse coordinates
leading to only five degrees of freedom on-shell. Clearly, we require
another three degrees of freedom on-shell. These must form a
representation of the little group $SO(4)$. If they are  belong to an
irreducible representation of $SO(4)$, this can only be a second rank
self-dual antisymmetric tensor. On-shell this corresponds to a second
rank antisymmetric gauge field whose field strength obeys a self
duality condition.  The dynamics of the fivebrane is significantly
more complicated than that of simple branes and will be given in
 section 5.4.
\par
Let us now consider the branes that can couple to the IIA string.
In section six we found that the IIA theory has gauge fields and
their duals of ranks $1,2,3,5,6,7$. Of these ranks 2 and 6 arise
in the $NS\otimes NS$ sector of  the string while those of ranks
$1,3,5$  and 7 arise in the
$R\otimes R$ sector of the string. This suggests the existence of
  $p$-branes for $p=1$ and 5 that can couple to the $NS\otimes NS$
sector of the IIA string and $p$-branes for
$p=0,2,4$ and 6 that can couple to the $R\otimes R$ sector of the IIA
string. The latter  are the
$D$-branes. Of these
  only the 1-brane can be a simple brane and it is this brane that
couples to the $A_{\mu\nu}$ gauge field of the IIA string. It is in
fact the IIA string itself and it is sometimes therefore called the
fundamental string.  The fivebrane is the straightforward dimensional
reduction of the five eleven-dimensional fivebrane that will be
discussed  in the next section.  In the literature  an 8-brane is also
discussed. This brane  is associated with the  massive supergravity
theory constructed in reference [174]. This theory  necessarily
contains a cosmological constant $c$ and so  has a term $\int
d^{10}x\  c\sqrt { -det g_{\underline n\underline m}}$ in the action.
However, we can write this term as
$\int d^{10}x\  c
\sqrt { -det g_{\underline n\underline
m}}F ^{\underline n_1\dots \underline n_{(10)}}
F_{\underline n_1\dots \underline n_{(10)}}$ where
$F_{\underline n_1\dots \underline n_{(10)}}$ is the curl of a
rank $9$ gauge field which  suggests the existence of an 8-brane
[185].
\par
Let us now turn to the IIB theory and its branes. The original gauge
fields in the IIB theory are of rank $2$ in the $NS\otimes NS$
sector and ranks $0,2,4$ in the $R\otimes R$ sector. If we include
their  dual gauge fields we have gauge fields of  ranks  $2,6$ in the
$NS\otimes NS$ sector and ranks $0,2,4,6,8$ in the $R\otimes R$
sector. We do not include two rank four gauge fields as their field
strength is the five rank self-dual field strength of the IIB theory.
This suggests that there exist
$1$ and
$5$-branes which couple to the
$NS\otimes NS$ of the IIB theory and $-1,1,3,5,7$-branes
which couple to the $R\otimes R$ sector. The latter are the
$D$-branes.  The 1-brane which couples to the rank two gauge field in
the
$NS\otimes NS$ sector  is the IIB string itself. The 5-brane
which couples to the $NS\otimes NS$ sector is
a more complicated object. Note that it is the threebrane that
couples to the rank four gauge field whose field strength satisfies a
self-duality property. As one might expect this $D$-brane possesses
a self-duality symmetry [136]. The $p=-1$ brane which couples to the
$R\otimes R$ sector occupies just a point in space-time and so is an
instanton.
\par
In fact, all the branes discussed above exist and the possible
branes in eleven and ten dimensions are listed in table 7.1. In this
table a "D" or "S" subsrcipt denotes the brane to be a
Dirichlet and simple brane respectively.

\centerline{\bf 7.1 Super Brane Scan}

$$\vbox{\tabskip=0pt \offinterlineskip

\halign to416pt{\strut#& \vrule# \tabskip=1em& \hfil#& \vrule#&

\hfil# \hfil& \vrule#&\hfil# \hfil& \vrule#&\hfil# \hfil& \vrule#&

\hfil# \hfil& \vrule#&\hfil# \hfil& \vrule#&\hfil# \hfil& \vrule#&\hfil#
\hfil& \vrule#&

\hfil# \hfil& \vrule#&\hfil# \hfil& \vrule# \tabskip=0pt
\cr \noalign{\hrule}

&&\qquad $ \backslash$ p&&&&&&&&&&&&&&&&&&&\cr

&&  Theory $\backslash$  \ &&&&&&&&&&&&&&&&&&&\cr
\noalign{\hrule}

&&M-theory&&&&&&$2_S$&&&&&&5&&&&&&&\cr

&&&&&&&&&&&&&&&&&&&&&\cr \noalign{\hrule}

&&IIA&&$0_D$&&$1_S$&&$2_D$&&&&$4_D$&&5&&$6_D$&&&&$8_D$&\cr

&&&&&&&&&&&&&&&&&&&&&\cr \noalign{\hrule}

&&IIB&&&&$1_S+1_D$&&&&$3_D$&&&&$5+5_D$&&&&$7_D$&&&\cr

&&&&&&&&&&&&&&&&&&&&&\cr \noalign{\hrule}}}$$

\par
As we have mentioned above, one can also search for the $p$-brane
solitons of the corresponding supergravity  theories. The p-brane
actions  then correspond to the low-energy motions of these solitons
and it has been shown that there exist  $p$-brane solitons for   all
the above superbranes.  We refer the reader to the lectures of G.
Gibbons in this volume  and the reviews of reference [176].
\par
We now discuss one further guide to determining which superbranes
occur in which theory.  Given a p-brane one can construct the following
current
$$j^{n\underline m_1\ldots \underline m_p}=\epsilon ^{nn_1\ldots n_p}
\partial _{n_1}X^{\underline m_1}\dots \partial _{n_p}X^{\underline
m_P}
\eqno(7.2.9)$$
which is obviously conserved. The charge associated
with this  current is given by
$$Z^{\underline m_1\ldots \underline m_p}= \int d^p\xi
j^{0\underline m_1\ldots \underline m_p}
\eqno(7.2.10)$$
where the integral is over the $p$ spatial coordinates of the brane.
Our previous discussion on the coupling of the p-brane to a
 $p+1$-form gauge background gauge field can be restated as  the
$p+1$-form  background gauge field couples to
the current $j^n\partial _n X^{\underline m}$.
When computing the space-time supersymmetry algebra in the presence
of the p-brane it turns out [180] that the above charge occurs as a
central charge. Hence, if a p-brane arises in a particular theory we
expect to find its p-form central charge in the corresponding
supersymmetry algebra. Clearly, if the supersymmetry algebra
does not admit a p-form central charge the p-brane cannot occur
unless further supersymmetry is broken. However, we can turn this
argument around and find out which central charges can occur in the
appropriate supersymmetry algebra and then postulate the existence of
a p-brane for each corresponding p-form central charge. Which central
charges can occur for a specified type of supercharge was discussed in
section 1.  For example,
in equation (1.5.5),  we found that the eleven-dimensional
supersymmetry algebra with one Majorana spinor, which is the
supersymmetry algebra appropriate to  eleven-dimensional
supergravity,  had a two-form and a three-form central charge. Hence
in M theory we expect from the above  argument to find a twobrane and
a fivebrane. This  agrees with our previous considerations.
\par
We close this section by giving some simple dimensional arguments
concerning the tensions of various branes. In eleven dimensions,
that is M theory we have only one scale the Planck scale $m_p$. As a
result,  up to constants, the tensions $T_2$ and $T_5$ of the 2-brane
and 5-brane can only be given by $T_2= (m_p)^3$ and
$T_5=(m_p)^6$ respectively. As we discussed in section 6, we can
obtain the IIA supergravity by dimensional
reduction of eleven-dimensional supergravity on a circle and it is
conjectured that IIA string theory can be obtained by
reduction of M theory in the same way. Hence for
the IIA theory in ten dimensions we
have two  scales, the Planck mass $m_p$ and the radius of
compactification of the circle $R_{11}$. Since the string coupling is
dimensionless it must be a function of $m_p R_{11}$ and we found  in
section 6 that  it is given by $g_s=  e^{<\sigma>}=(m_p
R_{11})^{3\over 2}$.
\par
From the 2-brane in eleven dimensions we can obtain a 2-brane in ten
dimensions by simply ignoring the dependence of the fields of the
eleven-dimensional 2-brane on $x^{10}$. We can also find  a
 1-brane, that is a string  if we simultaneously reduce and wrap
the 2-brane on the circle [175]. The tensions of these two objects in
ten  dimensions are therefore given by $T_2= (m_p)^3$ and
$T_1=R_{11}(m_p)^3$ respectively. The factor of $R_{11}$
occurs because  the energy per unit length of the 1-brane arises from
the energy of the 2-brane on the circle.  Similarly, from the fivebrane
in eleven dimensions we can obtain a fivebrane and a fourbrane  in ten
dimensions with tensions
$T_5= (m_p)^5$ and
$T_4=R_{11}(m_p)^5$. The 1-brane in ten dimensions is just the
fundamental string (i.e. the IIA string) and its tension
$T_1$  can be
identified as  the square of the string mass; that is $T_1=
R_{11}(m_p)^3 ={1\over 4\pi\alpha'}\equiv (m_s)^2$. It is instructive
to express the tensions in terms of the   variables $g_s$ and $m_s$
appropriate for the IIA string. Using
our relationship
$R_{11}= (m_p)^{-1}(g_s)^{2\over 3}$ we find that $R_{11}={g_s\over
m_s}$ and $m_p^3={m_s^3\over g_s}$. Substituting into the above
tensions we find that
$T_2={(m_s)^3\over g_s}$, $T_4={(m_s)^5\over g_s}$ and
$T_2={(m_s)^6\over g_s^2}$. We observe the characteristic inverse
coupling constant dependence that is typical of non-perturbative
solitons.   The 2 and 4-branes are Dirichlet branes
and have a
${1\over g_s}$ dependence  while the
fivebrane has a
${1\over g_s^2}$ dependence. This coupling constant dependence of the
tension  of Dirichlet branes is universal and is related to the
occurance of open strings in the theory.
\par
In fact, one can account
from M theory for  two of the remaining branes that are associated with
the IIA string.  The 0-brane and the 6-brane arise from the pp-wave
and Kaluza-Klein monopole solutions in eleven-dimensional supergravity
respectively. The eight brane is associated with massive supergravity
in ten dimensions and its connection to eleven dimensions is unclear.

\medskip
{\bf 7.3 Simple Superbranes}
\medskip
In this section we give the complete dynamics of simple
superbranes [151,157,173] which, we recall, depend only on the
embedding coordinates
$Z^{\underline N}= (X^{\underline n},
\Theta ^{\underline \alpha} )$.
The action for a simple super p-brane
is given by
$$A=A_1+A_2
\eqno(7.3.1)$$
The first term is given by
$$A_1= -T\int d^{p+1}\xi \sqrt{-det g_{mn}}
\eqno(7.3.2)$$
where
$$g_{mn}=\partial _nZ^{\underline N}\partial _m Z^{\underline M}
g_{\underline N \underline M },
\eqno(7.2.4)$$
$$ g_{\underline N \underline M }= E_{\underline N}^{\ \ \underline a}
E_{\underline M}^{\ \ \underline b}\eta _{\underline a \underline
b}
\eqno(7.3.5)$$
where   $ E_{\underline N}^{\ \ \underline A}$ is the superviebein of
the background superspace. The second term in the action
of equation (7.3.1) is given by
$$ A_2=  -{T\over (p+1)!}\int d^{p+1}\xi
\epsilon ^{n_1\ldots n_{p+1}}\partial_{n_1}Z^{\underline M_1}
E_{\underline M_1}^{\underline A_1}
\dots  \partial_{n_{p+1}}Z^{\underline M_{p+1}}E_{\underline
M_{p+1}}^{\underline A_{p+1}}B_{\underline A_1\dots \underline
A_{p+1}}
\eqno(7.3.6)$$
In this expression, $B_{\underline A_1\dots \underline A_{p+1}}$ is a
background superspace
$p+1$ gauge field  referred to superspace tangent indices. Its
corresponding    superspace $p+1$ form is
$$ B={1\over (p+1)!} E^{\underline A_{1}}\dots E^{\underline
A_{p+1}}B_{\underline A_1\dots \underline A_{p+1}}
=dZ^{\underline N_{p+1}}\dots dZ^{\underline N_{1}}
B_{\underline N_1\dots \underline N_{p+1}}
\eqno(7.3.7)$$
where $E^{\underline A}= dZ^{\underline N}E_{\underline
N}^{\ \ \underline A}$.
\par
 The geometry of the target superspace is described
as in section 4 using supervielbeins, but now also with the addition
of  the gauge field $B$.   The covariant objects are the torsions and
curvatures as  before, but now we have in addition the gauge field
strength of the gauge field $B$.  The corresponding superfield
strength $H$ is the exterior derivative,  in superspace of
course, acting on the superspace gauge field;
 i.e. $H=dB$.
The superspace torsions and curvatures  satisfy Bianchi identities as
in section four, but in addition we have the identity
$$ D_{\underline A_1}H_{\underline A_2\dots \underline A_{p+2}}
+T_{\underline A_1\underline A_2}^{\ \ \ \ \ \underline B}
H_{\underline B \underline A_3\ldots
\underline A_{p+1}}+ { \rm {super\ cyclic\  permutations}}=0
\eqno(7.3.8)$$
\par
In the case that the background superspace is flat then the
supervielbein are given as in equation (7.2.7). However, the gauge
field $B$  also has a non-trivial form.
 The resulting torsions all
vanish except for
$$ T_{\underline \alpha \underline \beta}^{\ \ \ \underline a}
=i {(\gamma^{\underline a}C^{-1})}_{\underline \alpha \underline
\beta},
\eqno(7.3.9)$$
and
$$
H_{\underline \alpha \underline \beta \underline a_1\ldots
\underline a_p}= -i(-1)^{{1\over 4}p(p-1)}
{(\gamma_{\underline a_1\ldots \underline a_p}c^{-1})}_{\underline
\alpha
\underline \beta}
\eqno(7.3.10)$$
In terms of forms we may express these results as
$$ T^{\underline a}= {1\over 2}i d\bar \Theta \gamma^{\underline
a}d \Theta
\eqno(7.3.11)$$
and
$$
H= {i\over 2 p!}E^{a_p}\dots E^{\underline a_{1}}
d\bar \Theta \gamma_{\underline a_1\dots \underline a_{p}}d\Theta
\eqno(7.3.12)$$
\par
Since $H$ is an exact form $dH=0$. Using this  on
 equation (7.3.12) and from the form of
$E^{\underline a}$  of equation (7.2.7) we find that $dH=0$
is equivalent to the condition [173]
$$d\bar \Theta \gamma_{\underline a} d\Theta
  d \bar \Theta\gamma^{\underline a \underline b_1\ldots
\underline b_{(p-1)}} d\Theta=0
\eqno(7.3.13)$$
Since $\Theta $ is Grassmann odd, $d\Theta $ is Grassmann even
and hence the above condition must hold for a Grassmann even
spinor  of the appropriate type that is complex, Weyl, Majorana or
Majorana-Weyl, but is otherwise arbitrary. This means in
effect that we
can discard the
$\Theta$ from the above equation provided we include a projector if
the spinor is Weyl, and symmetrise on all the free spinor indices
if it is Majorana and on the first and third and second and fourth
indices separately if it is complex. For example,  if $\Theta $ is
Majorana then we find the identity
$$(\gamma_{\underline a}C^{-1})_{(\underline \alpha
\underline \beta }
(\gamma^{\underline a \underline b_1\ldots
\underline b_{p-1}}C^{-1})_{\underline \delta \underline \epsilon)}
=0
\eqno(7.3.14)$$
\par
These identities can be reformulated using the appropriate Fierz
identity and it can then be shown [173] that they hold if  the
number of fermion and boson degrees of freedom of the brane are
equal.
We have already used this condition to find the
 simple branes  in ten and eleven dimensions.
\par
Finally it remains to discuss the $\kappa$-symmetry of the action
of equation
(7.3.1). The variation of the coordinates $Z^{\underline N}$
under this symmetry when referred to the tangent basis are given by
$$
\delta Z^{\underline N}E_{\underline N}^{\ \ \underline a}=0, \
\delta Z^{\underline N}E_{\underline N}^{\ \ \underline \alpha}=
(1+\Gamma)^{\underline \alpha }_{\ \underline \beta }\kappa^{\underline
\beta }
\eqno(7.3.15)
$$
where the local parameter $\kappa^{\underline \beta }$ is a world
surface scalar,  but a tangent space spinor. The matrix
$\Gamma$ is given by
$$\Gamma= {(-1)^{{1\over 4}(p-2)(p-1)}\over (p+1)! \sqrt{-det g_{nm}}}
\epsilon ^{n_1\ldots n_{p+1}}
\partial _{n_1}Z^{\underline N_1}
E_{\underline N_1}^{\underline a_1}\dots
\partial _{n_{p+1}}Z^{\underline N_{p+1}}
E_{\underline N_{p+1}}^{\underline a_{p+1}}
\gamma_{a_1\ldots a_{p+1} }
\eqno(7.3.16)$$
The matrix $\Gamma$ obeys the remarkably simple property
$\Gamma^2=1$ and as a result ${1\over 2}(1 \pm \Gamma)$ are
projectors. Clearly a $\kappa$ of the form  $\kappa={1\over
2}(1-\Gamma)\kappa^1$ leads to no contribution to the symmetry
to which therefore  only half of the components of $\kappa$ actually
contribute. The  remaining half
allows us to gauge away the  corresponding half of
$\Theta$. This property of $\kappa$ symmetry is
essential for getting the correct number of on-shell fermion
degrees of freedom namely 8.
In fact the action is only invariant under $\kappa$ symmetry  if the
background  fields   obey their equations of motion.
\par
For the case when the background superspace is flat
the above $\kappa$-symmetry
reduces to
$$\delta X^{\underline n}= {i\over 2}\bar \Theta\gamma^{\underline
n}\delta
\Theta ,\ \delta \Theta = (1+\Gamma)\kappa
\eqno(7.3.17)$$
\par
 The form of the redundancy discussed above in the $\kappa$-invariance
 leads to very difficult  problems when gauge fixing the
$\kappa$-symmetry and so quantizing the theory.  Imposing a gauge
condition on the spinor
$\Theta $ to fix the $\kappa$-symmetry can only fix
part of the symmetry and leaves unfixed that part of $\kappa$ which is
given by $\kappa^1$. A little thought shows that as a result
 the ghost action itself will have a local symmetry which must be
itself fixed. In fact,
 this process
continues and results in   an infinite number of  ghosts for
ghosts corresponding to the infinite set of invariances.
This would not in itself be a problem, but it has proved impossible
to find a Lorentz-covariant gauge-fixed formulation.
The problem of quantizing can also be seen from the viewpoint of
the Hamiltonian approach where one finds that the system possess
first and second class constraints that cannot be separated in a
Lorentz-covariant manner.
  The Green-Schwarz action, for the string  has so far
defied attempts to quantize it in a truly covariant manner.
\medskip
{\bf 7.4 D-branes Dynamics}
\medskip
As we discussed, D-branes   have a vector field $A_n$ living on their
world surface in addition to the embedding
coordinates $X^{\underline n},\ \Theta^{\underline
\alpha } $. The  action for the
D-branes is a generalization of that of the simple brane of equation
(7.3.1)  to  include this vector
 field $A_n$.
 In the absence of background fields other than the background metric
and  when $\Theta^{\underline
\alpha } =0$ the  action is just the Born-Infeld action
$$-T\int d^{p+1}\xi \sqrt{-det (g _{nm}+F_{nm}})
\eqno(7.4.1)$$
where $F_{mn}= 2\partial _{[m} A_{n]}$. The full action
was found in references [147-150] by using $\kappa$-symmetry. Its
precise form and the relationship to Dirichlet open strings can be
found in the lectures of Costas Bachas in this volume.
\medskip
{\bf 7.5 Branes in M Theory}
\medskip
As we have discussed above, there  exists a 2-brane and a 5-brane in
eleven dimensions. The 2-brane is a simple brane and so its dymanics
are those  given in section 7.3.
The dynamics of the fivebrane is the
subject of this section. As we discussed in section 7.2 the fivebrane
contains five scalar fields and a 16-component  spinor corresponding
to the  breaking of translations and supersymmetry by the fivebrane.
It also contains, living on its world surface,
an antisymmetric second-rank tensor gauge field whose  field strength
obeys a self-duality condition.  As this condition is
required to obtain the correct number of degrees of freedom, we only
need it to hold at the linearized level and it is an
important feature of the fivebrane dynamics that the self-duality
condition in  the full theory is a very non-linear condition on the
 third-rank  field strength.
\par
 As for other branes,
the bosonic  indices of the fields  on the fivebrane can be decomposed
into longitudinal and transverse indices (i.e. for the world
 target space indices $\underline n= (n,n')$) according to the
decomposition of the eleven-dimensional
 Lorentz group
$SO(1,10)$ into
$SO(1,5)\times SO(5)$. The corresponding decomposition of the eleven
dimensional  spin group $Spin (1,10)$ is  into $ Spin(1,5)
\times  Spin (5)$. These spin groups divided by $Z_2$
 are isomorphic to
their corresponding Lorentz groups in the usual way. In fact
$Spin (5)$ is isomorphic to  the group $USp(4)$ which is defined below
equation (1.3.18).
Thus the fivebrane possesses a $ Spin(1,5)
\times  Spin (5)$ or $ Spin(1,5)
\times  USp(4)$ symmetry.
\par
 We now  assign the fields of the
fivebrane to  multiplets of this symmetry. The five real scalars are
of course $Spin(1,5)$ singlets, but belong to the vector representation
of $SO(5)$,  $X^{n'}, n'=6,\ldots 10$.
This representation corresponds  to  the  second-rank
anti-symmetric tensor representation
$\phi_{ij},\ i,j=1,\ldots 4$,  of the isomorphic
$USp(4)$ group  which is traceless with respect to
the ant-symmetric metric
$\Omega^{ij}$ of this group. The gauge field $B_{ m  n}$ is
a singlet under $USp(4)$.
We recall from section one that in six dimensions one cannot have
Majorana spinors, but  can have symplectic Majorana spinors and
even symplectic Majorana-Weyl spinors. The fermions of the
fivebrane belong to
the four-dimensional vector representation of
$USp(4)$ and are $USp(4)$ symplectic Majorana-Weyl spinors and so obey
equation (1.3.17). Since they are Weyl we can work with their
Weyl projected components which take only four values as opposed
to the usual $8=2^3$ components for a six-dimensional spinor.
The spinor indices of the groups $Spin(1,5)$ and  $ USp(4)$
are denoted by  $\alpha,\beta,...=1,...,4$ and $i,j,...=
1,...,4$  respectively and  thus
the spinors carry the indices $\Theta _\alpha^i,\  \alpha =1,\ldots 4,
\ i =1,\ldots 4$.  The spinors therefore  have the required sixteen
real components.
\par
It is instructive to examine how this spinor arises from the
original eleven-dimensional spinor $\Theta ^{\underline \alpha}$ in
terms of which the fivebrane dynamics was first formulated.
Although we began with spinor
indices $\underline \alpha $ that  take  thirty-two  values
 we can, as above for all branes, split these indices into
two pairs of indices each taking sixteen values
${\underline
\alpha}=(\alpha,\alpha')$. In the final six-dimensional
expressions the spinor indices  are further written according
to the above decomposition of the spin groups and we take
$\alpha\rightarrow\alpha i$
and $\alpha'\rightarrow{}_{\alpha}^{i}$ when appearing as
superscripts and $\alpha\rightarrow\alpha i$
and $\alpha'\rightarrow{}^{\alpha}_{i}$ when appearing as subscripts.
It should be clear whether we mean $\alpha$ to be sixteen- or
four-dimensional
depending on the absence or presence of $i,j,...$ indices respectively.
For example, we will write  $\Theta ^{\alpha'}\to\Theta_{\alpha}^{i}$.
\par
The fields of the fivebrane belong to the so-called $(2,0)$ tensor
multiplet which transforms under $(2,0)$ six-dimensional supersymmetry.
The $(2,0)$ notation means  that the supersymmetry parameter is a
$USp(4)$ symplectic Majorana-Weyl spinor. By contrast, we
note that $(1,0)$ supersymmetry  means  that the supersymmetry
parameter is a
$USp(2)$ symplectic Majorana-Weyl spinor while, if the
parameter  is just a
$USp(4)$ symplectic Majorana spinor, the supersymmetry would be denoted
by $(2,2)$.

 The
classical equations of motion  of the fivebrane in the absence of
fermions and background fields are [137]
$$
G^{ m n} \nabla_{ m} \nabla_{ n} X^{a'}= 0\ ,
\eqno(7.5.1)
$$
and
$$
 G^{ m  n} \nabla_{ m}H_{ n p q}  = 0.
\eqno(7.5.2)
$$
where the world surface indices are $ m, n,  p=0,1,...,5$
and   $ a, b, c=0,1,...,5$ for world and  tangent indices
respectively.
The transverse indices are $a',b'=6,7,8,9,10$. We now define the
symbols that occur in the equation of motion. The usual
induced metric
for a $p$-brane is given, in static gauge and flat background
superspace,  by
$$
g_{ m  n} = \eta _{ m  n}+
\partial _{ m}X^{a'} \partial _{ n}X^{b'}\delta _{a' b'}\ .
\eqno(7.5.3)
$$
The covariant derivative in the equations of motion
is defined with the  Levi-Civita connection with respect to the metric
$g_{ m  n} $.
Its action on a vector field $T_{ n}$ is given by
 $$\nabla_{ m} T_{ n} = \partial _{ m} T_{ n}-
 \Gamma _{ m  n}^{\ \  p}T_{ p}
\eqno(7.5.4)$$
where
$$
\Gamma _{ m  n}^{\ \ \  p}
= \partial _{ m } \partial _{ n} X^{a'}
\partial _{ r} X^{b'}g^{ r  s}\delta _{a' b' }\ .
\eqno(7.5.5)
$$
We define the world surface vielbein associated with the above metric
in the usual way
$g_{ m n}=
e_{ m}^{\  a} \eta _{ a  b} e_{ n}^{\  b}$.
There is another  inverse metric $G^{ m n}$ which occurs in the
equations of motion and it
is related to the usual induced metric given above by the equation
$$
G^{ m n} = {(e^{-1})}^{\ \  m}_{\  c} \eta ^{ c  a}
m_{ a}^{\  d} m_{ d} ^{\  b} {(e^{-1})}^{\ \  m}_{\   b}\ .
\eqno(7.5.6)
$$
where the matrix $m$ is given by
$$
m_{ a}^{\  b} = \delta_{ a}^{\  b}
 -2h_{ a c d}h^{ b c d}\ .
\eqno(7.5.7)
$$
The field $h_{ a b c}$ is an anti-symmetric three-form
which is self-dual:
$$
h_{ a b c}=
{1\over3!}\varepsilon_{ a b c d
e f}h^{ d e f}\ ,
\eqno(7.5.8)
$$
but it is not the curl of a three-form gauge field. It is related to
the field
$H_{ m  n  p}=3\partial _{[m}B_{np]}$ which appears in the
equations of motion
and is the curl of a gauge field,  but
$H_{ m  n  p}$ is not self-dual.
The relationship between
the two fields is given by
$$
H_{ m  n  p}= e_{ m}^{\  a}
e_{ n}^{\  b} e_{ p}^{\  c} {({m }^{-1})}_{
c}^{\  d} h_{ a b d}\ .
\eqno(7.5.9)
$$
Clearly, the self-duality condition on $h_{ a b d}$
transforms into a
condition on $H_{ m  n  p}$ and vice versa
for the Bianchi identity $dH=0$.
\par
The fivebrane equations of
motion  were found in reference [137] for
arbitrary background fields
and to first-order in the fermions; we refer  the reader to
this reference   for the construction.
\par
The above fivebrane equations of motion were found using the
 embedding formalism. This formalism
including the superspace embedding condition of equation
(7.5.12) were first given
within the context of the superparticle in
reference [178].  Further work on
the  superparticle was carried out in references [195] and  [196].
The superembedding formalism was first applied to p-branes in
references [131,132, 197, 133] and to the M theory  fivebrane in [134].
The above fivebrane equations of motion were derived in  [137]. The
form of the fivebrane equations of motion and the relationship to the
D4-brane of the IIA theory was discussed  in reference [137].
Reference[138] contains a review of the embedding formalism.
The super embedding
approach was also used to find p-brane actions
and  equations of motion in references [141] and [198].
\par
We now briefly explain  the simple idea that underlies
this formalism. We consider  a target or background superspace
$\underline M$ in which a brane sweeps out a superspace $M$. On each
of these superspaces we have a set of  preferred frames or
supervielbeins. The frame vector fields on the target
manifold $\underline M$ and the fivebrane world surface
$M$ are denoted
 by $E_{\underline A} = E_{\underline A} ^{\ \underline
M}
\partial_{\underline M}$ and
$E_{A} = E_{A} ^{\ M} \partial_{M}$ respectively.
We recall that we  use the superspace index convention that
$\underline N,
\underline M, \ldots $ and $\underline A,\underline B,\ldots $
 represent the world and tangent indices of the  target superspace
$\underline M$
  while $ N,  M, \ldots $ and $ A, B,\ldots $ represent the world
and tangent space indices of the embedded superspace $M$.

Since
the supermanifold $M$ is embedded in the supermanifold
$\underline M$, the frame vector fields of $M$ must point
somewhere in $\underline M$. Exactly where they point is
encoded in the  coefficients $E_{A}^{\ \underline A}$ which
relate the vector fields $E_{A}$ and $E_{\underline A}$,
i.e.,
$$E_{A} = E_{A}^{\ \underline A}E_{\underline A}
\eqno(7.5.10)$$
Applying this
relationship to the coordinate $Z^{\underline M}$ we find the equation
$$
E_{A}^{\ \underline A}= E_{A}^{\ \ N}\partial_N Z^{\underline M}
E^{\ \underline A}_{\underline M}\ .
\eqno(7.5.11)
$$
It is now straightforward to express the torsion and
curvature tensors of $M$ in terms of those of $\underline M$
plus terms involving a suitable covariant derivative of
$E_{A}^{\ \underline A}$; one finds that
$$ \nabla_A E_B{}^{\underline C}-(-1)^{AB}\nabla_{B}
E_{A}{}^{\underline C}
+T_{AB}{}^C
E_C{}^{\underline C}
=(-1)^{A(B+\underline B)}E_{B}^{\ \underline B} E_{A}^{\ \underline A}
T_{\underline A\underline B}{}^{\underline C}
\eqno(7.5.12)$$
where the derivative $\nabla_A$ is covariant with respect to both
embedded and target superspaces, that is, it has connections which act
on both underlined and non-underlined
indices.
\par
The tangent space of the superspaces $\underline M$ and $M$
can be divided into their Grassmann odd and even
sectors, that is the odd and even pieces of $M$
are spanned by the vector fields $E_\alpha$ and
$E_a $ respectively and similarly for $\underline M$. The
superembedding formalism has only one assumption;   the odd tangent
space of $M$ should lie in the odd tangent space of $\underline M$.
This means that
$$E_{\alpha }^{\ \underline a}=0
\eqno(7.5.13)$$
To proceed one substitutes this condition into the relationships
of  equation (7.5.12) between the torsions
and  curvatures of $M$  in terms of $\underline M$
and analyses the resulting   equations in order of increasing
dimension. For example at dimension zero one finds the
equation
$$
E_{\underline a}^{\ \ \underline a}E_{\underline b}^{\ \ \underline
b}T_{\underline a\underline b}{}^{\underline c}
=T_{\underline a\underline b}^{\ \ \ \ c} E_{c}^{\ \underline c}
\eqno(7.5.14)$$
This procedure is  much the same as   for the  usual
superspace Bianchi identities for super Yang-Mills and supergravity
theories.  For the twobrane and fivebrane of M-theory this procedure
yields:

\item {(a)} the equations of motion for the fields of the brane,

\item {(b)} the equations of motion for the background fields,

\item {(c)} the geometry of the embedded manifold, that is its torsions
and curvatures.
\par

Finding the equations of motion of the supergravity background
fields means finding  the superspace constraints on the
target superspace torsions and curvatures.
\par
Although the embedding condition of equation (7.5.13) is very natural
in that, as we saw in section 4, all the geometry of superspace is
contained in the odd sectors of the tangent space of supermanifolds,
its deeper geometrical significance is  unclear.
However,  the power of this approach     became evident  once  it
was shown to  lead to the correct dynamics for the most
sophisticated brane, the M-theory fivebrane.   Although the
above results hold for many branes, they do not hold for all
branes unless the embedding condition is in general suplemented by a
further condition.
\par
The appearance of a metric in the equations of motion which is
different to the usual induced metric has its origins in the
fact that the natural metric that appears on the world
surface of the fivebrane has an associated  inverse vielbein  denoted
by
${(E^{-1})}_{ a}^{\  m}$ which
is related in the usual way through $G^{ m n} =
{(E^{-1})}_{ a}^{\  m}
{(E^{-1})}_{ b}^{\  n} \eta^{ a b}$.
The relationship between the two inverse vielbeins is
${(e^{-1})}_{ a}^{\  m} =(m^{-1})_{ a}^{\  b}
{(E^{-1})}_{ b}^{\ m}$.
\par
There is another formulation of the dynamics of the fivebrane
given in references [141] and [142]. Although this formulation
involves an action this is not necessarily  an advantage as has
been pointed out in reference [143]. Reference [141]  used the
superembedding formalism, but in a different way to that considered
in this section.
\par
The above equation of motion for the  fivebrane admits
two interesting solutions corresponding to onebranes (self-dual
strings) [144] and  threebranes [145].  The
threebrane solution has been used [146,177] to derive the complete low
energy effective action  of the
$N=2$ Yang-Mills theory. Indeed, the
fivebrane may self-intersect on a four manifold which  can be
considered as this 3-brane. Just like the
monopoles discussed at the beginning of this section we may consider
the moduli space of  threebranes. The low-energy motion of several
threebranes can then be described in terms of their moduli which now
depend on the world surface of the threebrane. The corresponding
action for the low energy motion of these threebranes is then a four
dimensional quantum field theory with $N=2$ supersymmetry which is
none other that the complete chiral effective action of the
spontaneously broken $N=2$ supersymmetry gauge theory. Thus  from the
{\bf classical} dynamics of the fivebrane we can deduce the chiral
effective action of the  spontaneously broken
$N=2$ Yang-Mills theory which includes an infinite number of instanton
corrections, only the first two of which  can be calculated
using known
instanton techniques.
\bigskip

\centerline{\bf 8. String Duality}
\medskip
In this section we use  our previous discussions on
supergravity theories and brane dynamics to give some of the
evidence for string dualities. We restrict our attention
to the relationships between the IIA and IIB strings and M theory
and only aim to give  some   introductory remarks. A  more
complete  discussion can be found in the lectures of Ashoke Sen in this
volume.
\medskip
Dualities in string theory can be of several different types. There
are dualities which relate one  string theory to another
and dualities which relate a given string theory to itself. We refer
to the latter as self-dualities. Dualities can also be classified by
whether they map the perturbative
regime of the theory   to  the same perturbative regime or whether
they map the perturbative regime to the    non-perturbative regime.
Roughly speaking, the perturbative and non-perturbative regimes
correspond to  the small and large coupling constant regimes of the
theory respectively.  Currently, our ability to calculate
systematically in  quantum field theories is limited to the
perturbative regime where  one can
extract meaningful answers by using perturbation theory using   the
coupling constant as the parameter. In fact, in the absence of a
special non-renormalization  theorem, such expansions, which are
derived from the path integral, are not  convergent.  However,  for
certain quantum field theories the first few terms give increasingly
more accurate results and one can by resumming the series find the
result to any desired accuracy as a matter of principle. To determine
the  range of values of the  coupling constant for which the theory is
in its pertubative regime is not always very straightforward
and it  is found
 by studying the  behaviour of the  amplitudes as a complex function
of the coupling constant.  In some theories there are no values of
the coupling conatant for which the perturbation series can be
resummed and so these theories have in effect no perturbative regime.
\par
In fact, many interesting phenomena are
outside the range of perturbation theory. Such  phenomena are often
related to the presence of instantons and solitons in the theory. It
is of course one of the outstanding problems of theoretical physics to
 understand properly  non-perturbative effects, such as quark
confinement.
\par
Dualities which map from the perturbative to the same perturbative
regime   of a theory
are mapping from a regime  of the theory which is relatively well
understood to the same region. As a result,   such dualities  are
straightforward to check. The best example of this type of  duality
in string theory is
 so called T-duality [183, 191, 192] which  occurs in string
theories that are compactified. The simplest  example of
$T$-duality occurs
for the closed  bosonic string theory  compactified on a circle of
radius
$R$.  One finds that the theory is invariant under the transformation
$R\to {\alpha'\over R}$. To be more precise, the mass spectrum  of the
physical states and their scattering amplitudes are invariant under
this transformation [184].
\par
More generally, we can consider the T-duality which occurs in a
   string theory which has been  compactified on    d of its
dimensions.    In string theory we can make separate
compactifications for the
  left and right modes  of the string and as a result
the relavent torus is
$T^{(d,d)}={{\bf R}^{(d,d)}\over
\Lambda^{(d,d)}}$  where
$\Lambda^{(d,d)}$ is the Lorentzian lattice that generates the torus.
The lattice is formed from the momenta in the left and right
directions $(p_L,p_R)$. This lattice is just  the momenta on the
torus, while the  Lorentzian signature
arises  from  the scalar product
encoded in the constraint $(L_0-\bar
L_0)\psi=0$. This constraint also implies the lattice is even.
Modular invariance of this  string theory
requires that
$\Lambda^{(d,d)}$  be   an even self-dual Lorentzian lattice.
Clearly, the resulting string
theory depends on the  torus, or equivalently
the self-dual lattice used in the compactification.
It turns out that all even self-dual Lorentzian lattices that can
occur in such  string compactifications are    uniquely classified by
the coset [190]
$$ O(d;{\bf R})
\otimes O(d;{\bf R})\backslash O(d,d;{\bf R})/ O(d,d;{\bf Z})
\eqno(8.1)$$
 Consequently, the resulting string theories are also classified by
this coset. Clearly, we can act with the
group  $O(d,d;{\bf Z})$ on the
basis  vectors of
a given lattice and it will take that  lattice to itself.
Hence we must also divide  out by this group.
The
$d^2$ parameters of the coset can be accounted for by the $d^2$
expectation values of  the metric and anti-symmetric tensor of the
string in the compactified directions [161].  The corresponding string
theory is invariant under the duality symmetry
$O(d,d;{\bf Z})$. The T-dualities map a given theory to itself
and are therefore an example of a self-duality.  The T-duality
symmetries are well understood and have been shown to hold to all
orders in string perturbation theory [184].
\par
Dualities which map from perturbative to non-perturbative regimes
 are not very well understood. The problem with verifying such
dualities is that  they relate quantities in the perturbative regime,
which  in general can be reliably  computed,  to quantities
 in the   non-perturbative regime which in general can not be
calculated. Therefore  such dualities are in general difficult to
test.  However, for
 certain supersymmetric theories,
 some properties of the theory  in the non-perturbative regime
are  reliably known.
One example is the low-energy effective action of a string
theory that possesses a supersymmetry with 32 supercharges. Such a
low-energy effective action is none other than a supergravity theory.
  We recall  that   supergravity theories can have an underlying
supersymmetry algebra that has  at most 32 supercharges. Furthermore,
for one of the maximal supergravity theories, the  action  is
essentially uniquely determined by the underlying supersymmetry
algebra. For example,  if we are
in ten dimensions then  a  supersymmetry algebra with 32
supercharges has either two Majorana-Weyl spinors of the same
chirality or two  Majorana-Weyl spinors
of the opposite  chirality and the corresponding unique supergravity
theories are   the  IIB  and
IIA supergravities respectively.   Therefore, given any
string theory which is  invariant under a space-time supersymmetry
that has 32 supercharges, then the low energy effective action  of this
string theory must be the unique maximal supergravity  in that
dimension for the particular maximal supersymmetry algebra involved.
Hence  if we suspect that two string theories  are related by a
duality
symmetry one test we can apply is to  find if the suspected duality
relates their  low energy effective actions    which are the unique
corresponding  maximal supergravities.  We can turn this approach
around and consider  which string theories  have low-energy  effective
theories (i.e supergravity theories) that can   be related by a
duality symmetry and then consider if these symmetries can be promoted
to string dualities.
\par
A further consequence of these considerations concerns string
theories that are invariant under a space-time supersymmetry with
32 supercharges and  that are self-dual. Since their low energy
effective actions
 in both the strong and weak coupling limits are the same low-energy
supergravity, it follows that this self-duality symmetry must be a
symmetry of the maximal supergravity theory. Thus to look for
self-duality symmetries we can examine the symmetries of the maximal
supergravities and wonder which to promote to a string duality.
\par
One other property that we can reliably establish in all coupling
constant regimes of a supersymmetric theory is the existence of the
BPS states discussed in section 3. Recall from there that
 these supermultiplets have fewer states than  a
supermultiplet with a generic mass and their existence relies on their
mass being equal to one of the central charges that appear in the
supersymmetry algebra. For these states to disappear as one changes
the coupling constant would require the abrupt existence of additional
degrees of freedom and this does not usually occur.
\par
Equipped with this strategy we now examine the relationships
between the maximal supergravity theories. For simplicity let us
consider the relationships between the maximal supergravities in
eleven, ten and nine dimensions. These are the eleven-dimensional
supergravity and the IIA and IIB theories in ten dimensions
and the single nine-dimensional maximal supergravity. Their
relationships are set out in figure 8.1. We recall from section
 6 that IIA supergravity can be obtained from eleven-dimensional
supergravity by reduction on a circle and that the    radius
$R$ of this circle and   the coupling constant  $g_s$ of the IIA
string theory are  related by equation (6.2.11). The IIB theory has
a supersymmetry algebra with two Majorana-weyl spinors of the
same chirality and so can not be obtained from
eleven-dimensional supergravity by dimensional reduction. It does,
however, possess  an $SL(2,{\bf R})$ symmetry. Finally, we found that
if we reduce either the IIA or IIB  theory to nine dimensions then we
obtain the same unique supergravity.
\par
Let us now consider what these relationships suggest for the
corresponding string theories. Let us begin by considering which
of the theories in ten dimensions could be self-dual. The obvious
candidate is the IIB theory whose low-energy effective action
has an $SL(2,{\bf R})$ symmetry [110]. It is natural to
consider this group or one of its subgroups to be a symmetry
of the IIB string theory.  The IIB supergravity has two second-rank
tensor gauge fields $B_{\mu\nu}^1$ and $B_{\mu\nu}^2$ which transform
into each other under $SL(2,{\bf R})$ as in equation (6.3.39). From
the viewpoint of the IIB string these  fields belongs to the
$NS\otimes NS$ and  $R\otimes R$ sectors respectively. As we
explained in section 7.2 these target space gauge fields couple to
the  1-branes which are the IIB string itself and
the Dirichlet 1-brane.
\par
Let us consider an object which is charged with
respect to both gauge fields $B_{\mu\nu}^1$ and $B_{\mu\nu}^2$  and
  denote the charges by
$q^1$ and
$q^2$   respectively. Since  the charges they carry are  proportional
to their corresponding field strengths they are  rotated into
each other under
$SL(2,{\bf R})$. Using   equation (6.3.39), we
conclude that under an
$SL(2,{\bf R})$ transformation the charges are changed according
to  [110]
$$
\left(\matrix{q^{1\prime}\cr q^{2 \prime}\cr}\right)\
=
\left(\matrix{a&b\cr c&d\cr
}\right)
\left(\matrix{q^1\cr
q^2\cr}\right)
\eqno(8.2)$$
However, the analogue of the Dirac quantization condition for
branes  [191]
 implies that  the charges are always quantized in integers once we
adopt suitable units. It is easy to see  that the maximal subgroup of
$SL(2,{\bf R})$ which preserves such a charge quantization is
$SL(2,{\bf Z})$.  The charge quantization condition can be expressed
as the statement that  one can only ever find whole
fundamental strings and D1-branes and  their bound states.
\par
Thus we can at most choose the group  $SL(2,{\bf Z})$
to be a   symmetry of IIB string theory. This symmetry
includes the transformation of equation (6.3.40) and so  takes
the   weak coupling regime
theory to  the  strong coupling regime of the IIB string.
\par
Clearly, in the IIB string theory there exist
elementary string states of unit charge with respect to the gauge
field $B_{\mu\nu}^1$, but no charge with respect to the other gauge
field $B_{\mu\nu}^2$. One such state has  charge
$\left(\matrix{1\cr 0\cr}\right)$.  Acting  with an
$SL(2,{\bf Z})$ transformation   we  find, using equation
(8.2),   states  with the charges $\left(\matrix{a \cr c \cr}\right)$
where $a$ and
$c$ are  integers and are the  charges
 with respect  $B_{\mu\nu}^1$ and
$B_{\mu\nu}^2$ respectively. While the states  charged with
respect to just $B_{\mu\nu}^1$ are the elementary string states which
occur in the perturbative domain, the $SL(2,{\bf Z})$
rotated states carry charge with respect to $B_{\mu\nu}^1$ and
$B_{\mu\nu}^2$ and are non-perturbative in nature. Indeed, the
elementary states of the IIB string cannot be charged with respect to
$B_{\mu\nu}^2$.
Hence,  the existence of an
$SL(2,{\bf Z})$ symmetry in the theory
  implies that there must exist  states with the above charges
for all integers $a$ and $c$ which are relatively prime, that is
have no common factor. The latter condition follows from the
$SL(2,{\bf Z})$  condition $ad-bc=1$, since in this case $ad-bc$
would contain a common factor and so could not equal one. It was
shown [164], by considering parallel Dirichlet 1-branes,  that these
states actually exist and this provides us with an important
consistency check on the conjectured
$SL(2,{\bf Z})$  symmetry in the string theory.
\par
It was in reference [152] that it was first suggested that
string theory could possess a duality symmetry that transformed
weak- to strong-coupling regimes. In particular,  it was known that
that the low-energy effective action for the heterotic string theory
compactified on
$T^6$ possessed  an $SL(2,{\bf R})$ symmetry  and the authors of
reference [152,153]  suggested that the $SL(2,{\bf Z})$ subgroup be a
symmetry of the the heterotic string theory
compactified on
$T^6$. By considering the action of this symmetry on BPS states,
 in references [152-154] and [155]
evidence for this conjecture was given. In reference [156], it was
suggested that the $SL(2,{\bf Z})$ subgroup of the
known $SL(2,{\bf R})$ [110] symmetry of the IIB low energy effective
action  be a symmetry of the full IIB theory. Indeed,
reference [156] conjectured that if we changed the field from
${\bf R}$ to ${\bf Z}$ in the known coset symmetry of the low
energy effective action (i.e. supergravity theory) then we will find a
symmetry of the corresponding string theory. To be precise it was
suggested that since the supergravity theories possessed scalars
which were contained in coset spaces $G/H$ that
the corresponding string theory have a symmetry $G({\bf Z})$ where
$G({\bf Z})$  is the group $G$ with its field changed form ${\bf R}$
to ${\bf Z}$.  The  coset space symmetries of the maximal supergravity
theories are listed in table 6.1. For the IIA or IIB string
compactified on a six-dimensional
torus, the low-energy effective
action is the $N=8$ supergravity in four dimensions and so we expect
to find a
$E_7({\bf Z})$ symmetry.   These references followed the pattern
that occurs in the $N=4$ Yang-Mills theory which was earlier given
in reference [193].
\par
Now let us turn our attention to the relationship between the
eleven-dimensional supergravity [106] and the IIA supergravity
[107-109]. As explained in section 6.2 the former theory
results from the latter if we compactify  on a circle [107-109]. We
found  in equation (6.2.18) that the IIA string coupling $g_s$ and
the radius of compactification $R_{11}$ are related by $g_s\propto
R_{11}^{3\over 2}$. Clearly, in the strong-coupling limit of the
IIA string i.e. as $g_s\to \infty$ the radius $R_{11}\to \infty$.
However, in this limit the circle of compactification becomes
flat and one expects to  recover an
eleven-dimensional theory whose low-energy limit is
eleven-dimensional supergravity. This realization
has lead to the conjecture [123],[124]  that the strong coupling limit
of  IIA theory defines a consistent theory called M-theory which
possesses eleven-dimensional Poinca\'e invariance  and
has eleven-dimensional supergravity as its  low-energy
effective action.
\par
We now summarise some of the evidence for the existence of M
theory. When compactifying the eleven-dimensional supergravity in
section 6.2 in addition to the massless
modes of the IIA supergravity theory we found Kaluza-Klein modes.
These    Kaluza-Klein modes have a  mass
${n\over R_{11}}$ which have   charge
${n\over R}= {n\over R_{11}}g_s^{{2\over 3}} $
with respect to the $U(1)$ gauge field $B_\mu$ in the IIA theory
that arises from the graviton in eleven dimensions. Since the
low-energy effective action of M theory is eleven-dimensional
supergravity,
 these Kaluza-Klein modes are also present in M
theory.  However,  if we are to believe  that the strong-coupling
limit  of IIA  string theory is M theory then these Kaluza-Klein modes
should be also be present in the IIA string. In fact, the Kaluza-Klein
states belong to massive supermultiplets that have fewer states than a
supermultiplet with a generic mass and so they are the so called BPS
states considered in section 3 whose mass is equal to their central
charge. As we discussed at the end of that section, the existence of
BPS states must  be present in the strong coupling regime
of a theory if they are present in the weak-coupling regime and
vice versa. As such, we should find the analogue of the Kaluza-Klein
states at all coupling constant regimes of the IIA string theory.
\par
Hence an essential test of the conjecture is to find the analogues
of the Kaluza-Klein states in the IIA theory. The first evidence
for the existence of these  states in the IIA string was
the realization that the IIA supergravity admitted solitonic states of
the correct mass and charge [123].
The elementary states, that is perturbative states of the IIA
string, are not charged with respect to  the gauge field $B_\mu$
since it belongs to the $R\otimes R$ sector of the IIA string.
However,  as  we discussed in section 6, Dirichlet 0-branes  do couple
to the
$B_\mu$ gauge field in the
$R\otimes R$  sector of the IIA string. The Dirichlet 0-branes are
non-perturbatitve in nature, but this identification is in accord
with fact that the masses of the Kaluza-Klein states are given by
${n\over R_{11}}=n {m_s\over g_s}$ when
expressed in terms of the string coupling constant. These states
becomes very large as the coupling constant becomes small, in a manner
typical of non-perturbative solitons. In the limit of large coupling
they become massless. The appearance of an infinite number of massless
modes indicates the transition to a theory in which a dimension has
become decompactified.
\par
The dynamics of a single such
Dirichlet 0-brane  was outline in section 7. In a flat background, it
consists of a supersymmetric generalization of Born-Infeld action  in
one dimension. At low-energy this theory in just the dimensional
reduction of $D=10$ $U(1)$ gauge theory to one dimension. Such a single
Dirichlet 0-brane   has the correct mass and charge to be identified
with the lowest mass  Kaluza-Klein state. It is thought that the
higher mass Kaluza-Klein states can be identified with bound states of
Dirichlet 0-branes at threshold, that is, bound states that have the
same energy as the lowest energy state of two particles. The
dynamics of several parallel Dirichlet 0-branes   is described by a
supersymmetric generalization of Born-Infeld theory which now carries
a $U(N)$ gauge group if $N$ is the number of parallel Dirichlet
0-branes. At low energy, this theory is   just a dimensional reduction
of $D=10$ Yang-Mills theory to one dimension. Just as one quantizes
the  point particle to discover that it corresponds to the
Klein-Gordon equation in quantum field theory and  quantizes a 1-brane
(i.e. string) to discover  its particle spectrum, one must
quantize this supersymmetric
 generalization of non-abelian Born-Infeld theory to discover the
states in the IIA string arising from the presence of the Dirichlet
0-branes. It is thought that this quantum mechanical system does
indeed have bound states that have the correct mass and charge to be
identified with the Kaluza-Klein states [182].

\par
The last similarity between the maximal supergravity theories
is the equivalence of the IIA and IIB supergravity theories when
compactified on  a circle [107]. Hence, if one compactifies the  IIA
and IIB string theories on circles of radius $R_A$ and $R_B$
respectively  one obtains string theories
in nine dimensions which
have the same low effective action, since the maximal supergravity
theory in this dimension is unique. This suggests that these two
 string theories in nine dimensions are really the same
theory. However, the variables in which the two theories are found
after compactification may be related in a non-trivial way.
In fact, the two IIA and IIB
string theories compactified  on circles of radius $R_A$ and $R_B$
respectively   are the same theory, but they are related by a
T-duality transformation such that $R_A\to {\alpha'\over R_B}$.
[186-188]. In the limit  $R_A\to \infty $ the theory decompactifies
and one recovers the IIA string in ten dimensions.
Just as for the reduction of the IIA theory, the radius of the circle
is related to a scalar field in the theory
in nine dimensions that
appears from the metric in ten dimensions and as a result each
value of the radius corresponds to  a point in the moduli space of the
 theory in nine dimensions . The limit in which
$R_A\to
\infty
$ is just a limit to a point in moduli space. Similarly, in the limit
$R_B\to \infty$, which is also the limit $R_A\to 0$, we recover the
IIB string theory in ten dimensions.
Hence, different limits in the moduli space of the nine
dimensional theory  lead to different theories in ten dimensions.
Starting with say the IIA theory in ten dimensions, one can compactify
on a circle, carry out a T-duality transformation and then take the
appropriate limit to recover  the IIB theory in ten dimensions.
Because of the need to take limits one cannot, in general, simply
carry out the T-dualtiy  directly in the ten-dimensional theories.
\par
Of course the IIA supergravity theory in nine dimensions can be
obtained from eleven-dimensional supergravity by compactification on
a torus and so one may expect that the IIA string in nine dimensions
can be obtained from compactification of M theory on a torus.
The resulting nine-dimensional theories will inherit the isometries
of the torus as symmetries. In fact it can be shown that
the $SL(2,{\bf Z})$ which acts on the basis vectors of lattice which
underlies the torus, can be identified with the $SL(2,{\bf Z})$
of the IIB theory after it has been compactified on a circle.
\par
The relations between M theory and the IIA and IIB  string theories
in ten dimensions and the one string theory in nine dimensions
discussed above  are  summarized in figure 8.2. The similarity, for
the reasons explained above,   between this table and talbe 8.1 for
the  corresponding supergravity theories  is obvious. It
is relatively straightforward, using similar arguments,  to
incorporate the heterotic and type I string  theories into these
pictures. We refer to the lectures of Ashoke Sen contained in this
volume for a more complete treatment of string duality.
\bigskip
$$\epsffile{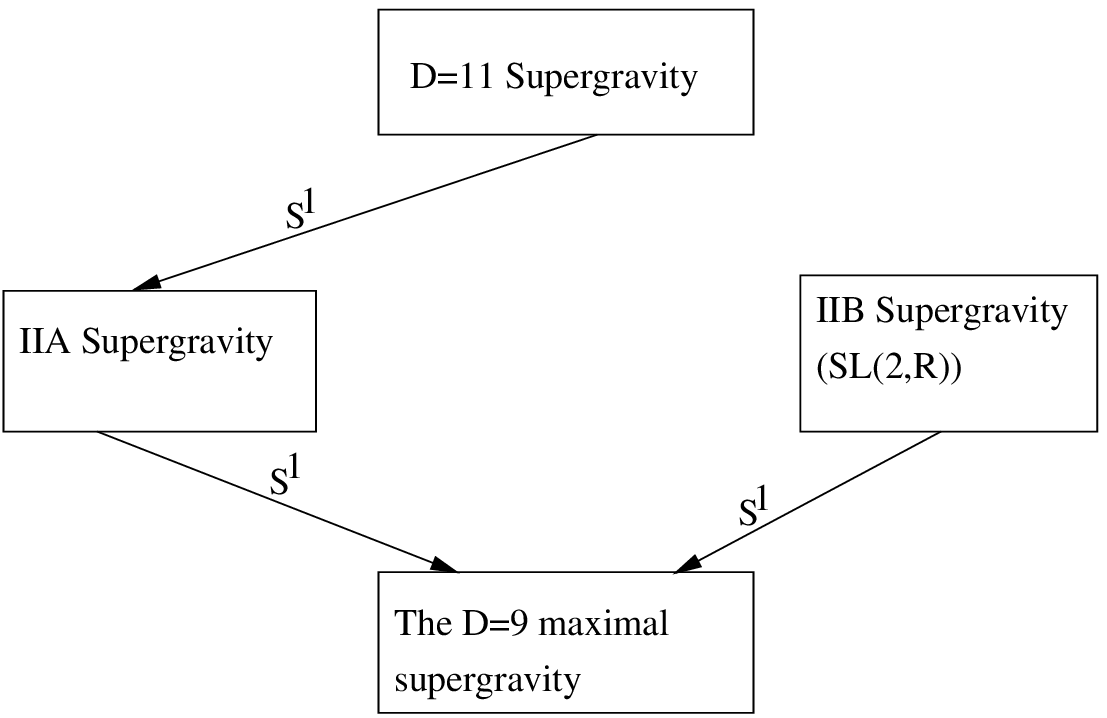}$$
\centerline {\bf {Relations between Maximal Supergravities}}
\bigskip
$$\epsffile{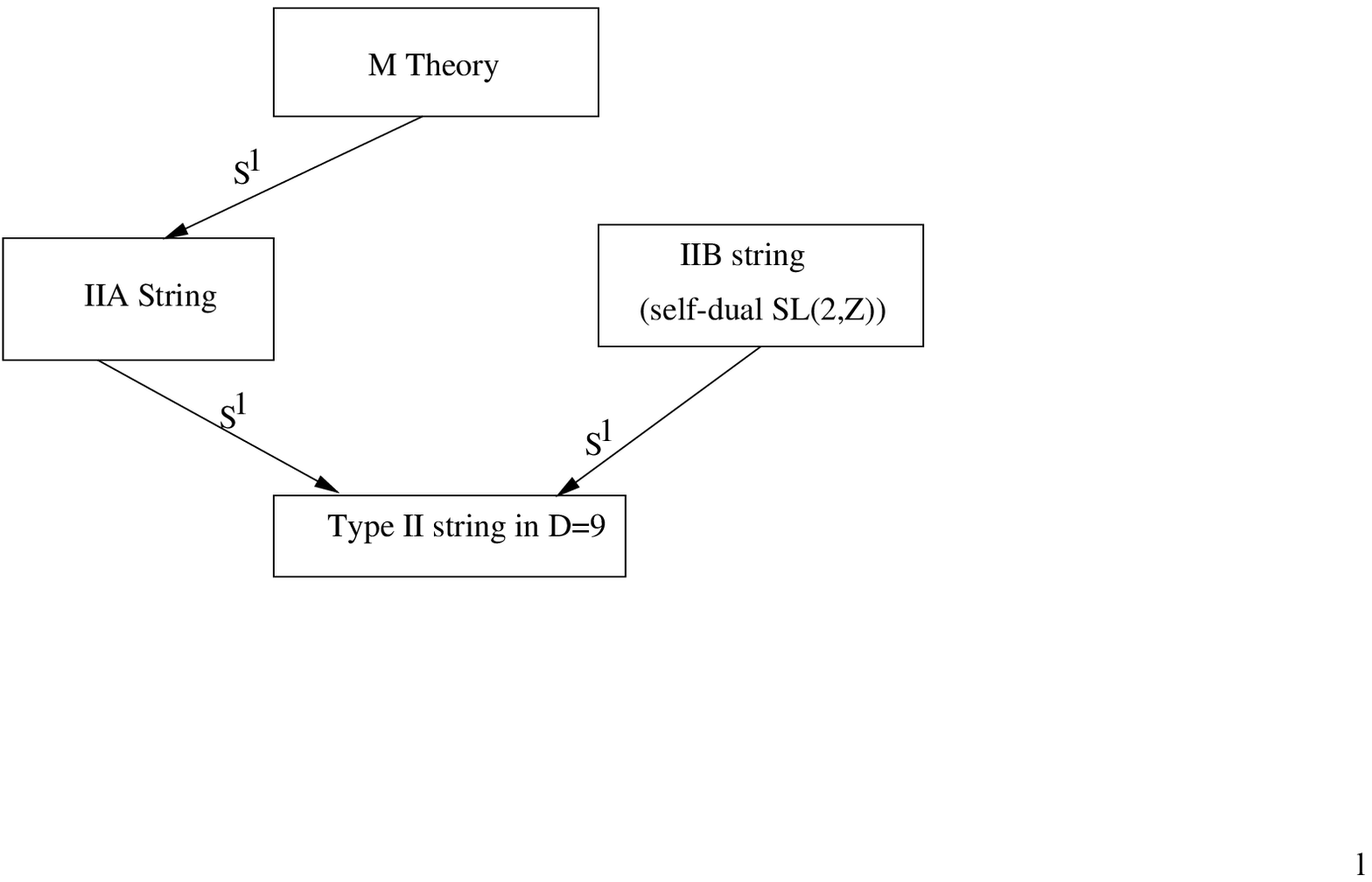}$$
\centerline {\bf {Relations between the IIA and
IIB string theories and M Theory}}

\bigskip
\centerline{\bf Acknowledgement}
\medskip
I wish to thank World Scientific Publishing for their kind
premission to reproduce some of the material from reference [0],
Neil Lambert for suggesting many  improvements, David Olive for
discussions  and Rachel George for help in preparing the manuscript.
\bigskip
\centerline {\bf {References}}
\medskip
\item {[0]} P. West, {\it Introduction to Supersymmetry and
Supergravity}, (1990),
 Extended and Revised Second Edition,  World Scientific Publishing.
\item {[1]} Y.A. Golfand and E.S. Likhtman, {\it JETP Lett.}
{\bf13}, 323 (1971).
\item {[2]} D.V. Volkov and V.P. Akulov, {\it Pis'ma Zh. Eksp.
Teor. Fiz.} {\bf16}, 621 (1972);
{\it Phys. Lett.} {\bf46B}, 109 (1973).
\item {[3]} J. Wess and B. Zumino, {\it Nucl. Phys.} {\bf B70},
139 (1974).
\item {[4]} S. Coleman and J. Mandula, {\it Phys. Rev.} {\bf159},
1251 (1967).
\item {[5]} R. Hagg, J. Lopuszanski and M. Sohnius, {\it Nucl.
Phys.} {\bf B88}, 61 (1975).
\item {[6]} P. van Nieuwenhuizen and P. West, {\it Principles of
Supersymmetry and Supergravity},
forthcoming book to be published by Cambridge University Press.
\item {[7]} P. Ramond, {\it Phys. Rev.} {\bf D3}, 2415 (1971); A.
Neveu and J.H. Schwarz, {\it Nucl. Phys.} {\bf B31}, 86 (1971); {\it
Phys. Rev.} {\bf D4}, 1109 (1971); J.-L. Gervais and B. Sakita, {\it
Nucl. Phys.} {\bf B34}, 477, 632 (1971); F. Gliozzi, J. Scherk and D.I.
Olive, {\it Nucl. Phys.} {\bf B122}, 253 (1977).
\item {[8]} J. Wess and B. Zumino, {\it Nucl. Phys.} {\bf B78}, 1
(1974).
\item {[9]} For a discussion of the Noether procedure in the
context of supergravity, see:
S. Ferrara, D.Z. Freedman and P. van Nieuwenhuizen, {\it Phys. Rev.}
{\bf D13}, 3214 (1976).
\item {[10]} S. Ferrara and B. Zumino, {\it Nucl. Phys.} {\bf
B79}, 413 (1974); A. Salam  and J. Strathdee, {\it Phys. Rev.} {\bf
Dll}, 1521 (1975).
\item {[11]} A. Salam and J. Strathdee,
{\it Nucl. Phys.} {\bf B80}, 499 (1974); M. Gell-Mann and Y. Neeman,
(1974) unpublished; W. Nahm, {\it Nucl. Phys.} {\bf B135}, 149
(1978). For a review, see: D.Z. Freedman in {\it Recent Developments
in Gravitation}, Carg\`ese (1978), eds. M. Levy and S. Deser (Gordon
and Breach, New York, 1979); S.  Ferrara and C. Savoy in {\it
Supergravity '81}, eds. S. Ferrara and J. Taylor (Cambridge University
Press, Cambridge, 1982).
\item {[12]} E.P. Wigner, {\it Ann.
of Math.} {\bf 40}, 149 (1939).
\item {[13]} P. van Nieuwenhuizen, {\it Phys. Rep.} {\bf 68}, 189
(198 1).
\item {[14]} D. Freedman, P. van Nieuwenhuizen and S. Ferrara,
{\it Phys. Rev.} {\bf D13}, 3214 (1976); {\it Phys. Rev.} {\bf D14},
912 (1976).
\item {[15]} S. Deser and B. Zumino, {\it Phys. Lett.} {\bf 62B},
335 (1976).
\item {[16]} K. Stelle and P. West, {\it Phys. Lett.} {\bf B74},
330 (1978).
\item {[17]} S. Ferrara and P. van Nieuwenhuizen, {\it Phys.
Lett.} {\bf B74}, 333 (1978).
\item {[18]} A. Chamseddine and P. West, {\it Nucl. Phys.} {\bf
B129}, 39 (1977).
\item {[19]} P. Townsend and P. van Nieuwenhuizen, {\it Phys.
Lett.} {\bf B67}, 439 (1977).
\item {[20]} J. Wess and B. Zumino, {\it Nucl. Phys.} {\bf B78}, 1
(1974).
\item {[21]} S. Ferrara in {\it Proceedings of the 9th
International Conference on General Relativity and Gravitation}
(1980), ed. Ernst Schmutzer.
\item {[22]} M. Sohnius, K. Stelle and P. West, in {\it Superspace
and Supergravity}, eds. S.W. Hawking and M. Rocek (Cambridge University
Press, Cambridge, 1981).
\item {[23]} A. Salam and J. Strathdee, {\it Phys. Lett.}
{\bf51B}, 353 (1974); P. Fayet, {\it Nucl. Phys.} {\bf B113}, 135
(1976).
\item {[24]} M. Sohnius, K. Stelle and P. West, {\it
Nucl. Phys.} {\bf B17}, 727 (1980); {\it Phys. Lett.} {\bf 92B}, 123
(1980).
\item {[25]} P. Fayet, {\it Nucl. Phys.} {\bf B113},
135 (1976).
\item {[26]} P. Breitenlohner and M. Sohnius, {\it Nucl. Phys.}
{\bf B178}, 151 (1981); M. Sohnius, K. Stelle and P. West, in {\it
Superspace and Supergravity}, eds. S.W. Hawking and M. Rocek
(Cambridge University Press, Cambridge, 1981). \item
{[27]} P. Howe and P. West, {\it N=1, d=6 Harmonic Superspace},
in preparation.
\item {[28]} G. Sierra and P.K. Townsend, {\it Nucl. Phys.} {\bf
B233}, 289 (1984); L. Mezincescu and Y.P. Yao, {\it Nucl. Phys.} {\bf
B241}, 605 (1984).
\item {[29]} F. Gliozzi, J. Scherk and D. Olive, {\it Nucl. Phys.}
{\bf B122}, 253 (1977); L. Brink, J. Schwarz and J. Scherk, {\it Nucl.
Phys.} {\bf B121}, 77 (1977).
\item {[30]} In this context, see: M. Rocek and W. Siegel, {\it
Phys. Lett.} {\bf 105B}, 275 (1981); V.0. Rivelles and J.G. Taylor,
{\it J. Phys. A. Math. Gen.} {\bf15}, 163 (1982).
\item {[31]} S. Ferrara, J. Scherk and P. van Nieuwenhuizen, {\it
Phys. Rev. Lett.} {\bf 37}, 1035 (1976); S. Ferrara, F. Gliozzi, J.
Scherk and P. van Nieuwenhuizen, {\it Nucl. Phys.} {\bf B117}, 333
(1976); P. Breitenlohner, S. Ferrara, D.Z. Freedman, F. Gliozzi, J.
Scherk and P. van Nieuwenhuizen, {\it Phys. Rev.} {\bf D15}, 1013
(1977); D.Z. Freedman, {\it Phys. Rev.} {\bf D15}, 1173 (1977).
\item {[32]} S. Ferrara and P. van Nieuwenhuizen, {\it Phys.
Lett.} {\bf 76B}, 404 (1978).
\item {[33]} K.S. Stelle and P. West, {\it Phys. Lett.} {\bf 77B},
376 (1978).
\item {[34]} S. Ferrara and P. van Nieuwenhuizen, {\it Phys.
Lett.} {\bf 78B}, 573 (1978).
\item {[35]} K.S. Stelle and P. West, {\it Nucl. Phys.} {\bf
B145}, 175 (1978).
\item {[36]} M. Sohnius and P. West, {\it Nucl. Phys.} {\bf B203},
179 (1982).
\item {[37]} R. Barbieri, S. Ferrara, D. Nanopoulos and K. Stelle,
{\it Phys. Lett.} {\bf 113B}, 219 (1982).
\item {[38]} E. Cremmer, S. Ferrara, B. Julia, J. Scherk and L.
Girardello, {\it Phys. Lett.} {\bf 76B}, 231 (1978).
\item {[39]} E. Cremmer, B. Julia, J. Scherk, S. Ferrara, L.
Girardello and P. van Nieuwenhuizen, {\it Nucl. Phys.} {\bf B147}, 105
(1979).
\item {[40]} E. Cremmer, S. Ferrara, L. Girardello and A. Van
Proeyen, {\it Nucl. Phys.} {\bf B212}, 413 (1983); {\it Phys. Lett.}
{\bf 116B}, 231 (1982).
\item {[41]} S. Deser, J. Kay and K. Stelle, Phys. Rev. Lett. 38,
527 (1977); S. Ferrara and B.  Zumino, {\it Nucl. Phys.} {\bf B134},
301 (1978).
\item {[42]} M. Sohnius and P. West, {\it Nucl. Phys.} {\bf B198},
493 (1982).
\item {[43]} S. Ferrara, L. Girardello, T. Kugo and A. Van
Proeyen, {\it Nucl. Phys.} {\bf B223}, 191 (1983).
\item {[44]} S. Ferrara, M. Grisaru and P. van Nieuwenhuizen, {\it
Nucl. Phys.} {\bf B138}, 430 (1978).
\item {[45]} B. de Wit, J.W. van Holten and A. Van Proeyen, {\it
Nucl. Phys.} {\bf B184}, 77 (1981); {\it Phys. Lett.} {\bf 95B}, 51
(1980); {\it Nucl. Phys.} {\bf B167}, 186 (1980).
\item {[46]} A. Salam and J. Strathdee, {\it Phys. Rev.} {\bf
Dll}, 1521 (1975); {\it Nucl. Phys.}{\bf B86}, 142 (1975).
\item {[47]} W. Siegel, {\it Phys. Lett.} {\bf 85B}, 333 (1979).
\item {[48]} R. Arnowitt and P. Nath, {\it Phys. Lett.} {\bf 56B},
117 (1975); L. Brink, M. Gell-Mann, P. Ramond and J. Schwarz, {\it
Phys. Lett.} {\bf 74B}, 336 (1978); {\bf 76B}, 417 (1978);
S. Ferrara and P. van Nieuwenhuizen, {\it Ann. Phys.} {\bf126}, 111
(1980); P. van Nieuwenhuizen and P. West, {\it Nucl. Phys.} {\bf
B169}, 501 (1980).
\item {[49]} M. Sohnius, {\it Nucl. Phys.} {\bf B165}, 483 (1980).
\item {[50]} P. Howe, K. Stelle and P. Townsend, {\it Nucl. Phys.}
{\bf B214}, 519 (1983).
\item {[51]} M. Grisaru, M. Rocek and W. Siegel, {\it Nucl. Phys.}
{\bf B159}, 429 (1979).
\item {[52]} E. Berezin, {\it The Method of Second Quantization}
(Academic Press, New York, 1960).
\item {[53]} A. Salam and J. Strathdee, {\it Nucl. Phys.} {\bf
B76}, 477 (1974); S. Ferrara, J. Wess and B. Zumino, {\it Phys. Lett.}
{\bf 51B}, 239 (1974).
\item {[54]} M. Grisaru, M. Rocek and W. Siegel, {\it Nucl. Phys.}
{\bf B159}, 429 (1979).
\item {[55]} J. Wess, Lecture Notes in Physics 77 (Springer,
Berlin, 1978).
\item {[56]} J. Gates and W. Siegel, {\it Nucl. Phys.} {\bf B147},
77 (1979).
\item {[57]} J. Gates, K. Stelle and P. West, {\it Nucl. Phys.}
{\bf B169}, 347 (1980).
\item {[58]} R. Grimm, M. Sohnius and J. Wess, {\it Nucl. Phys.}
{\bf B133}, 275 (1978).
\item {[59]} P. Breitenlohner and M. Sohnius, {\it Nucl. Phys.}
{\bf B178}, 151 (1981).
\item {[60]} P. Howe, K. Stelle and P. West, {\it Phys. Lett.}
{\bf 124B}, 55 (1983).
\item {[61]} P. Howe, K. Stelle and P. West, {\it $N=1, \  d=6$
Harmonic Superspace}, Kings College preprint.
\item {[62]} M. Sohnius, K. Stelle and P. West, in {\it Superspace
and Supergravity}, eds. S.W. Hawking and M. Rocek (Cambridge University
Press, Cambridge, 1981).
\item {[63]} A. Galperin, E. Ivanov, S. Kalitzin, V. Ogievetsky
and E. Sokatchev, Trieste preprint.
\item {[64]} L. Mezincescu, JINR report P2-12572 (1979).
\item {[65]} J. Koller, {\it Nucl. Phys.} {\bf B222}, 319
(1983); {\it Phys. Lett.} {\bf 124B}, 324 (1983).
\item {[66]} P. Howe, K. Stelle and P.K. Townsend, {\it Nucl.
Phys.} {\bf B236}, 125 (1984).
\item {[67]} A. Salam and J. Strathdee, {\it Nucl. Phys.} {\bf
B80}, 499 (1974).
\item {[68]} S. Ferrara, J. Wess and B. Zumino, {\it Phys. Lett.}
{\bf 51B}, 239 (1974).
\item {[69]} J. Wess and B. Zumino, {\it Phys. Lett.} {\bf 66B},
361 (1977); V.P. Akulov, D.V. Volkov and V.A. Soroka, {\it JETP Lett.}
{\bf 22}, 187 (1975).
\item {[70]} R. Arnowitt, P. Nath and B. Zumino, {\it Phys. Lett.}
{\bf 56}, 81 (1975); P. Nath and R. Arnowitt, {\it Phys. Lett.} {\bf
56B}, 177 (1975); {\bf 78B}, 581 (1978).
\item {[71]} N. Dragon, {\bf Z. Phys.} {\bf C2}, 62 (1979).
\item {[72]} E.A. Ivanov and A.S. Sorin, {\it J. Phys. A.
Math. Gen} {\bf 13}, 1159 (1980).
\item {[73]} That some representations do not generalize to
supergravity was noticed in: M. Fischler, {\it Phys. Rev.} {\bf D20},
1842 (1979).
\item {[74]} P. Howe and R. Tucker, {\it Phys. Lett.} {\bf 80B},
138 (1978).
\item {[75]} P. Breitenlohner, {\it Phys. Lett.} {\bf 76B}, 49
(1977); {\bf 80B}, 217 (1979).
\item {[76]} W. Siegel, {\it Phys. Lett.} {\bf 80B}, 224 (1979).
\item {[77]} W. Siegel, {\it Supergravity Superfields Without a
Supermetric}, Harvard preprint HUTP-771 A068, {\it Nucl. Phys.} {\bf
B142}, 301 (1978); S.J. Gates Jr. and W. Siegel, {\it Nucl. Phys.}
{\bf B147}, 77 (1979).
\item {[78]} See also in this context: V. Ogievetsky and E.
Sokatchev, {\it Phys. Lett.} {\bf 79B}, 222 (1978).
\item {[79]} R. Grimm, J. Wess and B. Zumino, {\it Nucl. Phys.}
{\bf B152}, 1255 (1979).
\item {[80]} These constraints were first given by: J. Wess and B.
Zumino, {\it Phys. Lett.} {\bf 66B}, 361 (1977).
\item {[81]} J. Wess and B. Zumino, {\it Phys. Lett.} {\bf 79B},
394 (1978).
\item {[82]} P. Howe and P. West, {\it Nucl. Phys.} {\bf B238}, 81
(1983).
\item {[83]} P. Howe, {\it Nucl. Phys.} {\bf B199}, 309 (1982).
\item {[84]} A. Salam and J. Strathdee, {\it Phys. Rev.} {\bf
Dll}, 1521 (1975).
\item {[85]} S. Ferrara and 0. Piguet, {\it Nucl. Phys.} {\bf
B93}, 261 (1975).
\item {[86]} J. Wess and B. Zumino, {\it Phys. Lett.} {\bf 49B}, 52
(1974).
\item {[87]} J. lliopoulos and B. Zumino, {\it Nucl. Phys.} {\bf
B76}, 310 (1974).
\item {[88]} S. Ferrara, J. lliopoulos and B. Zumino, {\it Nucl.
Phys.} {\bf B77}, 41 (1974).
\item {[89]} D.M. Capper, {\it Nuovo Cim.} {\bf 25A}, 259 (1975);
R. Delbourgo, {\it Nuovo  Cim.} {\bf 25A}, 646 (1975).
\item {[90]} P. West, {\it Nucl. Phys.} {\bf B106}, 219 (1976); D.
Capper and M. Ramon Medrano, {\it J. Phys.} {\bf 62}, 269 (1976); S.
Weinberg, {\it Phys. Lett.} {\bf 62B}, 111 (1976).
\item {[91]} M. Grisaru, M. Rocek and W. Siegel, {\it Nucl. Phys.}
{\bf B159}, 429 (1979).
\item {[92]} B.W. Lee in {\it Methods in Field Theory}, Les
Houches 1975, eds. R. Balian and J. Zinn-Justin (North Holland,
Amsterdam and World Scientific, Singapore, 1981).
\item {[93]} W. Siegel, {\it Phys. Lett.} {\bf 84B}, 193 (1979);
{\it 94B}, 37 (1980).
\item {[94]} L.V. Avdeev, G.V. Ghochia and A.A. Vladiminov, {\it
Phys. Lett.} {\bf 105B}, 272 (1981); L.V. Avdeev and A.A. Vladiminov,
{\it Nucl. Phys.} {\bf B219}, 262 (1983).
\item {[95]} D.M. Capper, D.R.T. Jones and P. van Nieuwenhuizen,
{\it Nucl. Phys.} {\bf B167}, 479 (1980).
\item {[96]} G. 't Hooft and M. Veltman, {\it Nucl. Phys.} {\bf
B44}, 189 (1972); C. Bollini and J. Giambiagi, {\it Nuovo Cim.} {\bf
12B}, 20 (1972); J. Ashmore, {\it  Nuovo Cim. Lett.} {\bf 4}, 37
(1972).
\item {[97]} P. Howe, A. Parkes and P. West, {\it Phys. Lett.}
{\bf 147B}, 409 (1984); {\it Phys. Lett.} {\bf 150B}, 149 (1985).
 \item {[98]} J.W. Juer and D. Storey, {\it Nucl. Phys.} {\bf
B216}, 185 (1983); O. Piguet and K. Sibold, {\it Nucl. Phys.} {\bf
B248}, 301 (1984).
\item {[99]} E. Witten, {\it Nucl. Phys.} {\bf B188}, 52
(1981).
\item{[100]} F. Gliozzi, D. Olive and J. Scherk,
{\it ``Supersymmetry,
Supergravity Theories and the Dual Spinor Model''},
Nucl. Phys. B122
(1977) 253.
\item{[101]} P. van Niewenhuizen, "Six Lectures on Supergravity",
in supergravity '81, edited by S. Ferrara and J. Taylor,
Cambridge University Press.
\item{[102]} T. Kugo and P. Townsend, {\it Supersymmetry and the
Division Algebras},  Nucl. Phys. B221, (1983) 357.
\item{[103]}  Barut and Racka,
{\it Theory of Group Representations}, World Scientific Publishing.
\item{[104]} B. Zumino, Journ. Math. Phys. 3 (1962) 1055.
\item{[105]} W. Nahm,
{\it ``Supersymmetries and their representations''},
Nucl.\ Phys.\ {\bf B135} (1978) 149.
\item{[106]} E. Cremmer, B. Julia and J. Scherk, Phys. Lett. 76B
(1978) 409.
\item{[{107}]} C. Campbell and P. West,
{\it ``$N=2$ $D=10$ nonchiral
supergravity and its spontaneous compactification.''}
Nucl.\ Phys.\ {\bf B243} (1984) 112.
\item{[{108}]} M. Huq and M. Namazie,
{\it ``Kaluza--Klein supergravity in ten dimensions''},
Class.\ Q.\ Grav.\ {\bf 2} (1985).
\item{[{109}]} F. Giani and M. Pernici,
{\it ``$N=2$ supergravity in ten dimensions''},
Phys.\ Rev.\ {\bf D30} (1984) 325.
\item{[{110}]} J, Schwarz and P. West,
{\it ``Symmetries and Transformation of Chiral
$N=2$ $D=10$ Supergravity''},
Phys. Lett. {\bf 126B} (1983) 301.
\item {[{111}]} P. Howe and P. West,
{\it ``The Complete $N=2$ $D=10$ Supergravity''},
Nucl.\ Phys.\ {\bf B238} (1984) 181.
\item {[{112}]} J. Schwarz,
{\it ``Covariant Field Equations of Chiral $N=2$
$D=10$ Supergravity''},
Nucl.\ Phys.\ {\bf B226} (1983) 269.
\item{[113]} S.W. MacDowell and F. Mansouri, Phys. Rev. Lett. 38
(1977) 739.
\item{[114]} T. W. B. Kibble, J. Maths. Phys. 2 (1961) 212.
\item{[115]} K. S. Stelle and P. West, {\it Spontaneously Broken de
Sitter Symmetry and the Gravitational Holonomy Group}, Phys. Rev. D21,
(1980) 1466.
\item{[116]} see for example M. Naimark and A. Stern,
Theory of Group representations, Springer-Verlag, 1982
\item{[117]} J. W. Van Holten and A. Van Proeyen, J. Phys. Gen.
(1982) 3763.
\item{[118]} M. Kaku, P. van Niewenhuizen and P. K.
Townsend,
{\it Properties of Conformal Supergravity}, Phys. Rev. D17 (1978)
3179.
\item{[119]} L. Brink, J. Scherk and J.H. Schwarz, {\it
``Supersymmetric Yang-Mills Theories''},
Nucl. Phys. {\bf B121} (1977) 77;
F. Gliozzi, J. Scherk and D. Olive, {\it ``Supersymmetry,
Supergravity Theories and the Dual Spinor Model''}, Nucl. Phys. {\bf
B122} (1977) 253.
\item{[120]} A.H. Chamseddine, {\it ``Interacting supergravity
in ten dimensions: the role of the six-index gauge field''},
Phys. Rev. {\bf D24} (1981) 3065;
E.\ Bergshoeff, M.\ de Roo, B.\ de Wit and P.\ van
Nieuwenhuizen, {\it ``Ten-dimensional Maxwell-Einstein
supergravity, its currents, and the issue of its auxiliary
fields''}, Nucl.\ Phys.\ {\bf B195} (1982) 97;
E.\ Bergshoeff, M.\ de Roo and B.\ de Wit, {\it ``Conformal
supergravity in ten dimensions''}, Nucl.\ Phys.\ {\bf B217} (1983)
143.
\item{[121]} G. Chapline and N.S. Manton,
{\it ``Unification of Yang-Mills
theory and supergravity in ten dimensions''}, Phys. Lett. {\bf
120B} (1983) 105.
\item{[122]} A.H.\ Chamseddine and P.C.\ West, {\it ``Supergravity as
a Gauge  Theory of Supersymmetry''}, Nucl.\ Phys.\ {\bf B129} (1977)
39.
\item{[123]} P.K. Townsend, {\it ``The eleven-th
dimensional supermembrane revisited''}, Phys.\ Lett.\ {\bf B350}
(1995) 184,{ hep-th/9501068}, {\it ``D-branes from  M-branes''},
Phys.\ Lett.\ {\bf B373}  (1996) 68, {hep-th/9512062
\item{[124]} E. Witten,
{\it ``String theory dynamics in various dimensions''},
Nucl.\ Phys.\ {\bf B443} (1995) 85, hep-th/9503124.
\item{[125]} M.B. Green and J.H. Schwarz,
{\it ``Superstring interactions''}, Phys. Lett. {\bf
122B} (1983) 143.
\item{[126]} P. West, {\it Representations of Supersymmetry}, in
Supergravity 81, 111 (1982) edited by S. Ferrara and J. Taylor.
\item{[127]} C. Callan, S. Coleman, J. Wess and B. Zumino, Phys. Rev.
177 (1969) 2247.
\item{[128]} N. Manton, Phys. Lett. 110B (1982) 54.
\item{[129]} M. Atiyah and N. Hitchen, {\it The geometry and
Dynamics of Magnetic Monopoles}, (1988) Princeton University Press.
\item{[130]} M. B. Green and J. H. Schwarz, Nucl. Phys. B243 (1984)
285.
\item{[131]} E. Bergshoff and E. Sezgin, {\it Twistor-like
Formulation of Super p-branes}, Nucl. Phys. B422 (1994) 329.
Hep-th/9312168,
E. Sezgin, {\it Space-time and worldvolume Supersymmetric Super
p-brane actions}, hep-th/9312168.
\item{[132]} I. Bandos, D. Sorokin and and D. Volkov, {\it
On the Generalized       Action Principle for Superstrings     and
Superbranes}, Phys. Lett. B352 (1995) 269, hep-th/9502141, I. Bandos,
P. Pasti, D. Sorokin,  M. Tonin  and D. Volkov, {\it Superstrings and
Supermembranes  in the Doubly Supersymmetric Geometrical Approach},
Nucl. Phys.  B446 (1995) 79, hep-th/9501113.
\item{[133]} P.S. Howe and E. Sezgin, {\it Superbranes}, hep-th/9607227
\item{[134]} P.S. Howe and E. Sezgin,{\it
D=11, p=5},Phys. Lett. {\bf B394} (1997) 62, hep-th/9611008
\item{[135]} J. Polchinski, Phys. Rev. Lett. 75 (1995) 4724.
\item{[136]} A. Tseytlin, {\it Self-duallity of the Born-Infeld
Action and the Dirichlet 3-brane of type IIB Superstring Theory},
Nucl. Phys. B469 (1996) 51;
M.B. Green and M. Gutperle, {\it Comment on Three-branes},
Phys. Lettl B377 (1996) 28.
\item{[137]} P.S. Howe, E. Sezgin and P.C. West, Phys. Lett.
{\it Covariant field
	  equations of the M theory five-brane},
{\bf B399} (1997) 49, hep-th/9702008.
\item{[138]} P.S. Howe, E. Sezgin and P.C. West, {\it The
Six-Dimensional  Self-Dual Tensor},
hep-th/9702111.
\item{[139]}  P.S. Howe, E. Sezgin,
and P.C. West, {\it Aspects of Superembeddings},
Contribution to the D.V. Volkov memorial volume, hep-th/9705093
\item{[140]} M.K.\ Gaillard and B.\ Zumino,
{\it Duality rotations for interacting fields},
Nucl.\ Phys.\ {\bf B193} (1981) 221
\item{[141]} I. Bandos, K Lechner, A. Nurmagambetov, P. Pasti and D.
Sorokin, and M.
Tonin, {\it Covariant action for the super fivebrane of M-theory},
hep-th/9701149.
\item{[142]} M. Perry and J.H. Schwarz, Nucl. Phys.
{\bf B489} (1997) 47,
hep-th/9611065; M. Aganagic, J. Park, C. Popescu, and J. H. Schwarz,
{\it Worldvolume action of the M-theory fivebrane},
hep-th/9701166.
\item{[143]} E. Witten, {\it The Five-Brane Effective
Action in M theory}, hep-th/9610234.
\item{[144]} P.S. Howe, N.D. Lambert and P.C. West, {\it The Selfdual
String Soliton}, hep-th/9709014.
\item{[145]} P.S. Howe, N.D. Lambert and P.C. West,
{\it The Threebrane Soliton
of the M-fivebrane}, hep-th/9710033.
\item{[146]} P.S. Howe, N.D. Lambert and P.C. West,
{\it Classical M-Fivebrane
Dynamics and Quantum $N=2$ Yang-Mills}, hep-th/9710034.
\item{[147]} M. Cederwall, A. Von Gussich, B.E.W. Nilsson and A.
Westerberg,  The
Dirichlet super-three-brane in ten-dimensional Type IIB supergravity, hep-
th/9610148.
\item{[148]} M. Aganagic, C. Popescu and J.H. Schwarz, D-brane actions with
local kappa
symmetry, hep-th/9610249.
\item{[149]} M. Cederwall, A. Von Gussich, B.E.W. Nilsson P. Sundell and A.
Westerberg,
The Dirichlet super p-branes in ten-dimensional Type IIA and IIB
supergravity, hep-
th/9611159.
\item{[150]} M. Aganagic, C. Popescu and J.H. Schwarz, Gauge-invariant and
gauge-fixed
D-brane actions, hep-th/9612080.
\item{[151]} E. Bergshoeff, E. Sezgin and P.K. Townsend, Properties of eleven
dimensional supermembrane theory, Ann. Phys. 185 (1988) 330.
\item{[152]} A. Font, L. Ibanez, D. Lust and FD. Quevedo, Phys. Lett. B249
(1990) 35.
\item{[153]} S.J. Rey, Phys. Rev. D43 (1991) 526.
\item{[154]} A. Sen, Phys. Lett. B303 (1993) 22; Int. J. Mod. Phys. A9
(1994) 3707.
\item{[155]} J. Schwarz and A. Sen, Nucl. Phys. B411 (1994) 35.
\item{[156]} 	C.M. Hull and P.K. Townsend,
{\it ``Unity of superstring  dualities''},
Nucl.\ Phys.\ {\bf B438} (1995) 109, hep-th/9410167.
\item{[157]} E. Bergshoeff, E. Sezgin and P.K. Townsend,
Phys. Lett. B189 (1987) 75.
\item{[158]} N. Seiberg and E. Witten, Nucl. Phys. {\bf B426} (1994) 19,
hep-th/9407087.
\item{[159]} E. Witten, Nucl. Phys. {\bf B500} (1997) 3,
hep-th/9703166.
\item{[160]} J. Strathdee, Int Journal of Modern Physics A,
Vol2, no 1
(1987) 273.
\item{[161]} K. Narain, H. Sarmardi and E. Witten, Nucl.
Phys. {\bf B279} (1987) 369.
\item{[162]} C. Montonen and D. Olive,
{\it ``Magnetic monopoles
as gauge particles?''},
Phys.\ Lett.\ {\bf 72B} (1977) 117.
\item{[163]} E. Witten and D. Olive,
{\it ``Supersymmetry algebras
that include topological charges''},
Phys.\ Lett.\ {\bf 78B} (1978) 97.
\item{[164]} E. Witten, {\it ``Bound states of strings and
$p$-branes''},  Nucl.\ Phys.\ {\bf B460} (1996) 335, hep-th/9510135.
\item{[165]} For a review see, J.A. Harvey
{\it ``Magnetic monopoles, duality and supersymmetry''},
hep-th/9603086.
\item{[166]} M. Green, J. Schwarz and E. Witten,
{\it ``Superstring theory''},
Vol.~1\&2, Cambridge University Press,  (1987).
\item{[167]} B.\ Julia, {\it ``Group Disintegrations''},
in {\it Superspace \&
Supergravity}, p.\ 331,  eds.\ S.W.\ Hawking  and M.\ Ro\v{c}ek,
Cambridge University Press (1981).
\item{[168]} S.\ Ferrara, J.\ Scherk and B.\ Zumino, {\it
``Algebraic Properties of Extended Supersymmetry''},
Nucl.\ Phys.\ {\bf B121} (1977) 393;
E.\ Cremmer, J.\ Scherk and S.\ Ferrara, {\it ``SU(4) Invariant
Supergravity Theory''}, Phys.\ Lett.\ {\bf 74B} (1978) 61;
B.\ de Wit,  {\it ``Properties of SO(8)-extended supergravity''},
Nucl.\ Phys.\ {\bf B158} (1979) 189;
B.\ de Wit and H.\ Nicolai, {\it ``N=8 Supergravity''}, Nucl.\ Phys.\
{\bf B208} (1982) 323.
\item{[169]} A.A. Tseytlin, {\it On the dilaton dependence of
Type II Superstring Action} hep-th/9601109.
\item{[170]} E.\ Cremmer, J.\ Scherk and S.\ Ferrara, {\it ``SU(4) Invariant
Supergravity Theory''}, Phys.\ Lett.\ {\bf 74B} (1978) 61.
\item{[171]} E. Cremmer and B. Julia,
{\it ``The $N=8$ supergravity theory. I. The Lagrangian''},
Phys.\ Lett.\ {\bf 80B} (1978) 48.
\item{[172]}  P. Ramond, {\it Dual theory for free fermions}, Phys.
Rev. {\bf	D3} (1971) 2415;
 A. Neveu and J.H. Schwarz, {\it
Factorizable dual model of
	pions}, Nucl. Phys. {\bf B31} (1971) 86.
\item{[173]} A. Achucarro, J.M. Evans, P. K. Townsend and
D. L. Wiltshire, {\it Super p-Branes} Phys. Lett. 198 (1987) 441.
\item{[174]} L. Romans, Phys. Lett. {\bf B169} (1986) 374}.
\item{[175]} M. J. Duff, P. S. Howe, T. Inami and K. S. Stelle, {\it
Superstrings in $D=10$ from Supermembranes in $D=11$.}, Phys. Lett.
191B (1987) 70.
\item{[176]} C.G. Callan, J.A. Harvey and A. Strominger,
{\it ``Supersymmetric string solitons''}, hep-th/9112030;
M.J. Duff, R.R. Khuri and J.X. L\"u,
{\it ``String solitons''},
Phys.\ Rept.\ {\bf 259} (1995) 213, hep-th/9412184;
M. Duff,
{\it ``Supermembranes''}, hep-th/9611203;
K.S. Stelle,
{\it ``Lectures on Supergravity p-Branes''},
lectures given at the 1996 ICTP Summer School in High Energy
Physics and Cosmology, Trieste, hep-th/9701088.
\item{[177]} N.D. Lambert and P.C. West, {\it Gauge Fields and M-Fivebrane
Dynamics}, hep-th/9712040.
\item{[178]} D. Sorokin, V. Tkach and D.V. Volkov, {\it Superparticles,
	twistors and Siegel symmetry}, Mod. Phys. Lett. {\bf A4} (1989) 901;
 D. Sorokin, V. Tkach, D.V. Volkov and A. Zheltukhin, {\it From
	superparticle Siegel supersymmetry to the spinning particle
proper-time
	supersymmetry}, Phys. Lett. {\bf B259} (1989) 302.
\item {[179]}  W. Siegel, {\it Hidden local supersymmetry in the
	supersymmetric particle action}, Phys. Lett. {\bf 128B} (1983) 397.
\item{[180]} A de Azcarraga, J. P. Gauntlett, J.M. Izquierdo and P.K.
Townsend, {\it Topological Extensions of the Supersymmetry Algebra
for Extended Objects}, 63 (1989) 2443.
\item{[181]} E.\ Cremmer, {\it Supergravities in 5
dimensions}, in {\it Superspace \& Supergravity}, p.\ 267, eds.\
S.W.\ Hawking  and M.\ Ro\v{c}ek, Cambridge University Press
(1981).
\item{[182]} S. Sethi and M. Stern, hep-th/9705046.
\item{[183]} T. H. Buscher, Phys. Lett. B194 (1987) 51, B201 (1988)
466.
\item{[184]} M. Rocek and E. Verlinde, Nucl. Phys. B373 (1992) 630
\item{[185]} E. Bergshoff, M. de Roo, M. B. Green, G. Papadopoulos and
P. K. Townsend, {\it Duality of Type II 7-branes and 8-branes},
Hep-th/9601150
\item{[186]} M. Dine, P. Huet and N. Seiberg,
{\it ``Large and small radius in string theory''},
Nucl.\ Phys.\ {\bf B322} (1989) 301.
\item{[187]} J. Dai, R.G. Leigh, J. Polchinski,
{\it ``New connections between string theories''},
Mod.\ Phys.\ Lett.\ {\bf A4} (1989) 2073.
\item {[188]} E. Bergshoeff, C. Hull and T. Ortin,
{\it  ``Duality in the type-II
superstring effective action''},
Nucl.\ Phys.\ {\bf B451} (1995) 547,
hep-th/9504081.
\item {[189]} P.K.\ Townsend and P.\ van Nieuwenhuizen, {\it
``Geometrical Interpretation of Extended Supergravity''}, Phys.\
Lett.\ {\bf 67B} (1977) 439
\item {[190]} K.S.~Narain,
{\it ``New heterotic string theories in
uncompactified dimensions $< 10$''},
Phys.\ Lett.\ {\bf 169B} (1986) 41;
K.S.~Narain, M.H. Samadi and E. Witten,
{\it ``A note on toroidal compactification of
heterotic string theory''},
Nucl.\ Phys.\ {\bf B279} (1987) 369.
\item {[191]} R. Nepomechie , Phys. Rev. D31, (1984) 1921;
C. Teitelboim, Phys. Lett. B167 (1986) 69.
\item {[192]} K. Kikkawa and M. Yamasaki, Phys. Lett. B149 (1984)
357.N. Sakai and I. Senda, Prog. Theor. Phys. 75 (1986) 692.
\item {[193]} A. Sen, Phys. Lett. B329 (1994) 217.
\item {[194]} H. Braden, "$N$-dimensional spinors: their properties
in terms of finite groups", Journal of Mathematical Physics 26
(1985) 613
\item {[195]} P.S. Howe and P.K. Townsend, {\it The massless
superparticle as
	Chern-Simons mechanics}, Phys. Lett. {\bf B259} (1991) 285;
 F. Delduc and E. Sokatchev, {\it Superparticle with extended
	worldline supersymmetry}, Class. Quantum Grav. {\bf 9} (1992) 361;
 A. Galperin, P. Howe and K. Stelle, {\it The superparticle and
	the Lorentz group}, Nucl. Phys. {\bf B368} (1992) 248.
 A. Galperin and E. Sokatchev, {\it A twistor-like D=10
	superparticle action with manifest N=8 world-line supersymmetry},
Phys. Rev. {\bf D46} (1992) 714, hep-th/9203051.
\item{[196]} F. Delduc, A. Galperin,
P.S. Howe and E. Sokatchev, {\it A
	twistor formulation of the heterotic D=10 superstring with manifest
	(8,0) worldsheet supersymmetry}, Phys. Rev. {\bf D47} (1993) 578,
	hep-th/9207050;  F. Delduc, E. Ivanov and E. Sokatchev, {\it
Twistor-like
	superstrings with $D = 3, 4, 6$ target-superspace and
	$N = (1,0), (2,0), (4,0)$ world-sheet supersymmetry},
	Nucl. Phys. {\bf B384} (1992) 334, hep-th/9204071.
\item {[197]} P. Pasti and M. Tonin, {\it Twistor-like Formulation of
the Supermembrane in D=11}, Nucl. Phys. {\bf B418} (1994) 337,
	hep-th/9303156.
\item {[198]} I. Bandos, {\it Generalized action principle and
geometrical
	approach for superstrings and super p-branes}, hep-th/9608094;
I. Bandos, P. Pasti, D. Sorokin and M. Tonin, {\it Superbrane
	actions and geometrical approach}, hep-th/9705064;
 I. Bandos, D.
Sorokin and M. Tonin, {\it Generalized action
	principle and superfield equations of motion for $D=10$ D$p$-branes},
	Nucl. Phys. {\bf B497} (1997) 275, hep-th/9701127.

\end